%% file: ms.tex
\documentclass[useAMS]{mn2e}
\usepackage{amsmath,amssymb,graphicx,deluxetable,soul,color}
\usepackage{subfigure,rotating,longtable}
\input{def}
\title[A study of X-ray flares II.]{A study of X-ray flares -II.  RS CVn type Binaries}
\author[Pandey \& Singh]{J. C. Pandey$^1$\thanks{jeewan@aries.res.in} and K. P. Singh$^2$\\
$^{1}$Aryabhatta Research Institute of Observational Sciences , Nainital - 263 129, India\\
$^{2}$Tata Institute of Fundamental Research, Mumbai - 400 005, India}

\date{Accepted 2011 September 6.  Received 2011 September 6; in original form 2011 January 11}

\pagerange{\pageref{firstpage}--\pageref{lastpage}} \pubyear{2011}

\def\LaTeX{L\kern-.36em\raise.3ex\hbox{a}\kern-.15em
    T\kern-.1667em\lower.7ex\hbox{E}\kern-.125emX}

\begin{document}

\maketitle

\begin{abstract}
	We present an analysis of seven flares detected from five RS CVn-type binaries (UZ Lib, $\sigma$ Gem, $\lambda$ And, V711 Tau and EI Eri) observed with XMM-Newton observatory. The quiescent state X-ray luminosities in the energy band of 0.3-10.0 keV of these stars were found to be $10^{30.7-30.9}$ \lum. The exponential decay time in all the sample of flares range from $\sim 1$ to 8 hrs. The luminosity at peak  of the flares in the energy band of 0.3-10.0 keV were found to be in the range of $10^{30.8} - 10^{31.8}$ \lum. The great sensitivity of the XMM-EPIC instruments allowed us to perform time resolved  spectral analysis during the flares and also in the subsquent quiescent phases. The derived metal abundances of coronal plasma  were found to vary during the flares observed from  $\sigma$ Gem, V771 Tau and EI Eri. In these flares elemental abundances found to be enhanced by factors of $\sim 1.3-1.5$ to the quiescent states. In most of the flares, the peak temperature was  found to be more than 100 MK whereas emission measure increased by factors of 1.5 - 5.5. Significant sustained heating was present in the majority of flares.  The loop lengths ($L$) derived for flaring structure were found to be of the order of $10^{10 -11}$ cm and are  smaller than the stellar radii (\rstar) i.e. L/\rstar $\lesssim 1$. The flare from $\sigma$ Gem showed  a high and variable absorption column density during the flare.

\end{abstract}


\section{Introduction}
\label{sec:intro}
Solar/stellar flares are known to be a manifestation of the reconnection of magnetic loops, accompanied by particle beams, chromospheric evaporation, rapid bulk flows or mass ejection, and heating of plasma confined in loops (e.g. Dennis \& Schwartz 1989; Martens \& Kuin 1989). Observationally, flares reveal themselves across the electromagnetic spectrum, usually characterized by a  rapid increase of radiation reaching a peak value that may occur at somewhat different times in different wavelength bands (e.g. Osten et al. 2000), followed by a gradual decay.  Solar X-ray flare classification schemes (Pallavicini et al. 1977) distinguish between `compact' flares in which a small number of magnetic loops lighten up on a time scale of minutes, and `two-Ribbon' flares evolving on a time scale of  several hours. The latter class involves complex loop arcades anchored in two roughly parallel chromospheric H$\alpha$ ribbons. These ribbons define the footpoints of the loop arcades. Such flares are energized by continuous reconnections of initially open magnetic fields above a neutral line at progressively larger heights, so that nested magnetic loops lighten up sequentially, and possibly also at different times along the entire arcade. The largest solar flares are usually of this type. Analogy of solar flare classification schemes to the stellar case is problematic as most stellar flares radiate several orders of magnitude more energy than a solar flare (G\"{u}del \& Naze 2009). However, a comprehensive  EXOSAT survey of stellar flares by Pallavicini et al. (1990) revealed examples of two different type of flares: (1) impulsive flares which are like compact solar flares, and (2) long decay flares which are like two-ribbon solar flares.  The impulsive flares have total energy release of $10^{30}$ \lum, and typically last for less than an hour. These are very likely to be confined to a single loop.  The long decay flares are much more energetic ($10^{32}$ \lum), of longer durations ($\geq 1$ h) and are believed to release their energy in an arcade of loops.

The flares produced by stars in  RS CVn-like binaries are usually detected only in the UV or X-rays. These UV and X-ray flares show extreme luminosities and very hot temperatures ($\gtrsim 10 $MK).  Flares from these stars present many analogies with the solar flares. However, they also show significant differences, such as the amount of energy released.
Modeling the dynamic behavior of X-ray flares allows us to constrain the properties of magnetic loops in ways that cannot be done from analysis of quiescent coronae that provide only a spatial and temporal average over some ensemble of structures. Therefore, study of stellar flares is a valuable tool for understanding stellar coronae, as these are dynamical events, which therefore embody different information than available from static, time-averaged observations (e.g. Pandey \& Srivastava 2009; Pandey \& Singh 2008;  Favata et al. 2005; Reale et al. 2004).
Analysis of flares in cool giants and subgiants gives us an opportunity to probe the structure of corona, which in principle can be significantly different from the dwarf star coronae.  In the cool giants and subgiants the gravity is considerably lower than for the cool main-sequence stars, therefore, yielding a larger scale height, and possibly allowing very extended coronae to develop (Ayres et al. 2003).

In a previous paper (Pandey \& Singh 2008), we have examined the characteristics of X-ray flares in  G-K dwarfs. The flares in these G-K dwarfs are similar to the solar arcade flares, which are as energetic as M dwarfs but are much smaller than the flares observed in dMe star, giants and pre-main-sequence analogous.  In this  paper, we present an analysis of archival data obtained from XMM-Newton observations of five active cool giants and subgiants in RS CVn-type binaries, namely UZ Lib, $\sigma$ Gem, $\lambda$ And, V711 Tau and EI Eri with the aim to extend the understanding of the spectral and the temporal characteristics of X-ray flares in them.

The paper is structured  as follows: in section 2, we give the basic properties of stars in our sample, in section 3, we describe the observational data sets that we have analysed and the methods of data reduction, section 4 contains our analysis and results and in section 5 we present our discussion and conclusions.

\section{Basic properties of the sample}
\label{sec:basic}
The basic parameters of the sample of RS CVns are given in Table \ref{tab:parameter} and described below.
\subsection{UZ Lib}
UZ Librae (=BD -08\deg ~3999) contains a giant primary in a close binary system with a period of 4.76 days. It is among the most active binary systems of the RS CVn class (Bopp \& Stencel 1981). Grewing et al.  (1989) presented IUE spectra to discover an ultraviolet excess which they attributed to a companion star with the effective temperature of 8000 \deg K and radius of 1 R$_\odot$. Ol{\'a}h et al. (2002) rediscussed the absolute dimensions of the system and concluded that the secondary is most-likely a low-mass cool star and that the ultraviolet excess can be explained by the active chromosphere of the primary.
\subsection{$\sigma$ Gem}
 $\sigma$  Gem (=HD 62044) is a long period (19.6045 days) RS CVn-type binary containing an active  K1 III type, red giant primary. The angular diameter of the giant primary is 2.3 mas with a corresponding stellar radius of 9.3 $R_\odot$ (Nordgren et al. 1999). The secondary in $\sigma$ Gem has not been detected at any wavelength, but restrictions to its mass and the low luminosity suggest that it is most likely a late-type main sequence star of under one solar mass (Duemmler et al. 1997). It is well studied system at all wavelengths (e.g. Singh et al. 1987; Osten \& Brown 1999; Padmakar \& Pandey 1999; Brown \& Brown 2006; Nordon et al. 2006.).
\subsection{$\lambda$ And}
$\lambda$ And (=HD 222107) consists of a G8 III-IV primary with a rotation period of 54 days and an unseen companion in an orbit with an orbital period of 20.5 days (Strassmeier et al. 1993); the system is therefore asynchronous. It  is one of the brightest long-period RS CVn's at X-ray wavelengths. In the absence of any firm clues as to the identity of the secondary, the primary component in $\lambda$ And is thought to be wholly responsible for the observed X-ray emission.
\subsection{V711 Tau }
V711 Tau (=HD 22468) is one of the most extensively studied members of the RS CVn class of magnetically active close binaries. It is a detached, non-eclipsing SB2 system, consisting of a K1 subgiant star, filling about 80\% of its Roche lobe and a G5 dwarf star on a 2.84 day orbit (Fekel 1983; Donati 1999). The G5V component is also rapidly rotating, but its convection zone is less extended, which makes it much less active than the subgiant. Therefore, the magnetic activity signatures of V711 Tau are dominated by the K1IV star.  V711 Tau has been extensively studied in the optical, the ultra-violet and the radio, while its extreme ultra-violet (EUV) and X-ray emission was analyzed in the context of surveys of coronal emission from RS CVn systems (Majer et al. 1986; Pasquini et al. 1989; Schmitt et al.  1990; Griffiths \& Jordan 1998).  Rotational modulation in the RGS X-ray light curve of V711 Tau has been detected  by Audard  et al. (2001), with a minimum flux occurring near the phase when the G5 star is in front (phase = 0.5), consistent with a previously reported correlation between binary phase and X-ray intensity by Agrawal \& Vaidya (1988).
\subsection{EI Eri}
EI Eridani = HD 26337  is a rapidly rotating ($v sin i = 51 km s^{−1}$) active, noneclipsing, single-lined spectroscopic binary star.  It  consists G5IV primary and M4-5V secondary with  orbital period of 1.947232 days (Washuettl et al. 2009). 

\begin{table}
\caption{General properties of stars in the sample}
\label{tab:parameter}
\begin{tabular}{llccl}
\hline
Name   & Spectral Type   & V    &Period&Distance \\
                 &                           & (mag)&(days)&(pc)\\
\hline
UZ Lib           &    K0III                  &  9.11          & 4.75           & 140.3 \\
$\sigma$ Gem     &    K1III                  &  5.40          & 19.604         &  37.5 \\
$\lambda$ And    &    G8III-IV               &  3.82          & 54.07          &  25.8 \\
V711 Tau         &  K1IV/G5V                 &  5.90          & 2.84           &  29.0 \\
EI Eri           &  G0IV/M4-5V               &  6.95          & 1.95           & 56.2  \\
\hline
\end{tabular}
\end{table}

\section{Observations and Data Reduction}
\label{sec:obs}
\begin{table*}
\centering
\caption{Log of observations with the XMM-Newton}
\label{tab:log}
\begin{tabular}{lclccc}
\hline
Star         &Rev.& Instrument      &Start time          & Exposure& offset \\
Name         &    &(mode,filter)    &date (UT)           & Time(s) & (')\\
\hline
UZ Lib       &127 &MOS1(PFW,medium) &2000-08-19(04:53:55)& 22616& 0.148\\
             &    &MOS2(PFW,medium) &2000-08-19(11:09:59)& 22563&      \\
             &    &PN(PFW,Medium)   &2000-08-19(05:34:59)& 23567&       \\
	     &210 &MOS1(PFW,Medium) &2001-01-31(19:30:24)& 27336& 0.095\\
             &    &MOS2(PFW,Medium) &2001-01-31(19:30:23)& 27347&      \\
             &    &PN(PFW,Medium)   &2001-01-31(20:11:21)& 24896&       \\
$\sigma$ Gem &243 &MOS1(PFW,thick)  &2001-04-06(16:40:29)& 49225& 0.068\\
             &    &MOS2(PFW,Thick)  &2001-04-06(16:30:52)& 49920&      \\
             &    &PN(PSW,Thick)    &2001-04-06(16:46:36)& 55811&       \\
$\lambda$ And&208 &MOS1(T,thick)    &2001-01-26(20:20:58)& 27944& 0.061\\
             &    &MOS2(PPW,Thick)  &2001-01-26(20:10:31)& 28541&      \\
             &    &PN(PSW,Thick)    &2001-01-26(20:24:00)& 28200&       \\
V711 Tau     &310 &MOS1(PPW,Thick)  &2001-08-18(03:32:01)& 11911& 0.228\\
             &    &MOS1(PPW,Medium) &2001-08-18(07:25:52)& 11682&     \\
             &    &MOS2(PPW,Thick)  &2001-08-18(03:32:01)& 11935&     \\
             &    &MOS2(PPW,Medium) &2001-08-18(07:25:51)& 11684&     \\
             &    &PN(PSW,Medium)   &2001-08-18(03:47:57)& 25016&     \\
	     &221 &MOS1(PFW,Thick)  &2001-02-22(16:40:45)&  9997& 0.092 \\
	     &    &MOS1(PPW,Thick)  &2001-02-22(19:36:35)& 23828&  \\
	     &    &MOS2(PFW,Thick)  &2001-02-22(16:40:43)& 10000&  \\
	     &    &MOS2(PPW,Thick)  &2001-02-22(19:36:33)& 23800& \\
             & 27*&PN(PSW,Medium)   &2000-02-01(09:17:29)& 14603&1.936\\
EI Eri	     &1769&MOS1(PPW,Medium) &2009-08-06(21:00:31)& 37613& \\
             &    &MOS2(TIMING)     &2009-08-06(21:00:26)& 37361& \\
             &    &PN(PSW,Medium)   &2009-08-06(21:05:40)& 37471& \\
\hline
\end{tabular}
\\
{\it Note:}  FPW: FullPrime Window PPW: PrimePartial Window, SW: Small window, T: timing, PSW: PrimeSmall Window \\
{* No MOS observation}
\end{table*}

  Stars in the sample were observed with the
XMM-Newton satellite using many different set-ups of the detectors. The XMM-Newton satellite is composed of three co-aligned X-ray telescopes
(Jansen et al. 2001) which observe a source simultaneously, accumulating photons in three CCD-based instruments: the twin MOS1
and MOS2 and the PN (Str\"{u}der et al. 2001; Turner et al. 2001), and with
all three detectors constituting the European Photon Imaging Camera (EPIC ). The EPIC instruments provide the imaging with a good angular resolution [point spread function
(PSF) = 6 arcsec full width at half-maximum (FWHM)] and spectroscopy in the energy range
from 0.15 to 15 keV with a moderate spectral resolution ($E/\Delta E \approx 20-­50$). Exposure time for each
star was in the range of 10 -­56 ks. A log of observations is provided
in Table \ref{tab:log}. Star V711 Tau was observed by XMM-Newton in 31 occasions. In 10 occasions the exposure times were less then 4.0 ks, therefore, we have not considered these observations for further analysis. In 12 occasions only RGS observations were made. In one occasion (obs ID:0116710901)  the only MOS CCDs were on and data is heavily piled up, therefore, is not useful for further analysis. In rest of eight  occasions, only in four occasions flares were observed (see Table 2 for detail on observations), in other four occasions (obs ID: 0116340601, 0116320801, 0129350201, 0134540601) light curves appear to be constant. However, during the obs ID 0129350201 a small part of rise phase of a flare was observed  at the end of the observations. The flare observed during the obs ID 0117890901 was analysed by Audard et al. (2001). EI Eri was also observed in four occasions.

The data were reduced with standard XMM-Newton Science
Analysis System (SAS) software, version 10.0 with updated calibration files. The preliminary processing of raw EPIC
Observation Data Files was done using the {\sc epchain} and {\sc emchain}
tasks which allow calibration both in energy and astrometry of the
events registered in each CCD chip and combine them in a single data file for MOS and PN detectors. The background contribution is particularly relevant at high energies where coronal sources
have very little flux and are often undetectable. Therefore, for further analysis, we have selected the energy range between 0.3 and
10.0 keV. Event list files were extracted for each observations using the SAS task {\sc evselect}.
Data from the three cameras were individually checked for the time intervals with high background when the total count rates (for single events of energy above 10 keV) in the instruments exceeded 0.35 and 1.0 counts s$^{-1}$ for the MOS and PN detectors, respectively. No data were found to be affected by high background proton flaring events.
The {\sc epatplot} task was used for checking the existence of pile-up in
 the inner regions of the X-ray images of all stars. Only the MOS data of the  star V711 Tau were
found to be affected by pile-up. X-ray light curves and spectra of
all target stars were generated from on-source counts obtained from
circular regions with a radius 40-50 arcsec around each source. However, for the star V711 Tau, X-ray light curves of MOS detector were
extracted with events taken from an annulus of 42 arcsec with an inner
radius of 12 arcsec to avoid the pile-up effect in the inner region.
The background was taken from several source-free regions on the detectors at nearly the same offset as the source and surrounding the source.

\begin{figure*}
\centering
\subfigure[UZ Lib]{\includegraphics[width=53mm,angle=-90]{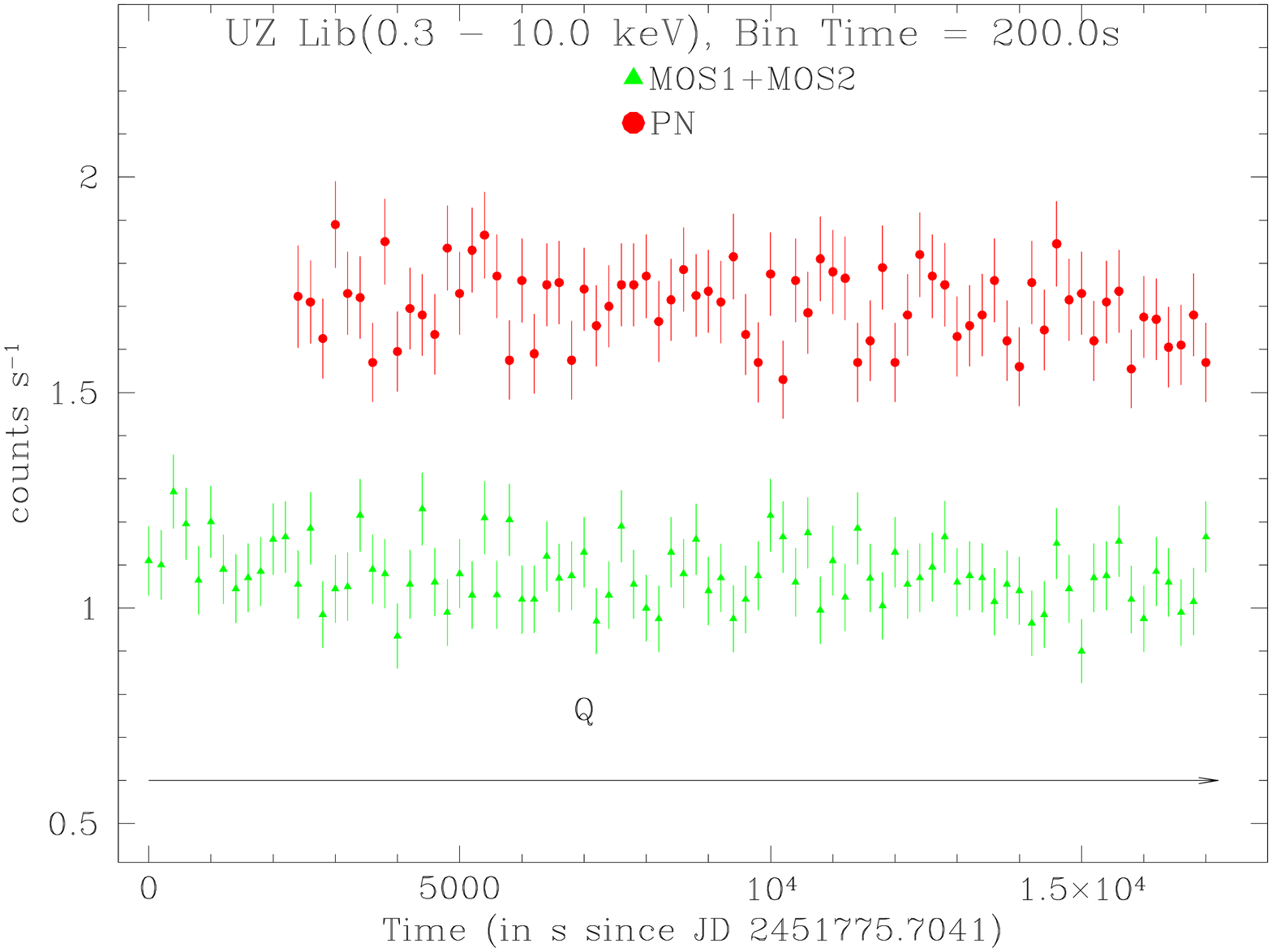}}
\subfigure[UZ Lib]{\includegraphics[width=53mm,angle=-90]{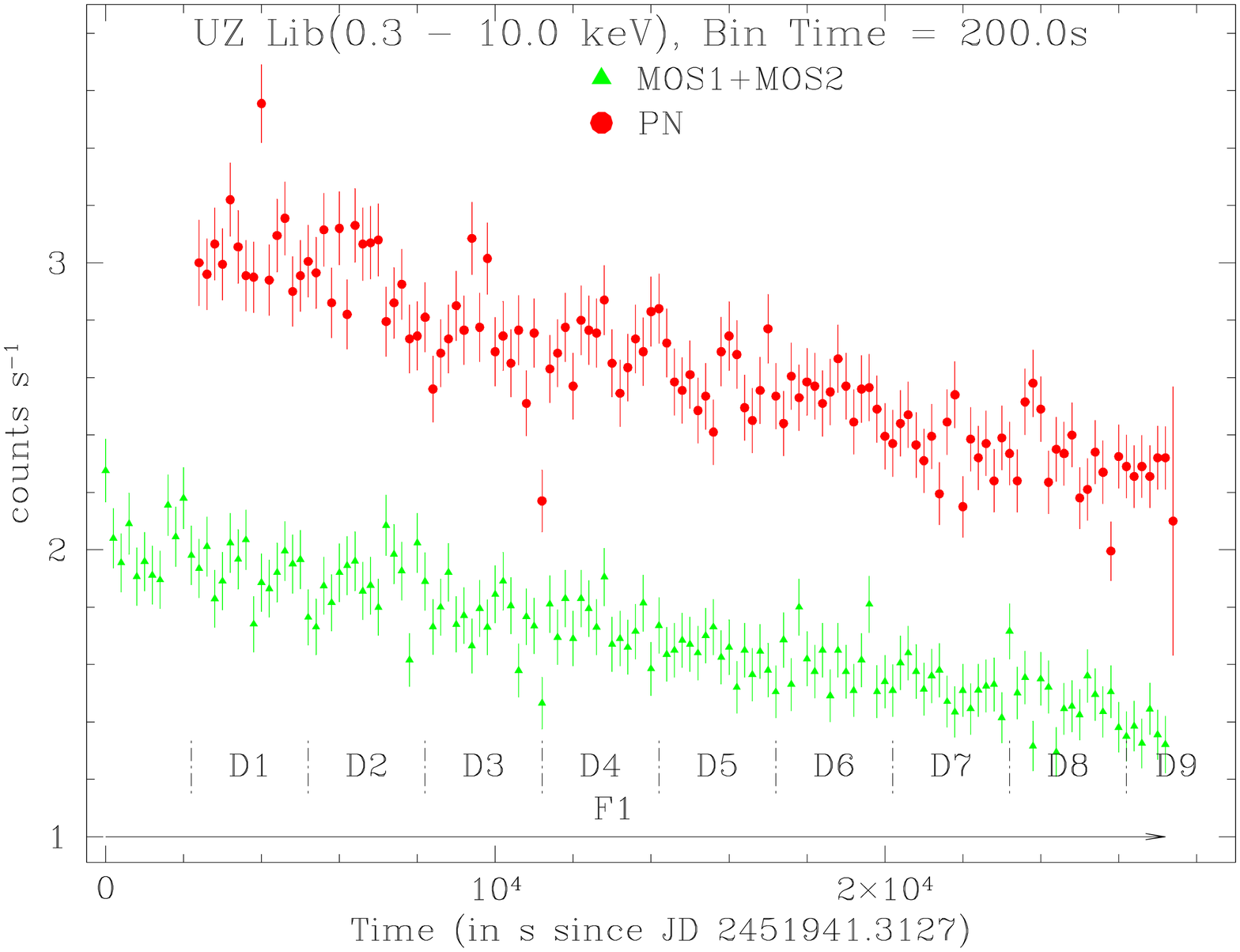}}
\subfigure[$\sigma$ Gem]{\includegraphics[width=53mm,angle=-90]{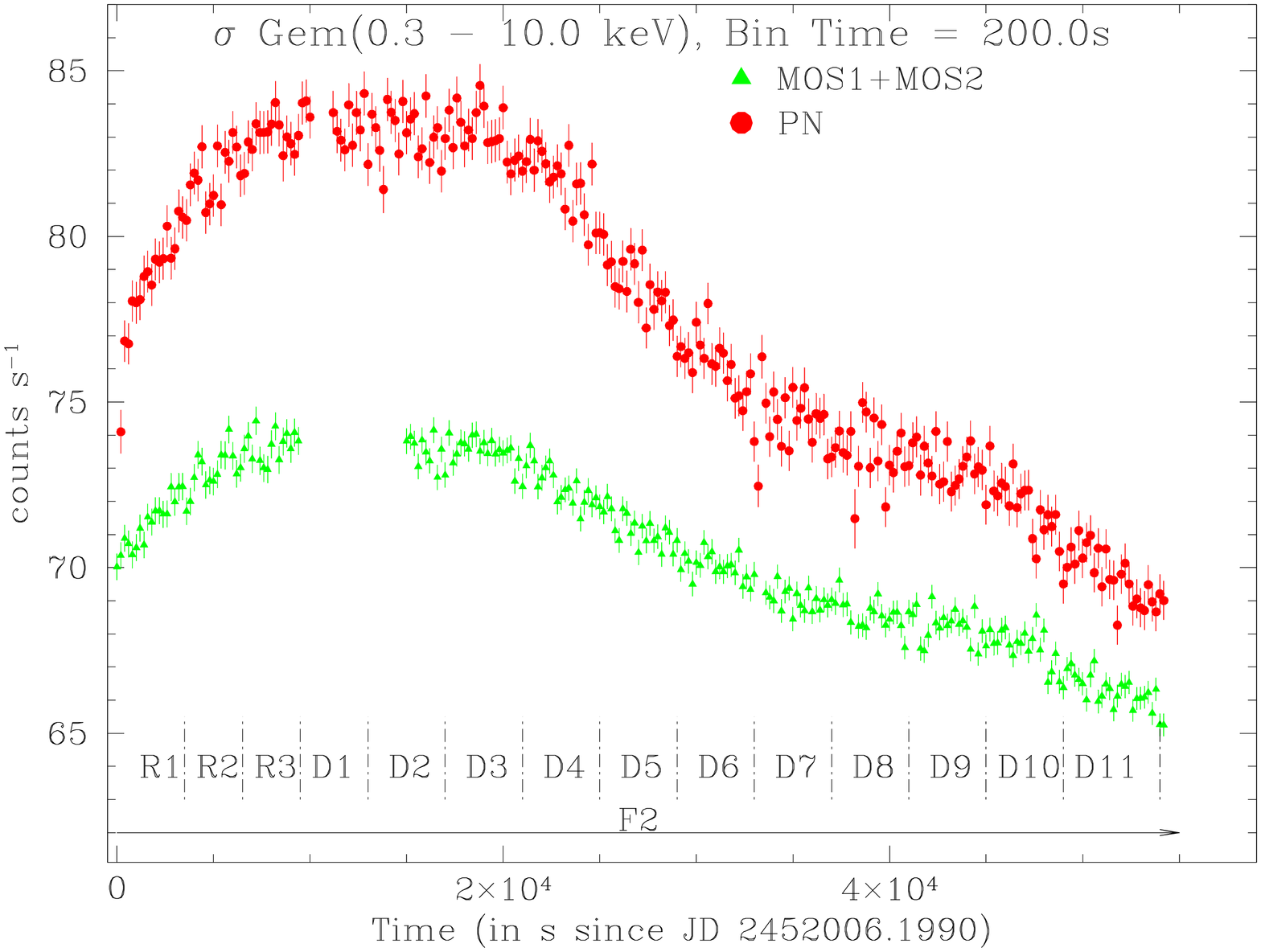}}
\subfigure[$\lambda$ And]{\includegraphics[width=53mm,angle=-90]{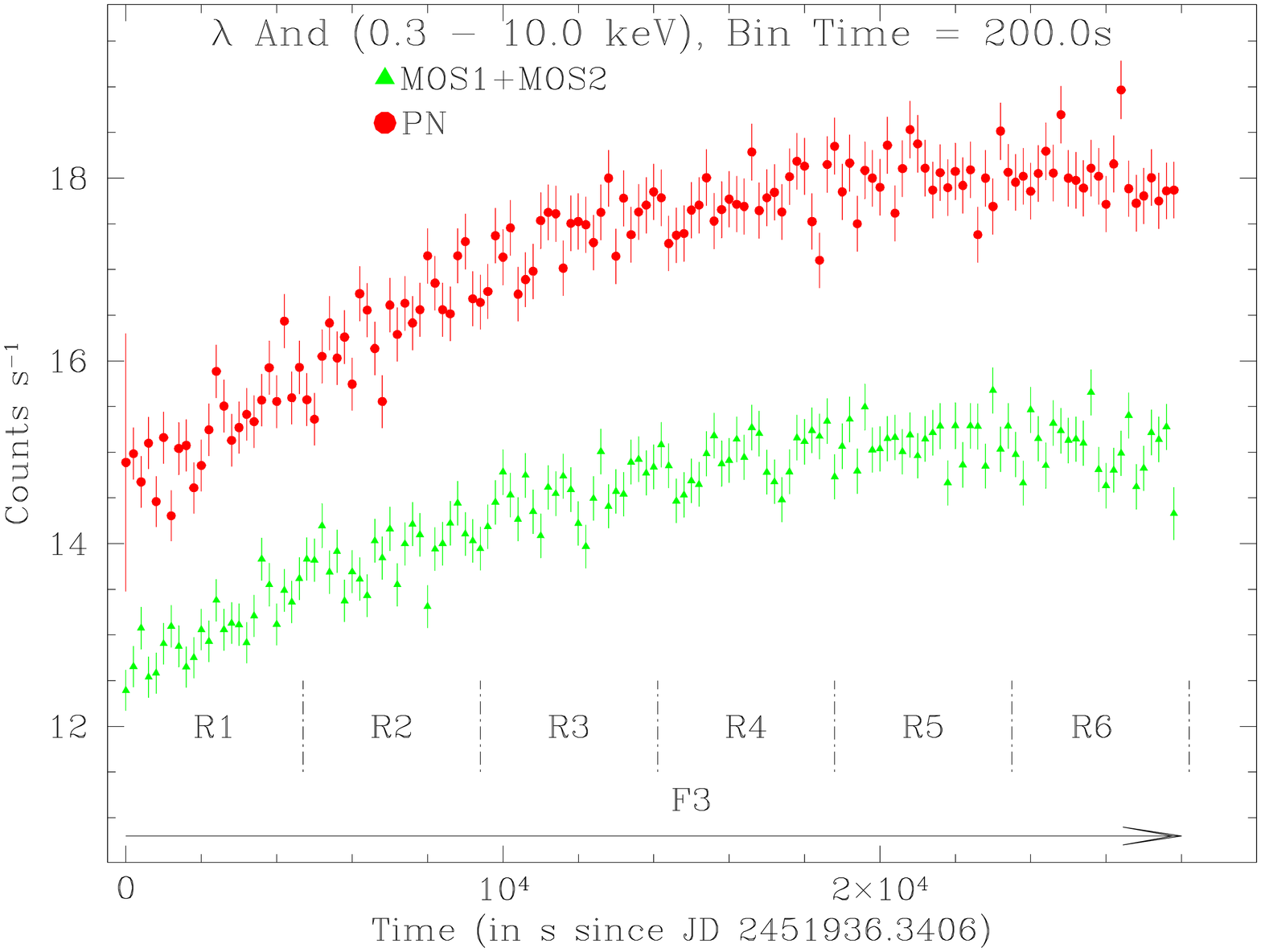}}
\subfigure[V711 Tau]{\includegraphics[width=53mm,angle=-90]{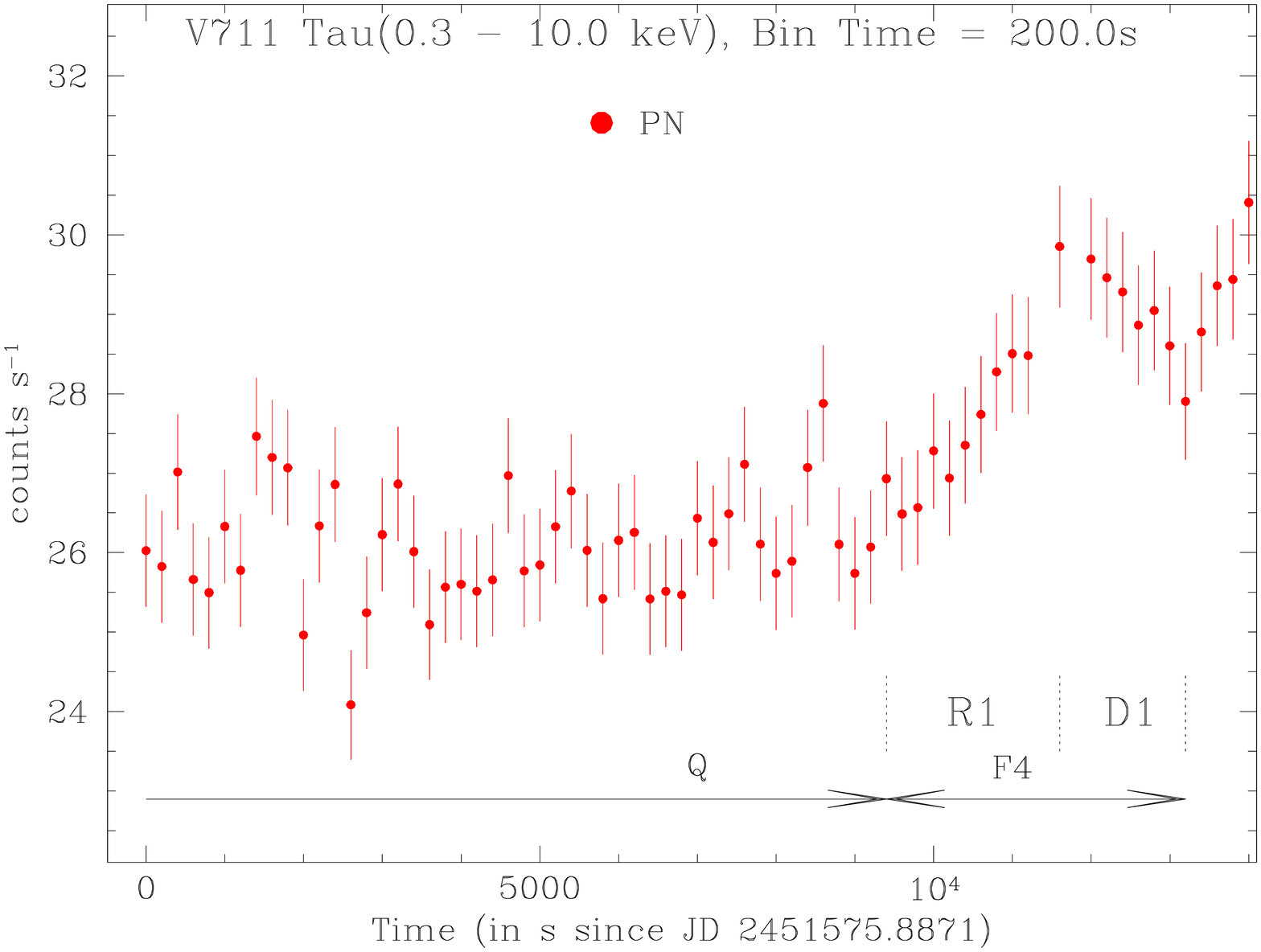}}
\subfigure[V711 Tau]{\includegraphics[width=53mm,angle=-90]{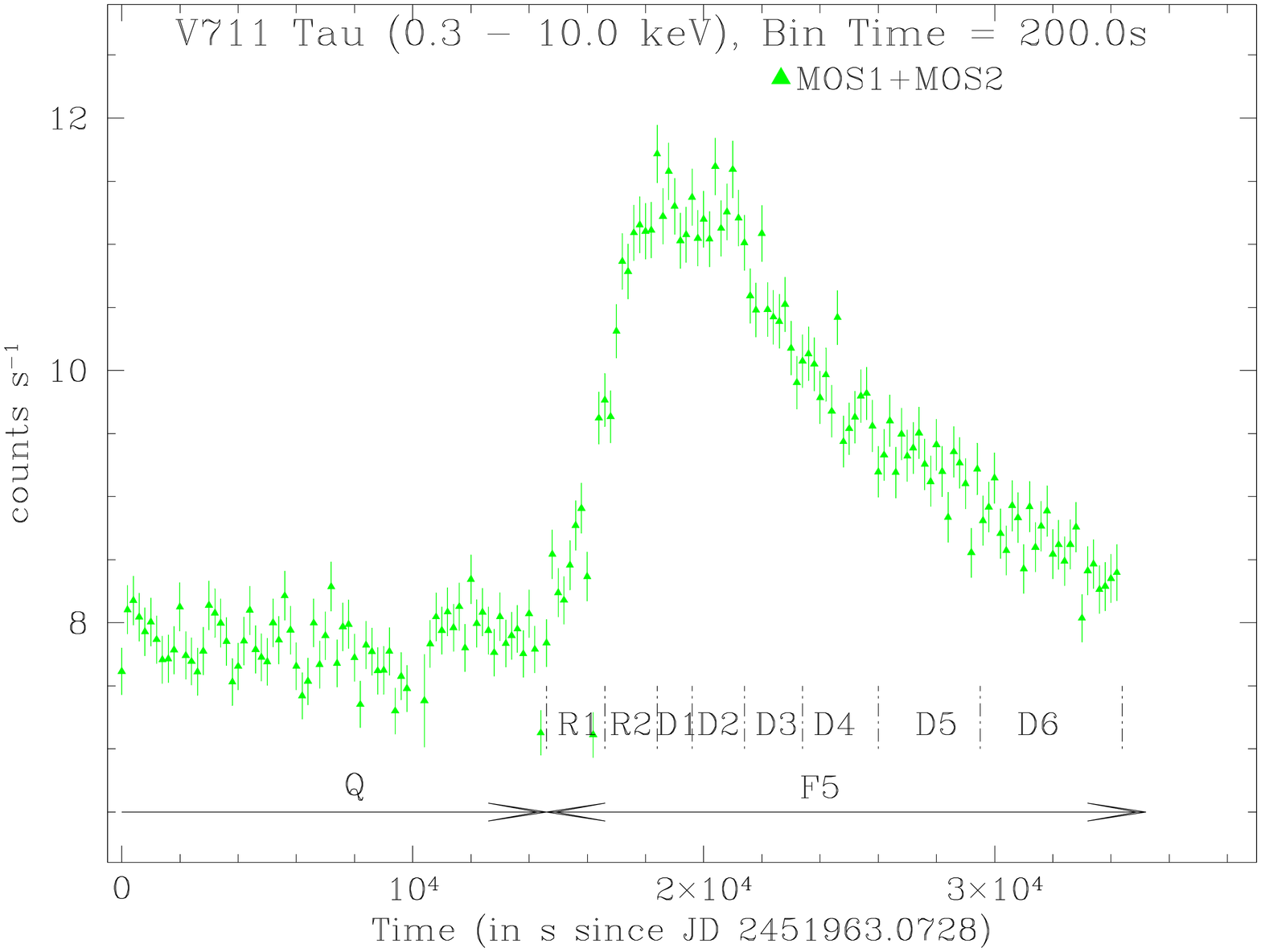}}
\subfigure[V711 Tau]{\includegraphics[width=53mm,angle=-90]{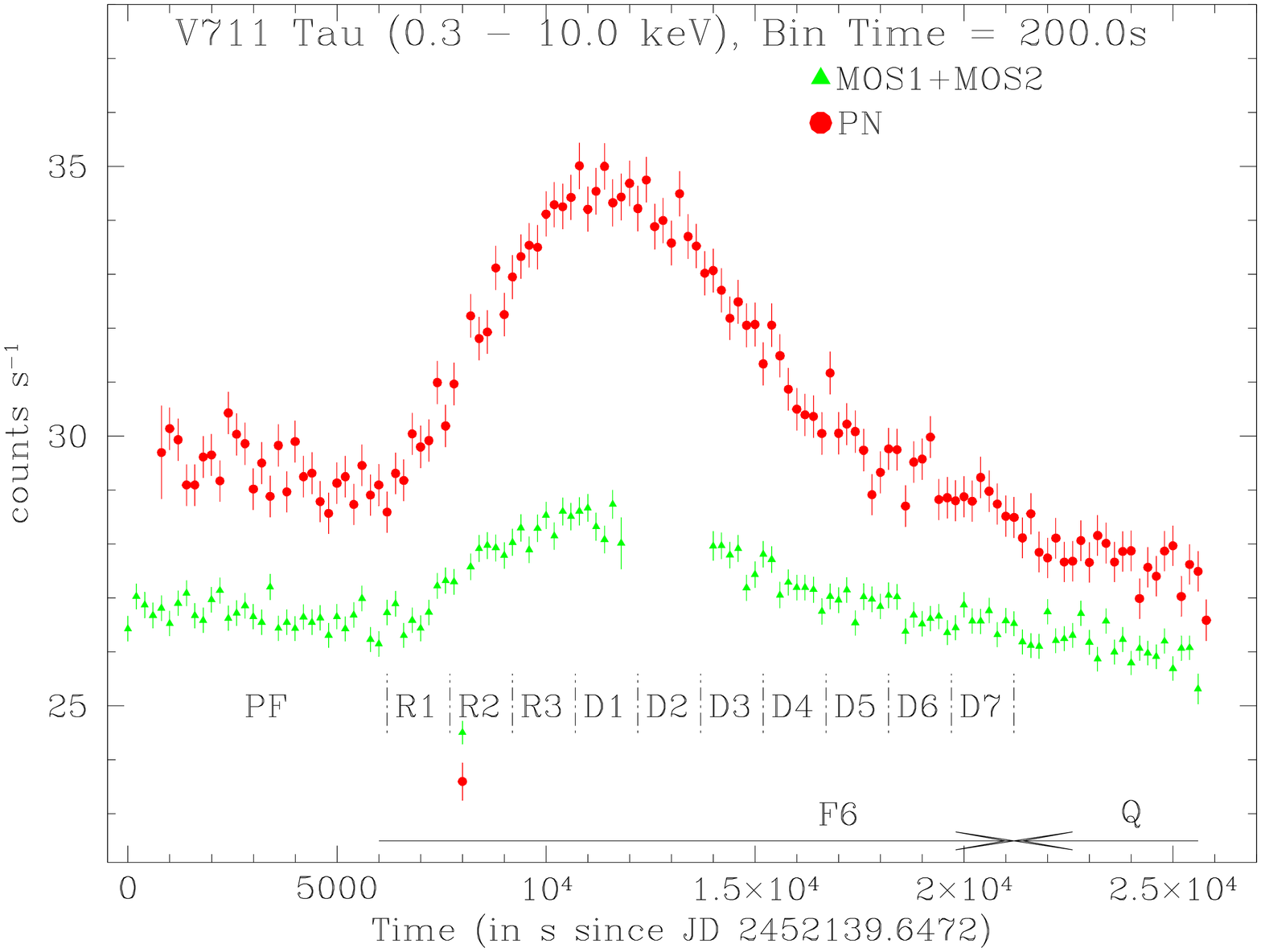}}
\subfigure[El Eri]{\includegraphics[width=53mm,angle=-90]{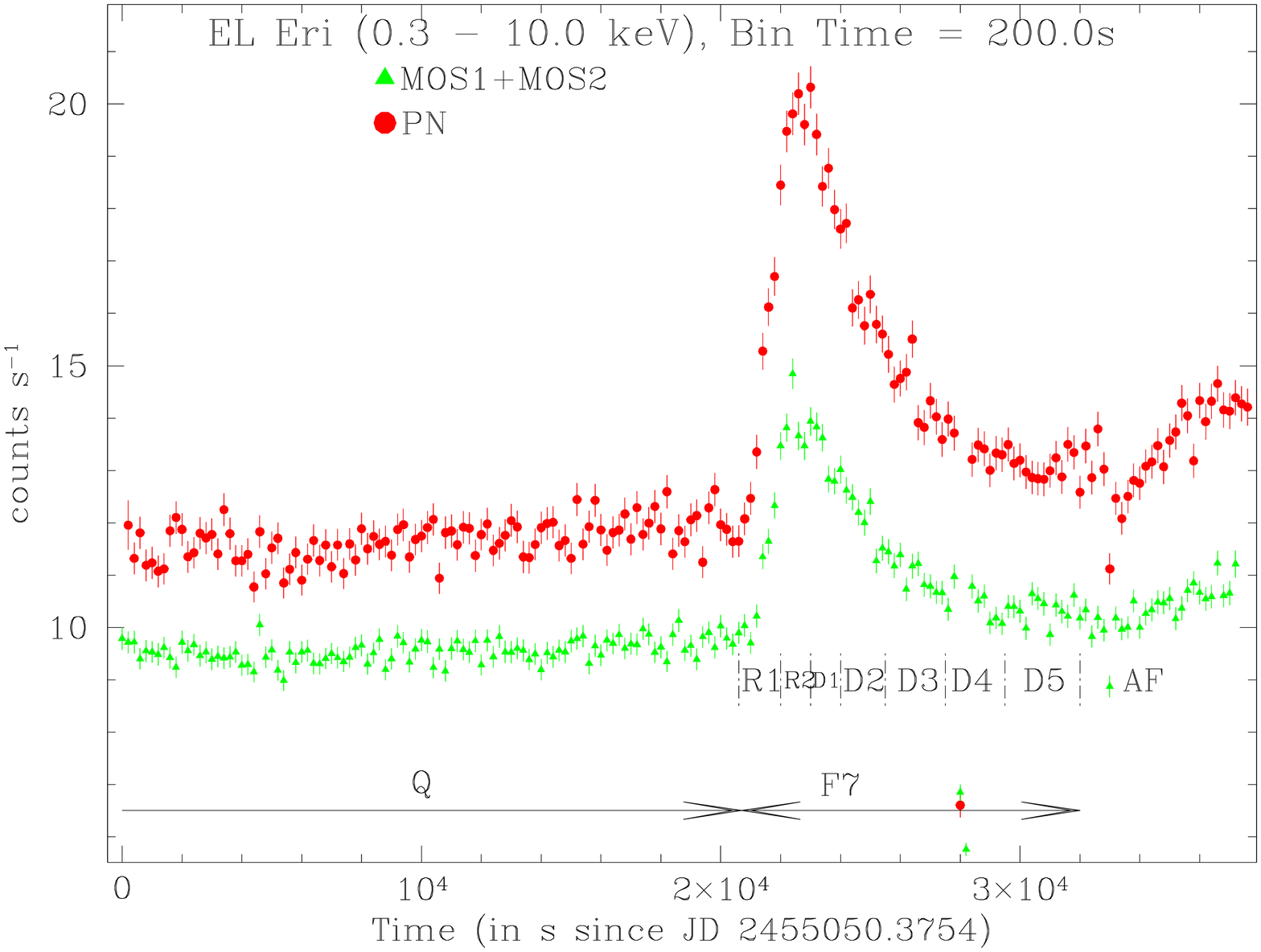}}
\caption{MOS and PN Light curves of the sample stars. The MOS light curves are scaled by adding 33, 2, 16 and 3.5  \cts for flares F2, F3, F6 and F7, respectively.}
\label{fig:mospnlc}
\end{figure*}

\section{Analysis and results}

\subsection{X-ray light curves}
X-ray light curves of the five stars in the energy band 0.3 -10.0 keV are shown in  Figure \ref{fig:mospnlc}. The MOS and PN light curves
are represented by solid triangles and circles, respectively. The temporal
binning of light curves is 200 s for all sources. The light curves of the stars UZ Lib (Fig. \ref{fig:mospnlc}b), $\sigma$ Gem (Fig. \ref{fig:mospnlc}c), $\lambda$ And (Fig. \ref{fig:mospnlc}d), V711 Tau (\ref{fig:mospnlc}e  to g)  and EI Eri (Fig. \ref{fig:mospnlc} h) show variability on a time
scale of hrs, most of which resembles flaring activity. The customary definition of a flare is a significant increase in intensity,
after which the initial or quiescent level of intensity is reached again.
Such patterns are observed in the light curves of the sample stars
shown in Fig. \ref{fig:mospnlc},
where  the
``flare regions'' are represented by arrows and marked by Fi, where i=1,2...7 refers to the flare number.
No flare like features were observed  during the satellite revolution of 127 in the light curves of the star UZ Lib (Fig. \ref{fig:mospnlc}a), therefore,  assumed to represent its quiescent state.
The mean quiescent level count rates of the other sources are taken from
regions that are free from flares and are  marked by Q in the Fig. \ref{fig:mospnlc}. For the stars $\sigma$ Gem and $\lambda$ And, the quiescent states were not observed.
Using the RGS data of flare F2 of $\sigma$ Gem, Nordon et al. (2006) found an overall 25 per cent flux increase from its quiescent state, which was observed 15 months earlier by Chandra X-ray observatory. Therefore, we have assumed that the flux during the quiescent state of $\sigma$ Gem was 25 per cent less than that of the peak flare flux.
The flux during the quiescent state of $\lambda$ And was derived from ASCA and ROSAT data by Ortolani et al. (1997) and is $\sim$ 31 per cent less than the peak value of the flare F3.
The duration of the flares and the flux values during quiescence and  at flare peak  are listed in Table \ref{tab:risedecay} for all the seven flares observed . The peak and quiescent state
count rates  as given in Table \ref{tab:risedecay} are converted into fluxes by using the WebPIMMS\footnote{http://heasarc.gsfc.nasa.gov/Tools/w3pimms.html},
where we have assumed a coronal plasma temperature of 1 keV.  The flux  at the peaks of flares was increased by 26 - 56 per cent from the quiescent states.

Peak count rates ($A_0$), and $1/e$ rise ($\tau_r$) and decay ($\tau_d$) times have been determined from the least-squares fit of the exponential function $c(t) = A_0 e^{(t_0-t)/\tau_{r,d}} + q$ to the count rate histogram.  Here, $t_0$ is the time at flare peaks and q is count rates during the quiescent state. The fitted values are given in Table \ref{tab:risedecay}. The exponential decay times were found in the range of 0.8 to 9 hrs. However, the 1/e rise  times of the observed flares were found  in the range of 0.3 to 6 hrs. Flare F2 of $\sigma$ Gem was found to be the strongest flare, where total energy ($E_{tot}$) released was $4.24\times10^{37}$ ergs.  All other flares were weaker by at least a factor of 10 (see Table \ref{tab:risedecay}).
The MOS data have less count rates than the PN but have similar characteristics to that of PN; therefore, for further analysis in this paper we use only the PN data. However, for the flare F5 from V711 Tau only MOS data were used as PN observations were  piled up.

\begin{figure*}
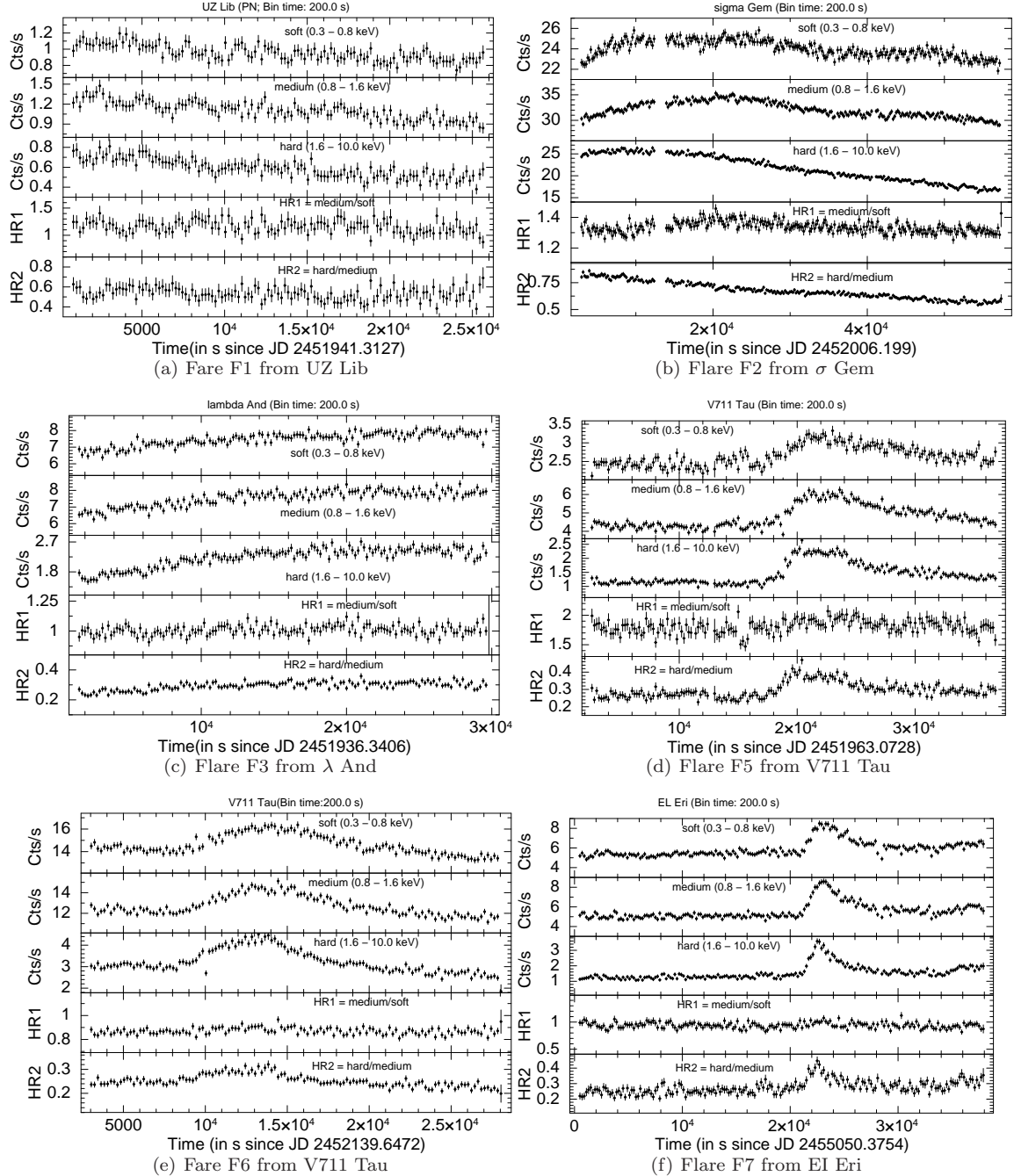

\centering
\subfigure[Fare F1 from UZ Lib]{\includegraphics[width=55mm,angle=-90]{UZLib_lc_3band.ps}}
\subfigure[Flare F2 from $\sigma$ Gem]{\includegraphics[width=55mm,angle=-90]{SigmaGem_lc_3band.ps}}
\subfigure[Flare F3 from $\lambda$ And]{\includegraphics[width=55mm,angle=-90]{LambdaAnd_lc_3band.ps}}
\subfigure[Flare F5 from V711 Tau]{\includegraphics[width=55mm,angle=-90]{V711Tau_lc1_3band.ps}}
\subfigure[Fare F6 from V711 Tau]{\includegraphics[width=55mm,angle=-90]{V711Tau_lc_3band.ps}}
\subfigure[Flare F7 from EI Eri]{\includegraphics[width=55mm,angle=-90]{EIEri_lc_3band.ps}}
\caption{PN light curves at three bands soft (0.3-0.8 keV), medium (0.8-1.6 keV), and hard(1.6-10.0 keV) and hardness ratio HR1 and HR2 curve, where HR1=medium/soft and HR2=hard/medium during flares from program stars}
\label{fig:pnlc_3band}
\end{figure*}

X-ray light curves were also divided separately into three energy bands namely soft (0.3-0.8 keV), medium (0.8-1.6 kev) and hard (1.6-10.0 keV). The boundaries of  the selected energy bands are chosen as the line-free regions of the low-resolution PN spectra. Hardness ratios HR1 and HR2 as defined by the ratio of medium to soft band, and hard to medium band count rate, respectively, were also derived. The soft, medium and hard bands intensity curves, and the hardness ratio (HR1 and HR2) curves  are shown in the sub-panels running from top to bottom in  Fig. \ref{fig:pnlc_3band} for all the stars.  The light curves in individual bands show significant variability in all sources.  The peak flare flux, quiescent state flux, and 1/e rise and decay times in the three energy bands are given in Table \ref{tab:rd3band}.  For the flares F1, F5, F6 and F7 the 1/e decay times in the soft band were found to be more than the decay times in the medium and the hard band. However, this difference between 1/e decay times in the different energy bands is within 1.6$\sigma$ level. The 1/e decay time for the flare F2 was found to be more in the hard band than in the soft and the medium bands.
The light curves in different energy bands, for $\sigma$ Gem, show different behavior. In the soft band the light curve shows a flat top, in the medium band it shows a narrower top, whereas, in the hard band the light curve  of $\sigma$ Gem hardly shows the rise phase of the flare.

 The ratio of flare peak flux to the quiescent state flux in the hard band
was found to lie between 1.4 and 2.5, which is more than that of the similar ratios in the medium and the soft band (see Table \ref{tab:rd3band}) for
all flares. In terms of the peak flux, the flare F2 of $\sigma$ Gem was stronger in the medium band than in the soft and hard bands.  The variation in the hardness ratio is indicative of changes in coronal temperature. An X-ray flare is well defined by a rise in the temperature and subsequent decay. Excluding flare F2, HR1 was found to be constant during all the flares . However, for most of the flares HR2 varied in the similar fashion to their light curves. This implies an increase in the temperature at flare peak and subsequent cooling. In the case of flare F2, the HR variation follows the trend of the hard band light curve.  No variation was observed in the HR2 of the flare  detected  from UZ Lib.

\begin{sidewaystable}
\caption{Flare duration and other parameters obtained from least square fitting of exponential function to the count rate histograms obtained with MOS and PN detectors  in the energy band of 0.3-10.0 keV}
 \label{tab:risedecay}
\begin{tabular}{lccccccccccccccc}
\hline
Object       & FN &  Flare  & $F_p$      & $F_q$         &\multicolumn{4}{c}{MOS}                            && \multicolumn{4}{c}{PN}  & \lx$^a$ & E$_{tot}$\\
\cline{6-9} \cline{11-14}
  Name       &    & duration&            &              & A$_0$         &\tr           &\td            & q    && A$_0$         &\tr              &\td             & q      & & \\
             &    &(ks)     &            &              &               & (s)          & (s)           &      &&               &(s)              &(s)             &  &     &\\
\hline
UZ Lib       &F1  & 27.0    &  0.82      &   0.45        &$2.08\pm0.02$ &  \nodata      &$21802\pm1059$ & 1.20  && $3.28\pm0.04$ & \nodata         &$21261\pm1107$  & 1.80  & 133.5     &36.05\\
$\sigma$ Gem &F2  & 53.0    & 20.97      &   15.00       &$41.2\pm0.12$ &$22036\pm1658$ &$30466\pm505 $ & 30.0  && $84.2\pm0.2 $ &$22012\pm1571$   &$32492\pm523 $  &62.3   & 800.3     &424.2\\
$\lambda$ And&F3  & $>28.2$ &  4.54      &   3.12        &$13.2\pm0.08$ &$19587\pm1034$ &\nodata        & 9.2   && $18.3\pm0.1 $ &$19728\pm995$    & \nodata        &12.6   &  26.2     &7.4\\
V711 Tau     &F4  & $ 3.8$  &  6.5       &   4.8         & \nodata      &\nodata        & \nodata       &\nodata&& $30.\pm0.1  $ &$1100\pm224$     &$2850\pm521$    &26.1   &  12.6     &0.47\\
             &F5  & 16.6    &  11.0      &   4.8         &$11.87\pm0.09$&$1600\pm127$   &$8667\pm289  $ & 7.73  && \nodata       & \nodata         &\nodata         &\nodata&  74.8     &12.4 \\
             &F6  & 20.0    &  8.67      &   4.8         &$12.4\pm0.1 $ &$6301\pm521  $ &$12388\pm631 $ & 8.7   && $35.5\pm0.1 $ &$5190\pm358     $&$ 6976\pm169$   &26.1   &  91.3     &18.28\\
EI Eri       &F7  &  9.0    &  3.71      &   1.93        &$11.2\pm0.3 $ &$1258\pm86   $ &$ 3756\pm203 $ & 6.05  && $20.3\pm0.3 $ &$1193\pm74      $&$ 3783\pm167$   &11.7   &  79.6     & 7.16 \\
\hline
\end{tabular}
{\it Note:} FN is flare name, A$_0$ is count rate at flare peak in untit of \cts,
\td ~is flare decay time, \tr ~is flare rise time, q is quiescent state count rate in unit of \cts, $F_p$ and $F_q$ 
are flux values at flare peak and quiescent state in unit of of \ften ~ \flu, respectively, \lx ~is
total X-ray luminosity during the flare in unit of 10$^{30}$ \lum and E$_{tot}$ is total X-ray energy emitted during the flare in unit of 10$^{35}$ erg. \\
{a}{\lx ~for each flare is integrated using the values given in Table 6 based on spectral fits}
\end{sidewaystable}

\begin{sidewaystable}
\caption{Parameters obtained from the fitting of an exponential function to the flare light curves observed with PN/MOS detector in the soft, medium and the hard bands.}
\label{tab:rd3band}
\begin{tabular}{lccccccccccccccc}
\hline
Object        & FN & \multicolumn{4}{c}{ Soft (0.3-0.8 keV)}     && \multicolumn{4}{c}{Medium (0.8-1.6 keV)}      && \multicolumn{4}{c}{Hard (1.6-10.0 keV)}   \\
\cline{3-6} \cline{8-11} \cline{13-16}
Name          &    &   \td(s)     &  \tr(s)      &  $F_p$      &$F_q$      &&  \td(s)      &  \tr(s)    &  $F_p$        &$F_q$      && \td(s)       & \tr(s)     &  $F_p$        &$F_q$  \\
\hline
UZ Lib	      & F1 &$29763\pm2726$& \nodata      &  1.3        &0.67       &&$25038\pm2365$&  \nodata   &   1.31        &   0.74    &&$27095\pm1662$& \nodata    & 0.73          & 0.29   \\
$\sigma$ Gem  & F2 &$15572\pm2736$&$4557\pm671$  &  25.4       &22.5       &&$17763\pm2874$&$7362\pm915$&   35.0        &  29.3     &&$25167\pm1068$& \nodata    & 25.8          & 16.1   \\
V711 Tau      & F5 &$9875\pm744 $ &$1645\pm288$  &  9.5        &6.5        &&$8921\pm430$  &$1433\pm148$&   12.0        &  7.3      &&$7673\pm336$  &$1085\pm77 $& 10.6          & 4.2    \\
              & F6 &$8731\pm1737$ &$6010\pm879$  &  16.2       &12.6       &&$5836\pm801$  &$4446\pm677$&   14.4        &  11.2     &&$4323\pm453$  &$4279\pm544$&  4.1          & 2.6    \\
EI Eri        & F7 &$4201\pm361 $ &$1227\pm127$  &  3.7        &2.7        &&$3486\pm247$  &$1062\pm76 $&   9.0         &  6.0      &&$3385\pm199$  &$ 760\pm55 $&  5.0          & 3.7    \\
\hline
\end{tabular}
\\
{\it Note:} FNMisce is flare Name, $F_p$  and $F_q$ are the fluxes at flare peak and during the  quiescent state in units of $10^{-12}$ \flu, respectively.  \td  ~and \tr ~are flare decay and rise times, respectively. 
\end{sidewaystable}


\subsection{Colour-colour diagrams}

\begin{figure*}
\centering
\subfigure[Flare F1 from UZ Lib]{\includegraphics[width=35mm,angle=-90]{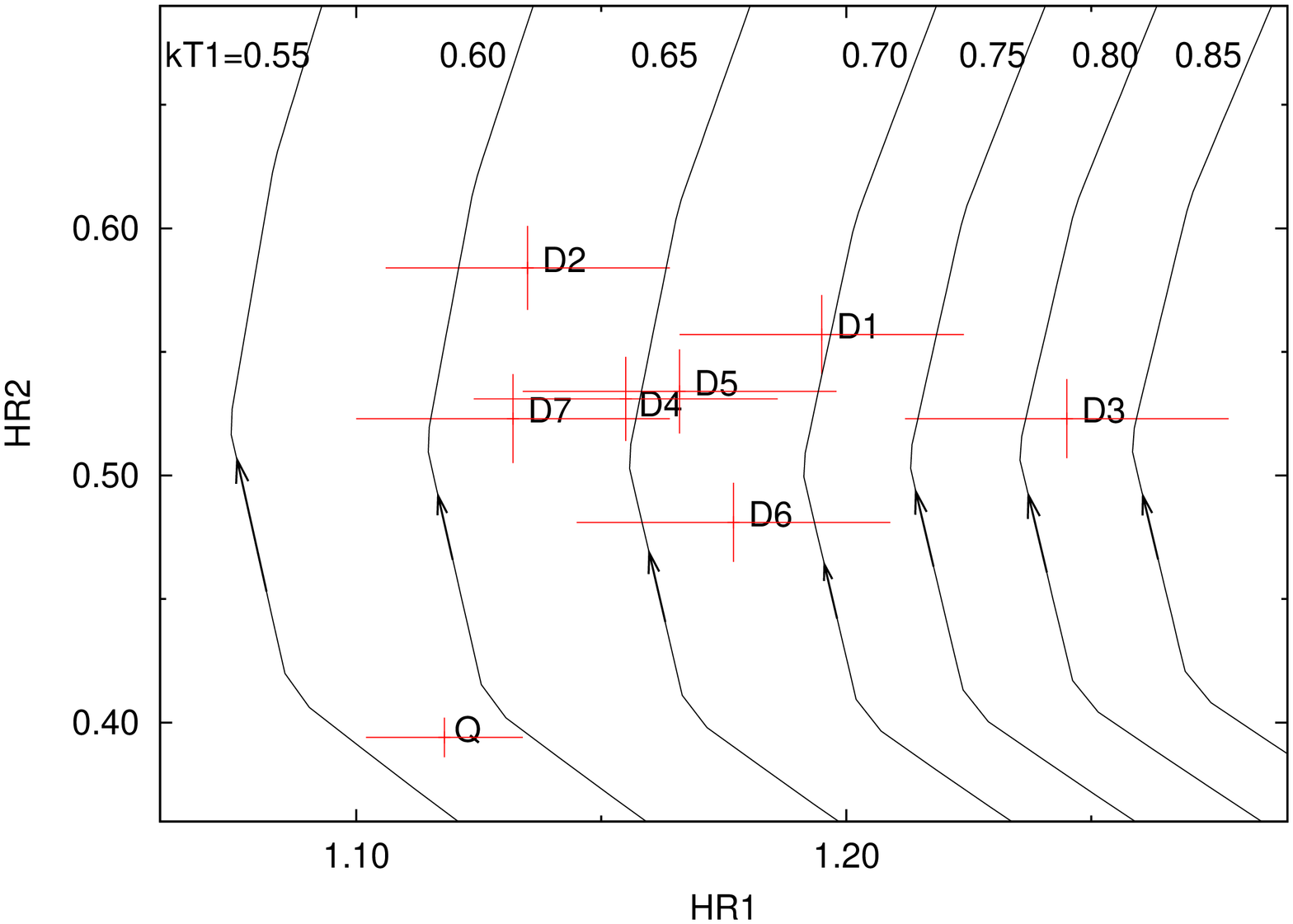}}
\subfigure[flare F2 from $\sigma$ Gem]{\includegraphics[width=35mm,angle=-90]{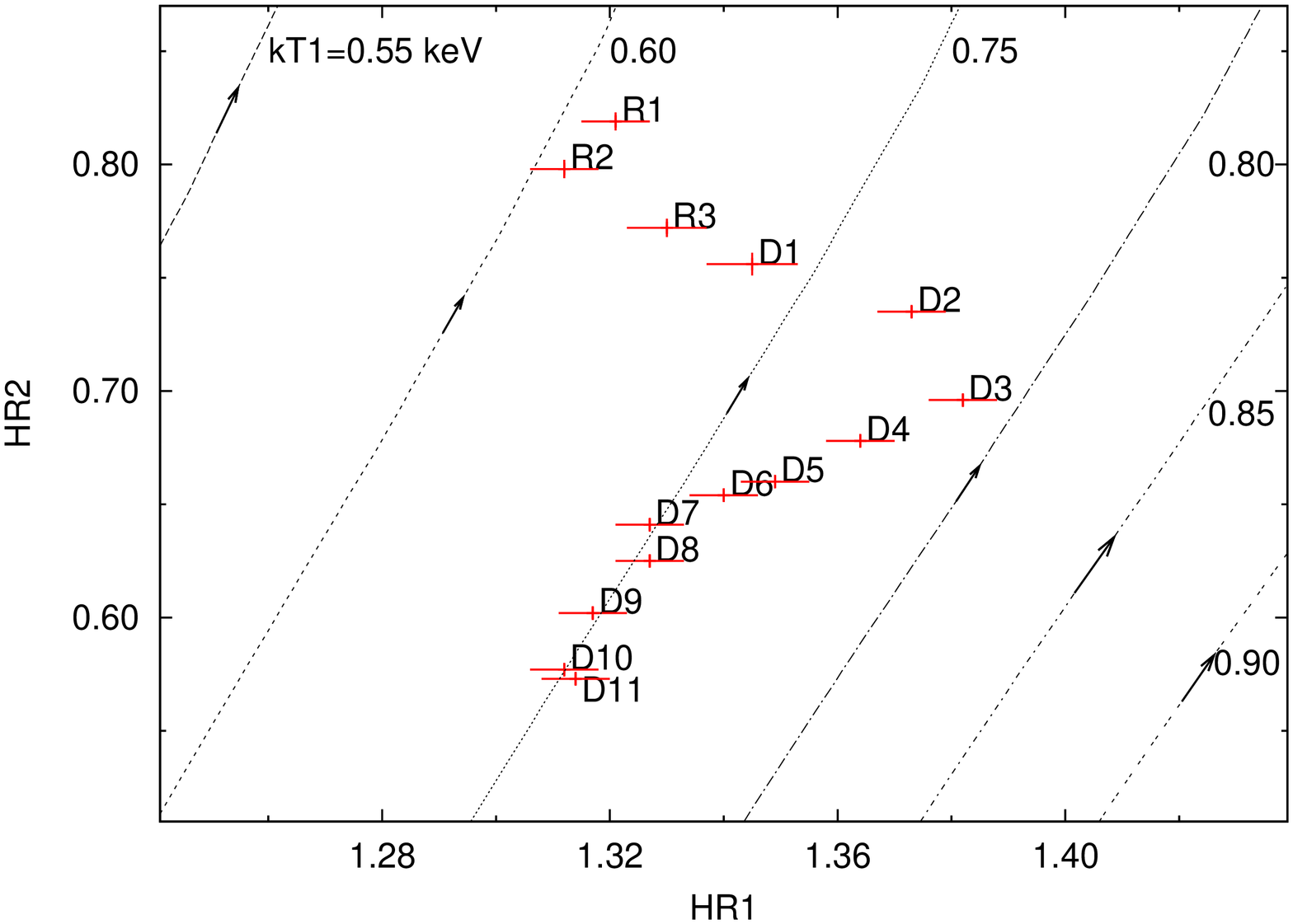}}
\subfigure[Flare F3 from $\lambda$ And]{\includegraphics[width=35mm,angle=-90]{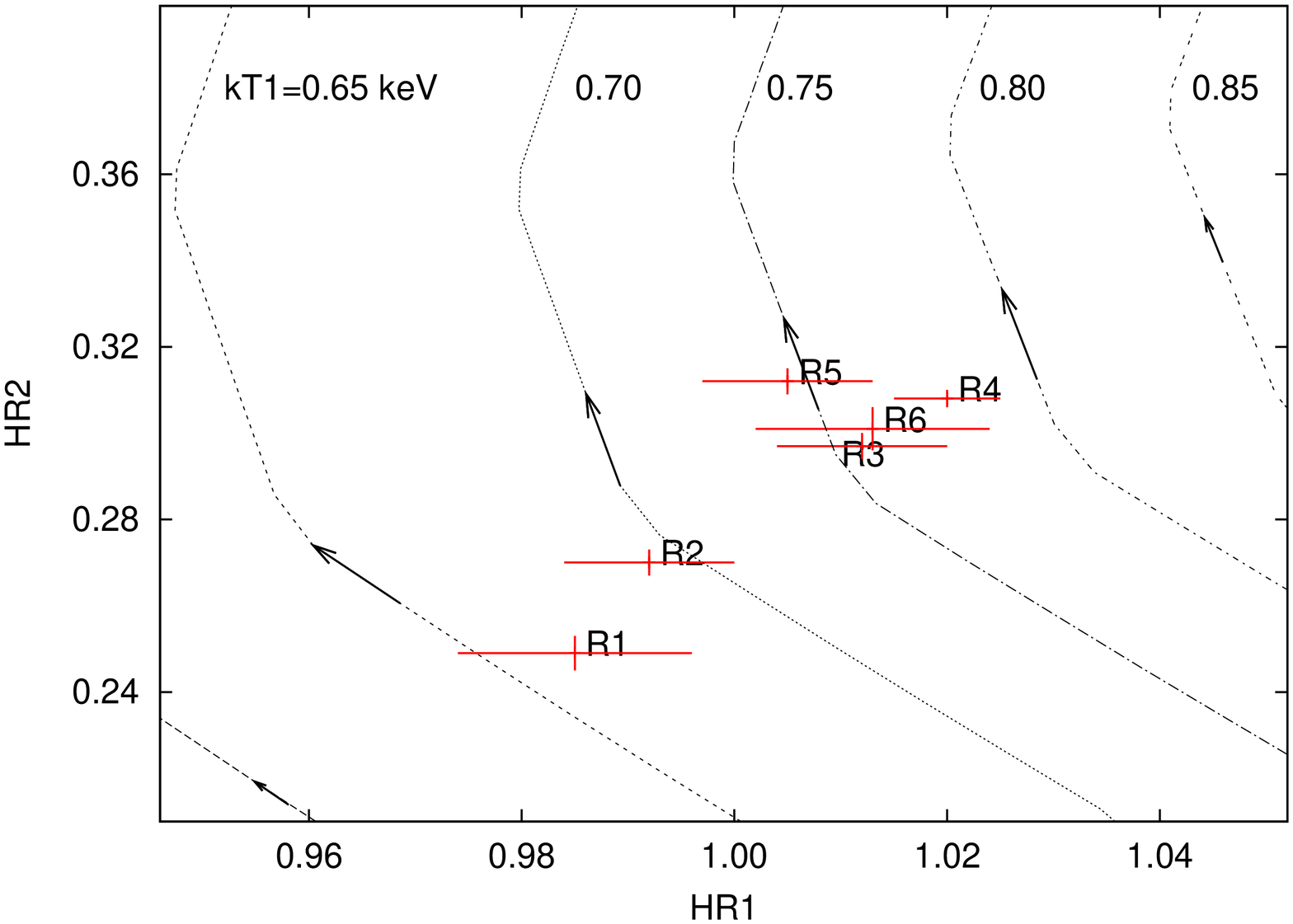}}
\subfigure[Flare F5 from V711 Tau]{\includegraphics[width=35mm,angle=-90]{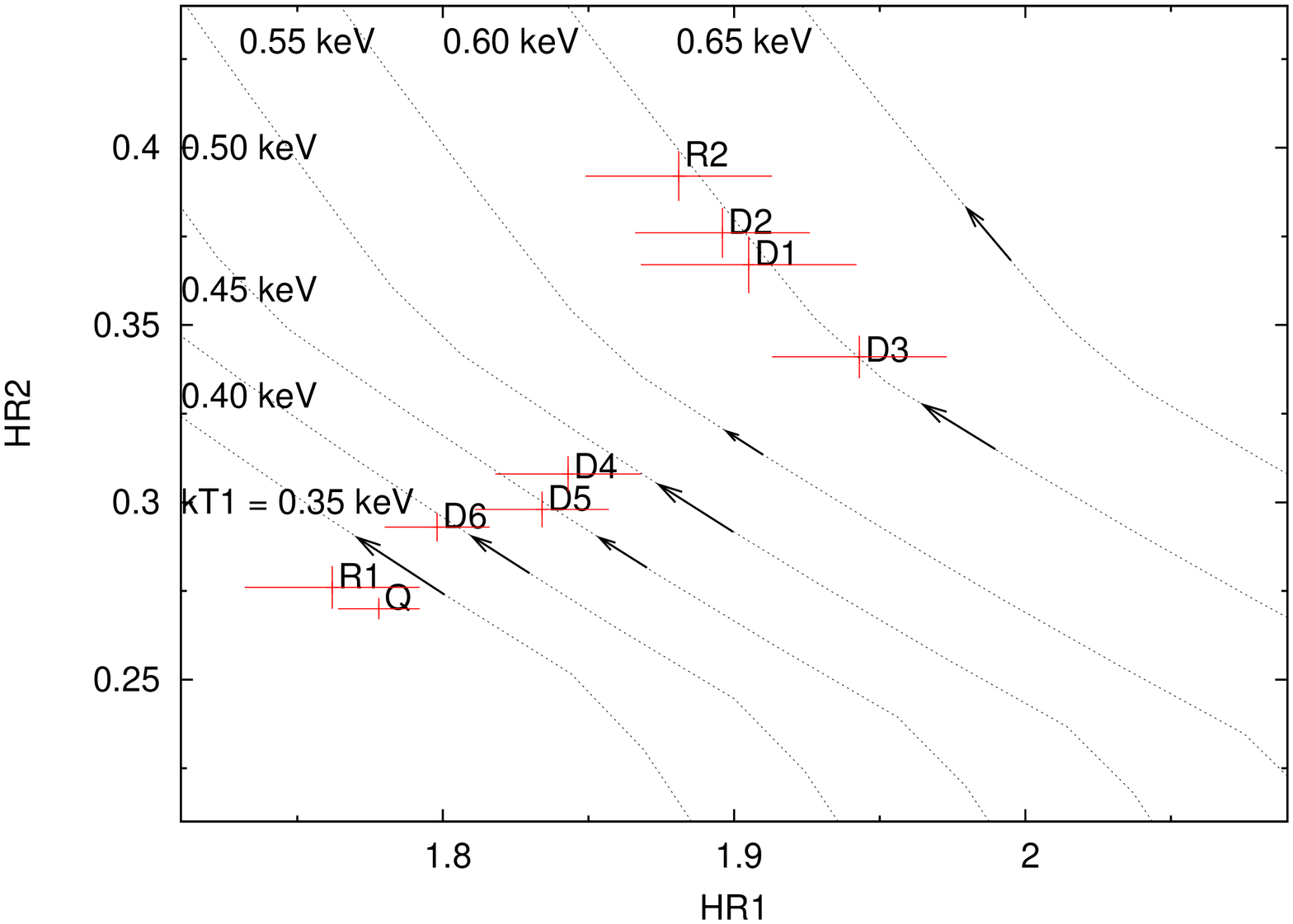}}
\subfigure[Flare F6 from V711 Tau]{\includegraphics[width=35mm,angle=-90]{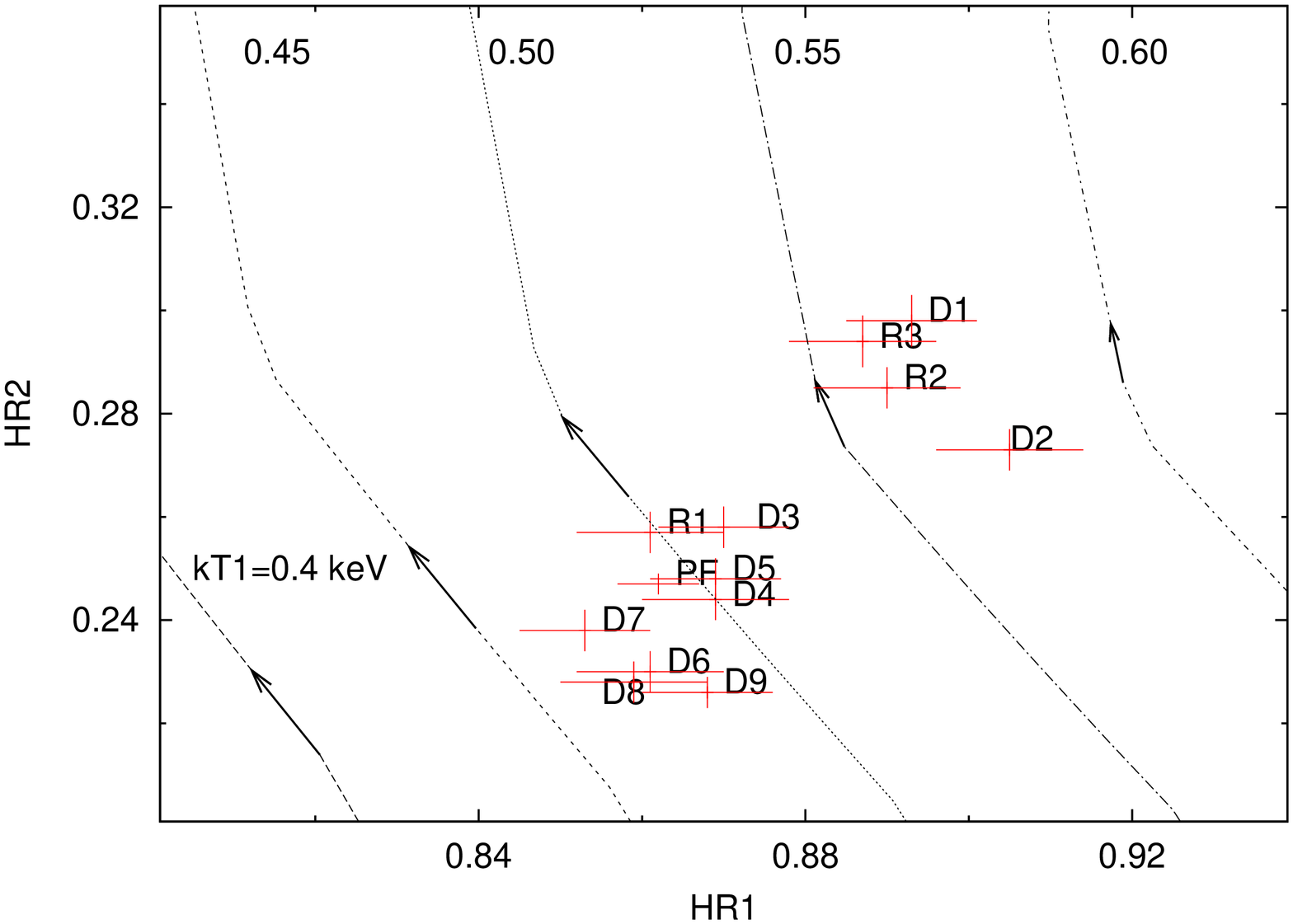}}
\subfigure[Flare F7 from EI Eri]{\includegraphics[width=35mm,angle=-90]{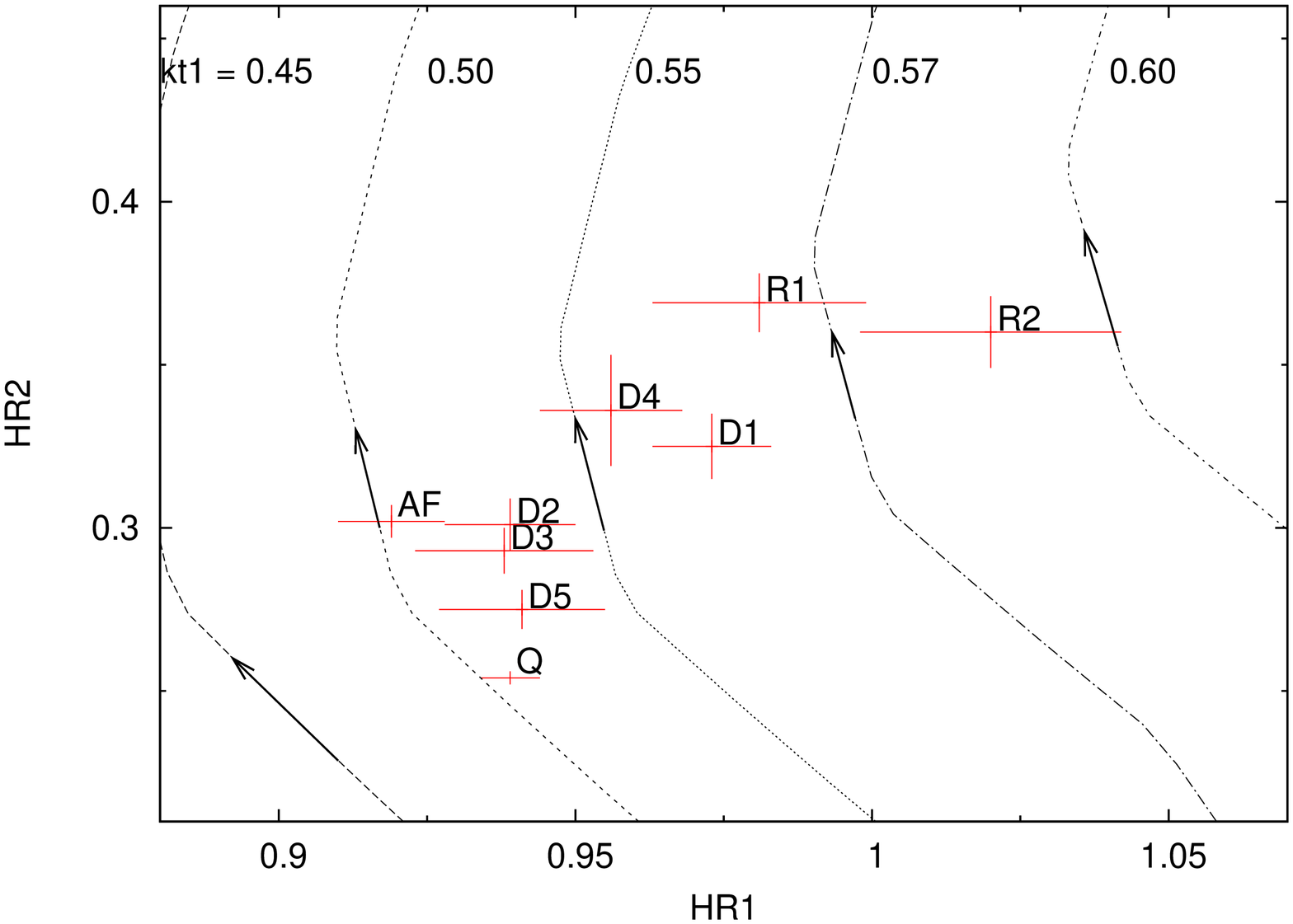}}
\caption{Colour-colour diagram for the various time segments as shown in Fig. \ref{fig:mospnlc}. Here Q represents quiescent state, Ri(i = 1,2,3) represents rise phase and Di(i=1,2,.....) decay phase of the flare. The curves overplotted are for model simulation where two-temperature plasma is assumed to predict the colours. The increasing directions of kT2 is shown by arrows.}
\label{fig:hr}
\end{figure*}

Hardness ratios plotted in the form of colour-colour (CC) diagrams can reveal information about spectral variations and serve as a guide for a more detailed spectral analysis.
 For this purpose,  the light curves and  the hardness ratio curves of the flares were divided into different time segments covering their rising and decaying phases as shown in Fig. \ref{fig:mospnlc}, where Ri(i=1,2..) represents the  rising phase, Di(i=1,2,..) represents the decay phase.
 The HR1 and HR2 values obtained from the PN data for each  of the flare segments  are plotted in Figs. \ref{fig:hr} (a-f).  To understand the observed behavior of HR1 and HR2 in terms of simple spectral models, we generated the soft, medium and hard bands count rate using the {\it ``fakeit''} provision in the XSPEC, using the most recent Canned Response Matrices downloaded from http://xmm.vilspa.esa.es/external/xmm\_SW\_cal/calib/epic \_files.html, and simple two temperature plasma models. The two temperature {\sc apec}  models were mixed such that the emission measure were in the ratio EM$_1$/EM$_2$ ranges from  0.5 to 2.0.  The abundances (0.1 to 0.3) and hydrogen column density ($10^{20}$ cm$^{-2}$) were kept fixed for each set of the EM$_1$/EM$_2$.  Further by keeping  kT$_1$ fixed at different values ranging from 0.35 to 0.9 keV in steps of 0.05 and varying kT$_2$ from 1.0 to 5.5 keV such that kT$_2 > $kT$_1$, we generated count rates in the  soft, medium and hard bands. These predicted count rates were used to generate the hardness ratios.  The families of curve thus generated are over-plotted on the data points in Fig. \ref{fig:hr} to understand the CC diagram where  kT$_2$ increases from bottom to top of each curve shown by arrows.

For flare F1, most of the flare segments intersect the CC diagram near kT1= 0.6 and 0.7 keV, indicating that the temperature during the flare remains constant. The best match CC diagram for the flare F1 is shown in Fig. \ref{fig:hr} (a), where we assumed  $EM_1/EM_2 = 0.5$ and $Z = 0.2Z_\odot$.
Fig. \ref{fig:hr} (b) shows the CC diagram for  a set of $EM_1/EM_2 = 1$ and $Z = 0.3Z_\odot$ for the star $\sigma$ Gem, where the flare peak part D2/D3 intersects the  generated CC curves near kT1 = 0.8 keV, however, the bottom part D11 intersects the curve near kT1 = 0.75 keV. This implies that at the flare top plasma temperature is high than the flare bottom.   A similar pattern was also seen for all the flares observed (see Figs. \ref{fig:hr}c-f), where the top of the flare is located near the CC curve for the high temperature component whereas the colours near the end of the flare intersects the CC curve corresponding to the lower temperature.  It appears that the quiescent states of the stars are cooler than the corresponding flare states. The temperature is high at the flare peak, and the plasma temperature decreases as flare decays.   For flare F3, F6 and F7 the simulated CC diagrams are well matched with the observed one for the parameters kT$_1$= 0.4-0.8 keV, kT$_2$ = 1.0 - 3.0 keV, Z= 0.1Z$_\odot$ and EM$_1$/EM$_2$ = 1.0.  However, for the  the flare F5 the   observed CC diagram are well matched with the simulated families of CC curve for a set of $EM_1/EM_2 = 0.5$ and $Z= 0.2Z_\odot$.

\subsection{The Quiescent coronae}
\label{sec:quiescent}
\begin{figure}
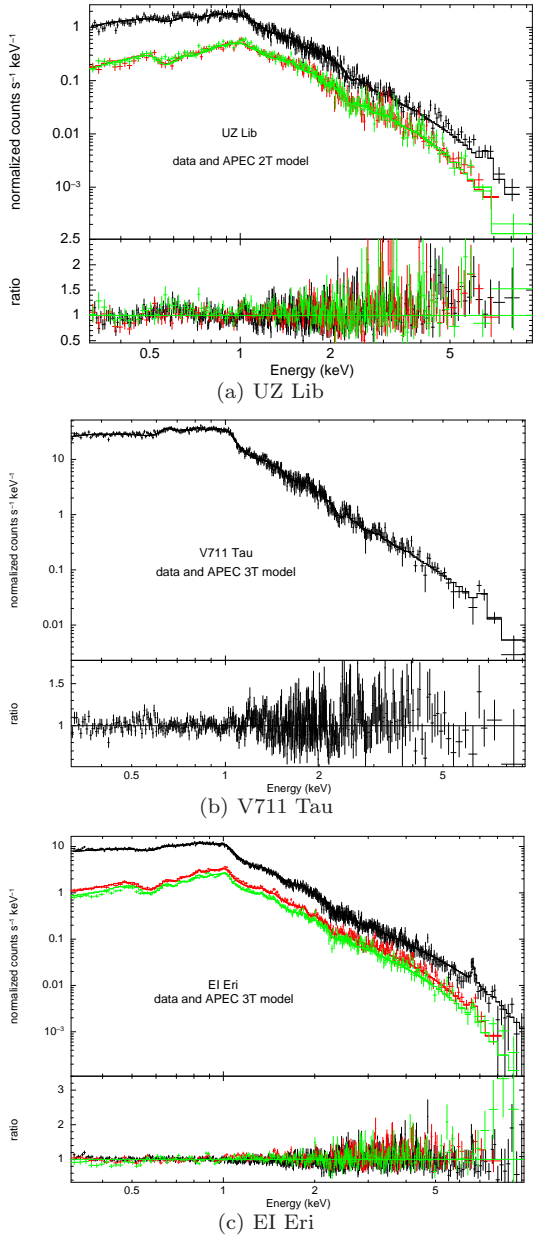

\subfigure[UZ Lib]{\includegraphics[width=50mm,angle=-90]{UZLib_Spec_Q.ps}}
\subfigure[V711 Tau]{\includegraphics[width=50mm,angle=-90]{V711Tau_Spec_Q.ps}}
\subfigure[EI Eri]{\includegraphics[width=50mm,angle=-90]{EIEri_Spec_Q.ps}}
\caption{Quiescent state spectra of MOS and PN data with APEC 2T/3T model in upper
subpanels of each graph. $\chi^2$ contributions in terms of ratio are given in the lower subpanel of the each graph.}
\label{fig:qspec}
\end{figure}

X-ray spectra for each star during the quiescent state were analysed
using XSPEC, version 12.5 (Arnaud 1996).  Spectral analysis of EPIC data was performed in the energy band between 0.3 and 10.0 keV.  Individual spectra were binned so as to have a minimum of 20 counts per energy bin.  The EPIC spectra of the stars were fitted with a single (1T), two (2T) and three (3T) temperature collisional plasma models known as {\sc apec} (Smith et al. 2001), with variable elemental abundances. Abundances ($Z$) for all the elements in the {\sc apec} were varied together. The interstellar hydrogen column density ($N_H$ ) was left free to vary. For all the stars, no 1T or 2T plasma models with solar photospheric (Anders \& Grevesse 1989) abundances ($Z_\odot$) could fit the data, as unacceptably large values of $\chi^2_\nu$ $=  \chi^2/\nu$ (where $\nu$ is degrees of freedom) were obtained. For the quiescent state of UZ Lib, acceptable {\sc apec} 2T fits  ($\chi^2_\nu$=1.21) were achieved only when the abundances were allowed to depart from the solar values.  Gondoin (2003) had also analysed the same data of UZ Lib  and achieved the best fit by using the 4T {\sc apec} model with $\chi^2_\nu$ of 1.18. It appears that there is not much improvement in $\chi^2_\nu$  while fitting with 4T {\sc apec} model. Therefore, we assume that  quiescent corona of UZ Lib is best represented by 2T plasma.  For the quiescent state of V711 Tau and EI Eri, only 3T plasma models with subsolar abundances were found to be acceptable. The best-fitting plasma models along with the significance of the residuals in terms of ratio are shown in Fig. \ref{fig:qspec} for the stars UZ Lib, V711 Tau and EI Eri. Table \ref{tab:q} summarizes the best-fitting values obtained for the various parameters along with the minimum  $\chi_{\nu}^2$  and the 90 per cent confidence error bars estimated from the minimum  $\chi^ 2 + 2.71$(Arnaud 1996).

\begin{figure}
\subfigure[]{\includegraphics[width=55mm,angle=-90]{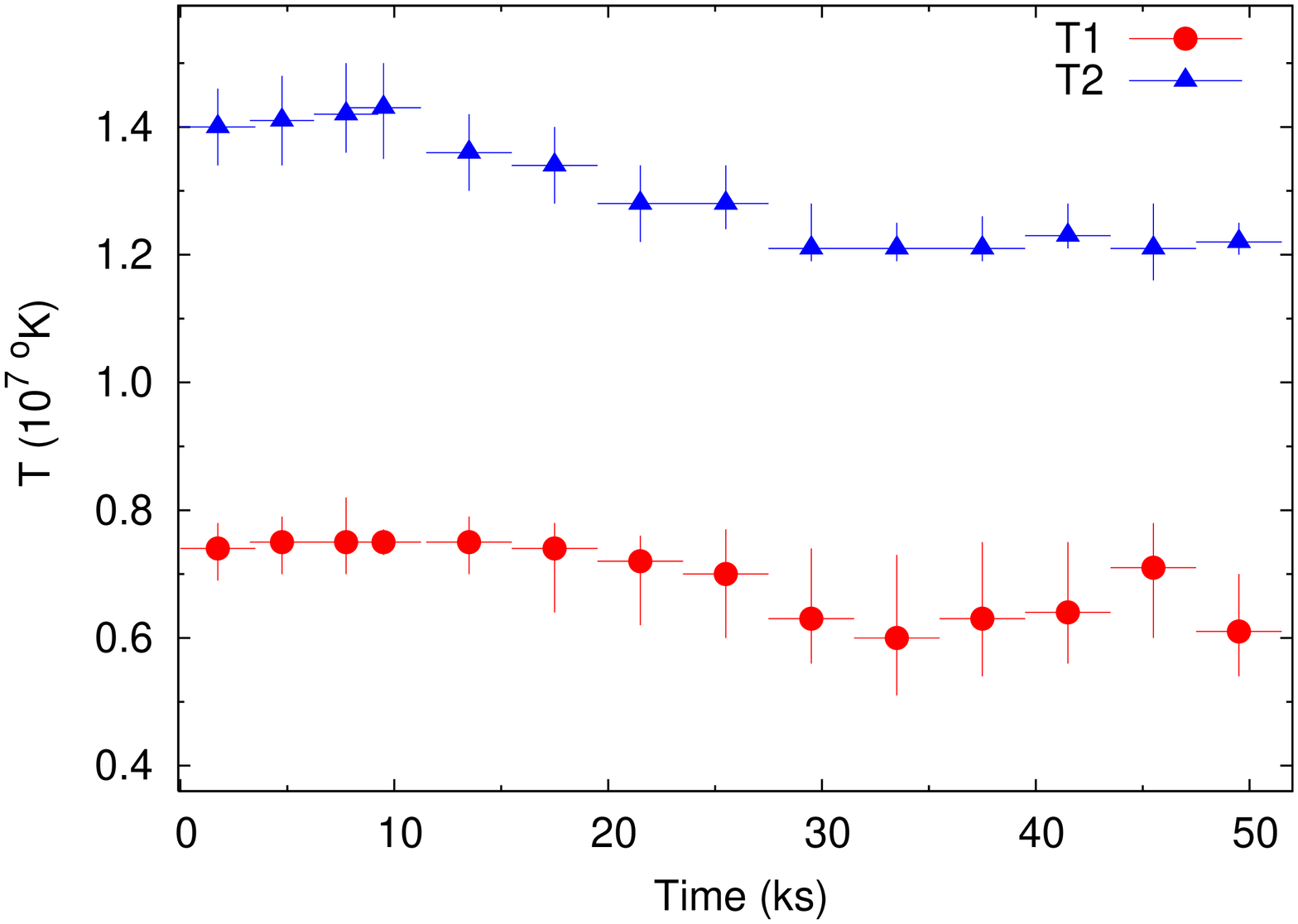}}
\subfigure[]{\includegraphics[width=55mm,angle=-90]{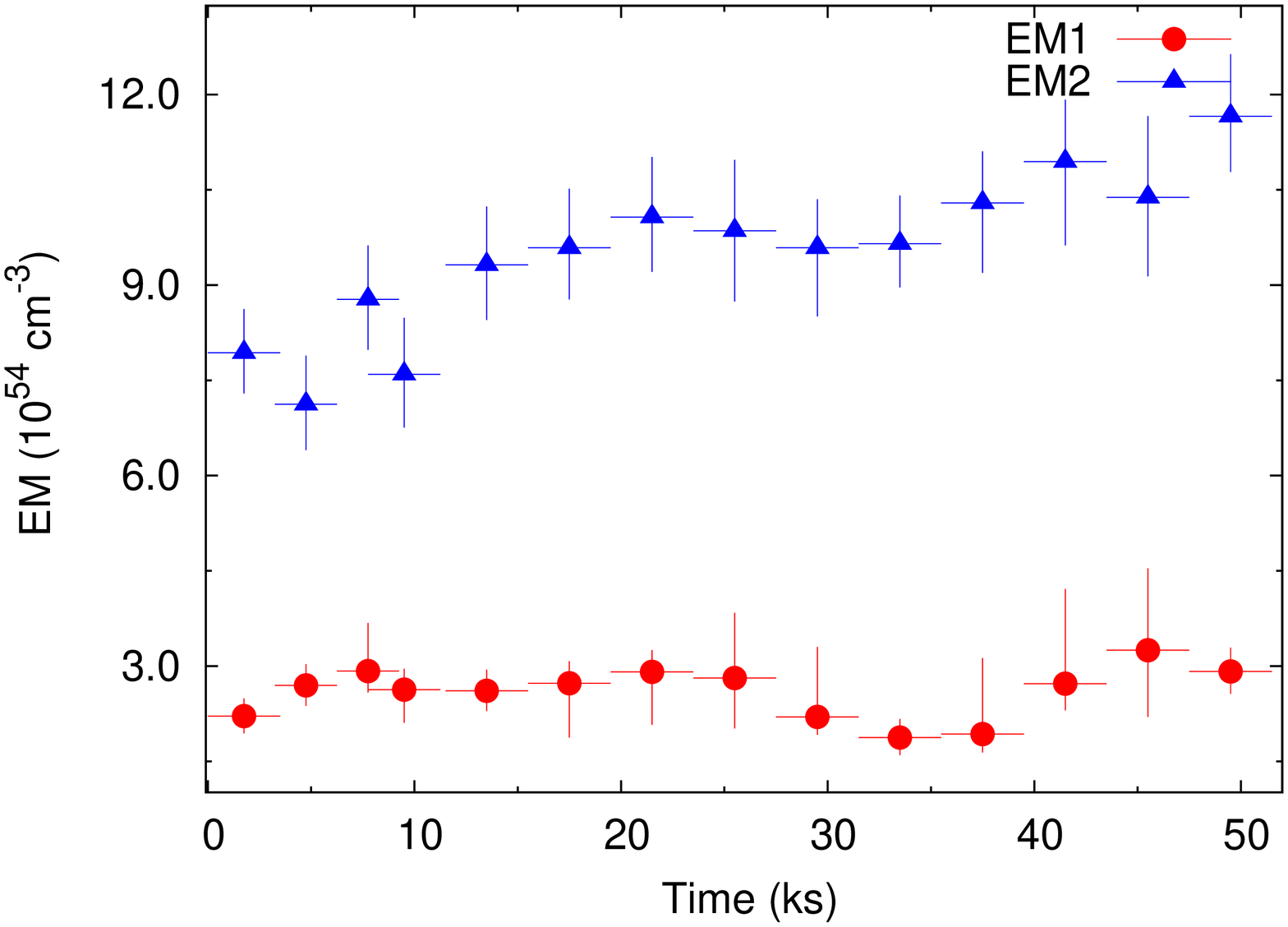}}
\caption{Variation of (a) kT$_1$ and kT$_2$ and (b)EM$_1$ and EM$_2$ during the flare F2 from $\sigma$ Gem.}
\label{fig:sigma_q}
\end{figure}

\begin{figure}
\subfigure[]{\includegraphics[width=5.5cm,angle=-90]{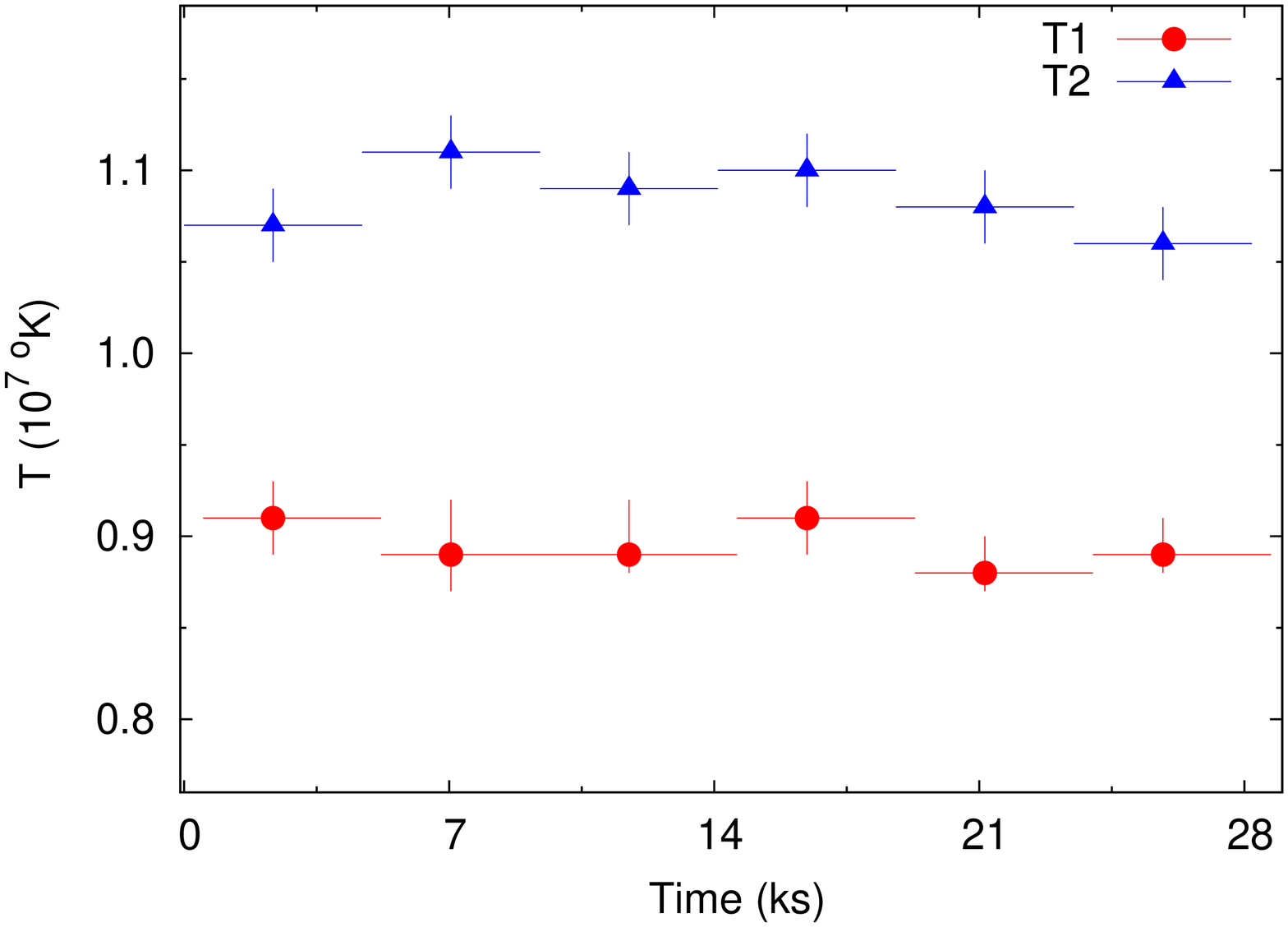}}
\subfigure[]{\includegraphics[width=5.5cm,angle=-90]{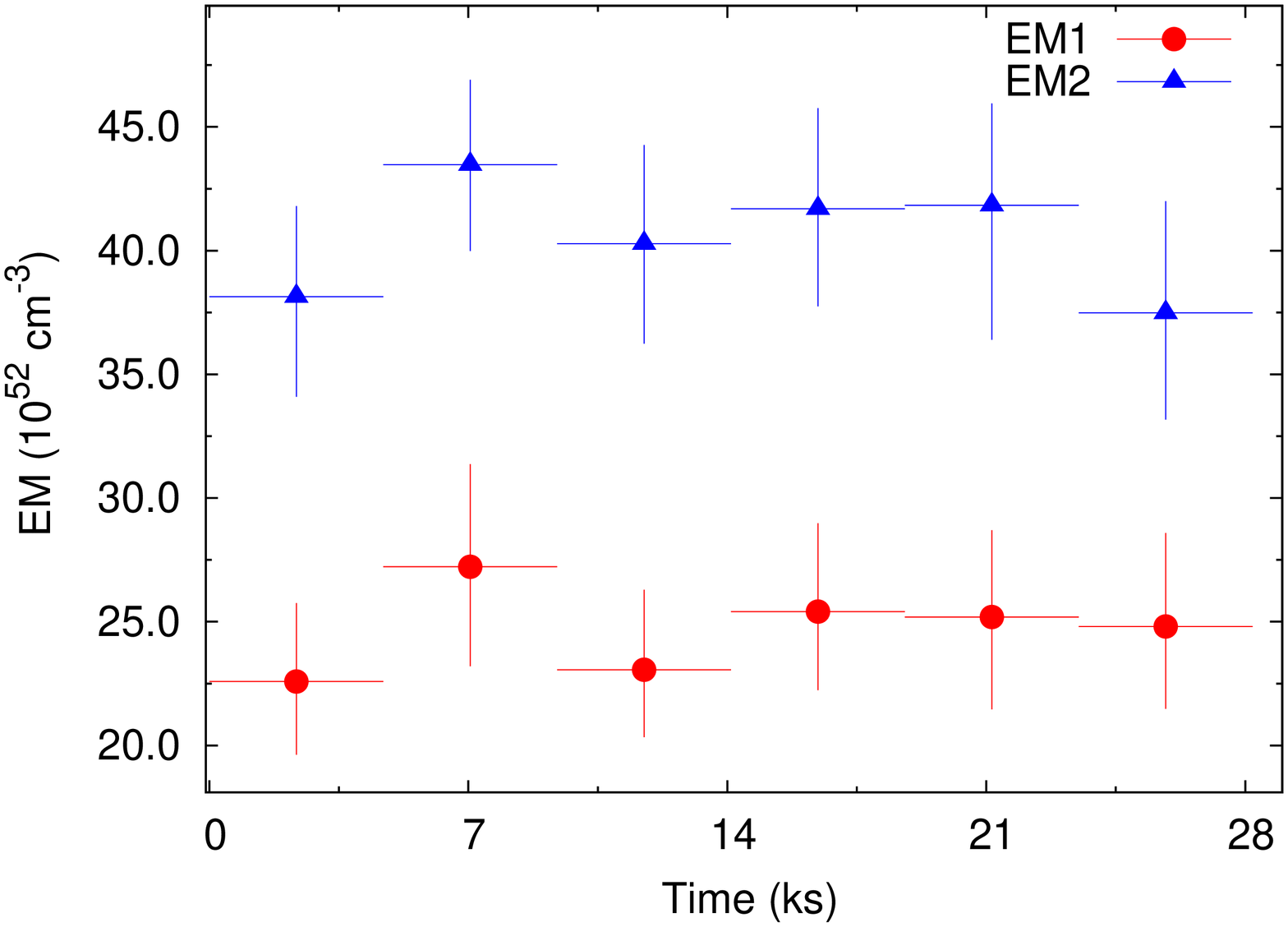}}
\caption{Variation of (a) kT$_1$ and kT$_2$ and (b)EM$_1$ and EM$_2$ during the flare F3 from $\lambda$ And.}
\label{fig:lambda_q}
\end{figure}

The quiescent states of $\sigma$ Gem and $\lambda$ And were not observed during the XMM-Newton observations.  The 3T plasma models were fitted to the X-ray spectra of each flare segment (for details see section 4.4.2 and 4.4.3) of these two stars. In the spectral fitting, all the  parameters were left free to vary.  Figs. \ref{fig:sigma_q} and \ref{fig:lambda_q}  show the variation of  temperatures (T$_1$ and T$_2$) and emission measures (EM$_1$ and EM$_2$) corresponding to the cool components during the flares for the stars $\sigma$ Gem and $\lambda$ And, respectively. The temperatures and emission measures corresponding to the cool components are found to be constant during the flare state of both stars. Therefore, it appears that the temperatures and emission measures corresponding to the cool components are related to the quiescent coronae of these stars.
 The average values of the $N_H$, Z, $kT_1$, $kT_2$, $EM_1$ and $EM_2$ for these two stars are also given in Table \ref{tab:q}. The average value of $EM_{1,2}$ and $kT_{1,2}$ are determined using the following formulae
 \begin{displaymath}
 EM = \frac{1}{n} \sum_{i=1}^n EM_i ~~~~~~{\rm {and}}~~~~~~~
 kT = \frac{1}{EM}\sum_{i=1}^n EM_i kT_i
 \end{displaymath}

\noindent
Here, n is total number of flare segments. The derived values of quiescent state temperatures and emission measures for the star $\sigma$  Gem and $\lambda$ And  are found to be similar to that derived from the ROSAT data (Shi et al. 1998; Ortolani et al. 1997).

The derived values of $N_H$ from the spectral analysis were found
to be lower than that of the total galactic {\sc Hi} column density (Dickey \& Lockman 1990) towards the direction of that star.  The cool and hot temperatures of quiescent coronae  of these stars were found in the range of 0.3-0.8 keV and 0.9 - 2.0 keV, respectively. The corresponding EM$_2$/EM$_1$ during the quiescent states of these stars were found in the range of 1 - 4.

\begin{table*}
\caption{Spectral parameters derived for the quiescent emission of each program star from the analysis of PN and MOS spectra  accumulated during the time interval 'Q' as shown in Fig. 1, using {\sc apec} 2T/3T models.}
\label{tab:q}
\begin{tabular}{lcccrcrcccr}
\hline
Object        & N$_H$ & Z & kT$_1$& EM$_1$~& kT$_2$& EM$_2$~& kT$_3$ & EM$_3$~&\lx & $\chi_\nu^2$ (DOF)\\
\hline
UZ Lib        &5.0$_{-0.5}^{+0.5}$ & 0.15$_{-0.02}^{+0.03}$ & 0.80$_{-0.02}^{+0.02}$ &  3.9$_{-0.6}^{+0.7}$   & 1.95$_{-0.06}^{+0.06}$ &  8.1$_{-0.3}^{+0.3}$   &     ...           &   ...               &8.69 & 1.21(931)\\
$\sigma$ Gem* &2.3$_{-0.2}^{+0.2}$ & 0.26$_{-0.02}^{+0.02}$ & 0.59$_{-0.07}^{+0.07}$ & 26.0$_{-8.0}^{+6.0}$   & 1.18$_{-0.07}^{+0.07}$ & 94.8$_{+5.6}^{-8.0}$   &     ...           &   ...               &...  & ...\\
$\lambda$ And*&3.0$_{-0.6}^{+0.6}$ & 0.11$_{-0.01}^{+0.01}$ & 0.34$_{-0.01}^{+0.02}$ & 24.7$_{-3.3}^{+3.6}$   & 0.93$_{-0.02}^{+0.02}$ & 40.5$_{+4.2}^{-4.0}$   &     ...           &   ...               &...  & ...\\
V711 Tau      &1.8$_{-0.5}^{+0.5}$ & 0.16$_{-0.02}^{+0.02}$ & 0.36$_{-0.01}^{+0.01}$ & 27.0$_{-3.1}^{+3.4}$   & 0.95$_{-0.02}^{+0.02}$ & 43.8$_{-5.1}^{+5.5}$   &2.5$_{-0.2}^{+0.3}$& 17.6$_{-3.1}^{+2.9}$& 5.95& 1.19(545)\\
EI Eri        &1.2$_{-0.4}^{+0.4}$ & 0.17$_{-0.02}^{+0.02}$ & 0.42$_{-0.02}^{+0.02}$ & 20.8$_{-2.4}^{+2.4}$   & 0.93$_{-0.02}^{+0.02}$ & 45.6$_{-5.5}^{+5.5}$   &2.5$_{-0.2}^{+0.2}$& 25.9$_{-3.1}^{+3.1}$& 7.11& 1.01(655)\\
\hline
\end{tabular}
\\
{\it Note:}
$N_H$ is in unit of $10^{20}$ cm$^{-2}$, Z is the best fit values of abundances, temperatures (kT) are in keV, emission measures (EM) are in $10^{53}$ cm$^{-3}$, and X-ray luminosity (\lx) is in $10^{30}$ \lum.  $\chi_\nu^2$ is the minimum reduced $\chi^2$ and DOF stands for degrees of freedom.\\
{*} Average values of parameters corresponding to the cool components from the best fit 3-T plasma model to each flare segment.
\end{table*}


\subsection{Spectral evolution of X-ray flares}
\label{sec:flare_evol}
The spectra of the different time intervals shown in Fig. \ref{fig:mospnlc} were analysed to trace the spectral changes during the flare. The temporal evolution of the spectral parameters of the flares detected from the targeted stars are shown in Figs. \ref{fig:uzlibfevol} - \ref{fig:v711taufevol} and are described below.

\subsubsection{Flare F1 from UZ Lib}
\begin{figure}
\includegraphics[width=90mm]{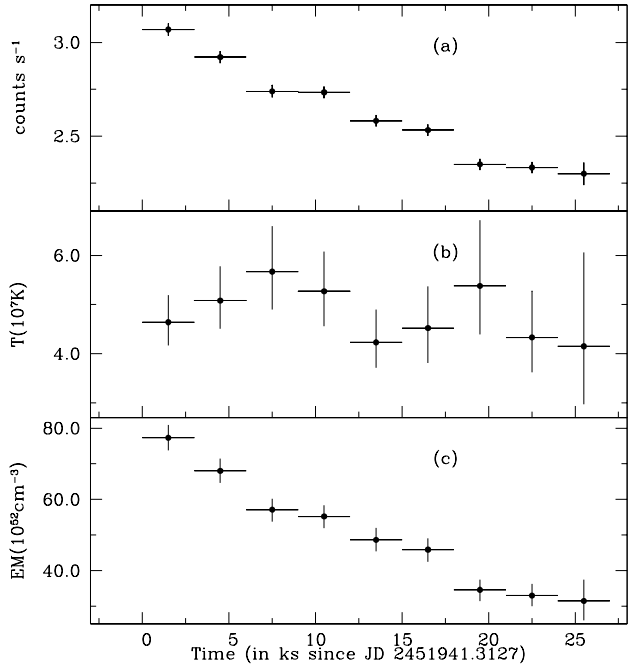}
\caption{Variation of (a) count rate, (b) temperature and (c) emission measure  during the XMM-Newton observations of the flare F1 from UZ Lib.  }
\label{fig:uzlibfevol}
\end{figure}

The flare F1 was divided into nine time bins.  To study the flare emission only, we have performed 3T spectral fits of
the data, with the quiescent emission taken into account by including
its best-fitting 2T model as a frozen background contribution. This is
equivalent to considering the flare emission subtracted of the quiescent
level, and allows us to derive one `effective' temperature and one EM of the flaring plasma.
Initially, the $N_H$ and $Z$ were left free to vary in the spectral fitting  of all flare segments and were found to be similar to that of quiescent state. Therefore, in the next stage of spectral fitting both were kept fix.  Fig. \ref{fig:uzlibfevol} shows the temporal evolution of temperature and emission measures along with X-ray light curves of the flare F1.
The temperature was found to be constant during the flare F1.   However,  the corresponding emission measure was found to decreased  by a factor of $\sim 2.5$ along to the decay path.  No firm conclusion can be drawn for the evolution of temperature and emission measure as rise and peak phases of the flare were not observed.  During the flare F1, the star was found $\sim2$ times more X-ray luminous than its quiescent state (see Tables \ref{tab:q} and \ref{tab:f}).

\subsubsection{ Flare F2 from $\sigma$ Gem}
As mentioned in \S \ref{sec:quiescent}, no quiescent state was observed for the star $\sigma$ Gem. We have made fourteen time bins  for flare F2 to study the flare evolution.  The temperature and the corresponding emission measure of the hottest component from the best fit 3T plasma model were found to vary along the flare, therefore, appear to be related to the flare emission.
 As shown in Fig. \ref{fig:sigmagemfevol}(b), the $N_H$ increased during the rise phase and decreased during the  decay phase of the flare F2. The value of $N_H$  reached  $2.7\times10^{20}$ cm$^{-2}$, which is a factor of $\sim2$ more than the minimum observed value.  The temperature and the $Z$ appear to be highest at the very beginning of the flare (R1) and decrease thereafter (see Figs. \ref{fig:sigmagemfevol} c and d). The highest value of $Z$ were found to be $0.32Z_\odot$, which is $\sim 1.5$ more than the minimum value observed. During the flare F2, the temperature evolved before the emission measure did. It peaked during the rise phase R1, whereas the emission measure peaked during the decay phase of the flare F2.  The emission measure was found to vary along the flare intensity. The temperature and emission measure were increased by  factors  of $\sim 1.8$ and $\sim 1.3$ times, respectively.  The highest luminosity during the flare F2 was $10^{31.81}$ \lum, which is three to six times more than that observed from EXOSAT (Singh et al. 1987) and ROSAT (Yi et al. 1997) observations.

\subsubsection{ Flare F3 from $\lambda$ And}
In the flare F3, only the rise and the peak phases  were observed.  Therefore, an approach similar  to that applied for flare F2 was applied to the spectral analysis for five time bins of  this flare. As shown in Fig. \ref{fig:lambdaandfevol} (b) and (c), the $N_H$  and $Z$ were found to be constant during the rising phase of the flare F3. The evolution of the temperature and the emission measure corresponding to the  hottest component in the best fit 3T plasma model  are shown in  Figs. \ref{fig:lambdaandfevol} (d) and (e), respectively. The emission measure was found to increase as flare rises. However, the temperature peaked during the flare segment 'R2' and  decreased as flare rises.  It appears that the temperature had peaked before the emission measure. The luminosity reached a value of $4.6\times10^{30}$ \lum, which is $\sim$ 25\% more than that observed $\sim 5$ years before by ASCA and and $\sim 10$ years before by ROSAT during its quiescent state( see Ortolani et al. 1997).

\subsubsection{ Flares F4, F5 and F6 from V711 tau}
Three flares (F4,F5 and F6) were observed during three different observations of V711 Tau. The flare F4 was a shortest duration flare among all observed flares. Therefore, only two time bins were made during this flare.  However, the flares F5 and F6 were divided into 8 and 10 time bins, respectively. For the flares F4, F5 and  F6 of V711 Tau, we have fitted the 4T {\sc apec} model to the spectral data of each time bin, with the quiescent emission taken into account by including its best-fitting 3T model as a frozen background contribution. The $N_H$ and $Z$ were left free to vary during the spectral fitting. For the flare F4, the values of $N_H$ and $Z$ were found similar to that of quiescent state, therefore, were not allowed to vary during the spectral fitting. The temperature and the corresponding emission measure were found to peaked during the decay phase of the flare F4 (see Table \ref{tab:f}).  The  variation of the temperature, the emission measure, $Z$ and $N_H$  along with count rate for the flare F5 and F6  are shown in Fig. \ref{fig:v711taufevol1} and Fig. \ref{fig:v711taufevol}, respectively. The $N_H$ was found to be constant with in 1$\sigma$ level during the flares F5 and F6. The values of $N_H$ during flares F5 and F6 were also found to be similar to that of quiescent state.  The $Z$ were found to vary during the flares F5 and F6. During the flare F5, the $Z$  increased by factor of 1.5. However, this increment  is well within $1.7\sigma$ level from the quiescent state.  In case of flare F6, $Z$ varied in similar fashion to that of count rates and increased by a factor of $\sim 1.5$ from the quiescent state. The emission measure varied along with the rise and decay phase of the light curve for the flares F5 and F6 (see Figs. \ref{fig:v711taufevol1}a,e and \ref{fig:v711taufevol}a,e). However, the temperature peaked before the emission measure. The temperature and emission measure during the flare F5 increased by factors of $\sim3$ and $\sim2$, respectively.  However, ratio of temperature and  emission measure at flare peak and flare bottom for the flare F6 were found to be 2.8 and 2.6, respectively. The X-ray luminosities at the flare peaks  were found 1.1, 1.8 and 1.5 times more than that of quiescent state for the flares F4, F5 and F6, respectively.

\subsubsection{Flare F7 from EI Eri}
Fig. \ref{fig:eierifevol} shows the spectral evolution of the flare F7  of
the star EI Eri.  Similar to flares observed from V711 Tau, we have fit the
4T plasma model keeping fixed 3T model parameters from quiescent state.
$N_H$ and $Z$ were left free to vary during the spectral fitting. We did 
not observe any significant variation in $N_H$ during the flare F7.  As
shown in Fig. \ref{fig:eierifevol} and table \ref{tab:f}, $Z$, temperature
and emission measure were found to vary during the flare. The $Z$  reached
a value of $0.22\pm0.01Z_\odot$ at flare peak and thereafter decreased to a
value of $0.16\pm0.01Z_\odot$ at the  bottom of the flare light curve. In the
flare F7, temperature was peaked before the emission measure.  During the
flare F7 temperature and emission measure were increased by a factor of
$\sim3$ and $\sim5.5$, respectively.  The X-ray luminosity reached  a value
of $10^{31.1}$ \lum, which is $\sim 2$ times more than that of the quiescent
state.


\begin{table}
\scriptsize\addtolength{\tabcolsep}{-5pt}
\caption{Spectral parameters for each temporal segment of the flares as marked by vertical lines in Fig. \ref{fig:mospnlc}.}
\label{tab:f}
\begin{tabular}{rcccccccc}
\hline
Object (LS) & FS & $N_H$ & Z &kT& EM~&\lx & $\chi_\nu^2$ (DOF) \\
\hline
UZ Lib       (F1)& D1 & \nodata             &  \nodata                  & 4.0$_{-0.4}^{+0.5}$    & 77$_{-3}^{+3}$          & 17.98 & 0.99(315)\\
                 & D2 & \nodata             &  \nodata                  & 4.4$_{-0.5}^{+0.6}$    & 68$_{-3}^{+3}$          & 17.24 & 1.09(311)\\
                 & D3 & \nodata             &  \nodata                  & 4.9$_{-0.7}^{+0.8}$    & 57$_{-3}^{+3}$          & 16.20 & 0.97(289)\\
                 & D4 & \nodata             &  \nodata                  & 4.5$_{-0.6}^{+0.7}$    & 55$_{-3}^{+3}$          & 15.73 & 1.07(291)\\
                 & D5 & \nodata             &  \nodata                  & 3.6$_{-0.5}^{+0.6}$    & 49$_{-3}^{+3}$          & 14.28 & 1.05(283)\\
                 & D6 & \nodata             &  \nodata                  & 3.9$_{-0.6}^{+0.7}$    & 46$_{-3}^{+3}$          & 14.12 & 0.95(278)\\
                 & D7 & \nodata             &  \nodata                  & 4.6$_{-0.8}^{+1.2}$    & 35$_{-3}^{+3}$          & 13.13 & 0.91(267)\\
                 & D8 & \nodata             &  \nodata                  & 3.7$_{-0.6}^{+0.8}$    & 33$_{-3}^{+3}$          & 12.54 & 1.08(260)\\
                 & D9 & \nodata             &  \nodata                  & 3.6$_{-1.0}^{+1.7}$    & 31$_{-5}^{+6}$          & 12.29 & 1.09(89)\\
$\sigma$ Gem (F2)*& R1 & 1.6$_{-0.2}^{+0.2}$ & 0.32$_{-0.02}^{+0.02}$    & 6.4$_{-0.2}^{+0.2}$    & 346$_{-6}^{+6}$         & 64.48 & 1.07(1428)\\
                 & R2 & 1.7$_{-0.2}^{+0.2}$ & 0.29$_{-0.02}^{+0.02}$    & 5.9$_{-0.2}^{+0.2}$    & 363$_{-6}^{+6}$         & 65.40 & 1.05(1345)\\
                 & R3 & 2.0$_{-0.2}^{+0.2}$ & 0.28$_{-0.02}^{+0.02}$    & 5.7$_{-0.2}^{+0.2}$    & 357$_{-7}^{+7}$         & 65.18 & 1.17(1332)\\
                 & D1 & 2.2$_{-0.2}^{+0.2}$ & 0.30$_{-0.02}^{+0.02}$    & 5.0$_{-0.1}^{+0.1}$    & 377$_{-8}^{+7}$         & 64.12 & 1.11(1367)\\
                 & D2 & 2.7$_{-0.2}^{+0.2}$ & 0.28$_{-0.02}^{+0.02}$    & 4.9$_{-0.1}^{+0.1}$    & 370$_{-8}^{+8}$         & 62.80 & 1.15(1385)\\
                 & D3 & 2.6$_{-0.2}^{+0.2}$ & 0.27$_{-0.01}^{+0.02}$    & 4.6$_{-0.1}^{+0.1}$    & 365$_{-8}^{+6}$         & 61.09 & 1.18(1361)\\
                 & D4 & 2.4$_{-0.2}^{+0.2}$ & 0.27$_{-0.02}^{+0.02}$    & 4.7$_{-0.1}^{+0.1}$    & 339$_{-7}^{+6}$         & 58.96 & 1.10(1354)\\
                 & D5 & 2.4$_{-0.2}^{+0.2}$ & 0.25$_{-0.01}^{+0.02}$    & 4.4$_{-0.1}^{+0.1}$    & 332$_{-5}^{+4}$         & 56.00 & 1.13(1308)\\
                 & D6 & 2.4$_{-0.2}^{+0.2}$ & 0.26$_{-0.02}^{+0.02}$    & 4.37$_{-0.08}^{+0.11}$ & 325$_{-5}^{+4}$         & 54.05 & 1.07(1288)\\
                 & D7 & 2.2$_{-0.2}^{+0.2}$ & 0.26$_{-0.02}^{+0.02}$    & 4.19$_{-0.08}^{+0.08}$ & 320$_{-4}^{+4}$         & 52.49 & 1.17(1263)\\
                 & D8 & 2.2$_{-0.2}^{+0.2}$ & 0.24$_{-0.02}^{+0.02}$    & 4.00$_{-0.08}^{+0.08}$ & 315$_{-5}^{+4}$         & 51.34 & 1.07(1249)\\
                 & D9 & 2.4$_{-0.2}^{+0.2}$ & 0.23$_{-0.02}^{+0.02}$    & 3.95$_{-0.08}^{+0.09}$ & 307$_{-6}^{+5}$         & 49.86 & 1.16(1223)\\
                 &D10 & 2.3$_{-0.2}^{+0.2}$ & 0.23$_{-0.02}^{+0.02}$    & 3.86$_{-0.09}^{+0.09}$ & 292$_{-5}^{+5}$         & 48.11 & 1.05(1199)\\
                 &D11 & 2.6$_{-0.2}^{+0.2}$ & 0.22$_{-0.02}^{+0.02}$    & 3.6$_{-0.08}^{+0.08}$  & 286$_{-5}^{+5}$         & 46.38 & 1.16(1166)\\
$\lambda$ And(F3)*& R1 & 2.9$_{-0.7}^{+0.6}$ & 0.11$_{-0.01}^{+0.01}$    & 3.4$_{-0.4}^{+0.6}$    &  8$_{-1}^{+1}$          & 3.74  & 1.22(513)\\
                 & R2 & 3.6$_{-0.7}^{+0.6}$ & 0.09$_{-0.01}^{+0.01}$    & 4.6$_{-0.7}^{+0.8}$    &  8$_{-1}^{+1}$          & 4.15  & 1.15(551)\\
                 & R3 & 2.9$_{-0.6}^{+0.6}$ & 0.11$_{-0.01}^{+0.01}$    & 3.7$_{-0.4}^{+0.5}$    & 12$_{-1}^{+1}$          & 4.47  & 1.06(576)\\
                 & R4 & 3.2$_{-0.6}^{+0.6}$ & 0.10$_{-0.01}^{+0.01}$    & 3.8$_{-0.4}^{+0.5}$    & 12$_{-1}^{+1}$          & 4.58  & 1.13(587)\\
                 & R5 & 3.1$_{-0.7}^{+0.6}$ & 0.11$_{-0.01}^{+0.02}$    & 3.4$_{-0.4}^{+0.4}$    & 14$_{-1}^{+2}$          & 4.62  & 1.07(582)\\
                 & R6 & 2.9$_{-0.6}^{+0.6}$ & 0.11$_{-0.01}^{+0.01}$    & 2.9$_{-0.2}^{+0.2}$    & 17$_{-2}^{+2}$          & 4.61  & 1.20(587)\\
V711 Tau     (F4)& R1 & \nodata             & \nodata                   & $< 3.3$                & 2.2$_{-0.7}^{+0.7}$     & 6.1   & 0.95(511)\\
                 & D1 & \nodata             & \nodata                   & 3.9$_{-0.9}^{+1.4}$    & 6.1$_{-0.9}^{+0.9}$     & 6.5   & 1.12(523)\\
             (F5)& R1 & 1.1$_{-0.4}^{+0.6}$ & 0.205$_{-0.011}^{+0.013}$ & 2.9$_{-0.9}^{+1.3}$    &  9.5$_{-2.1}^{+3.2}$    & 7.84  & 1.32(246)\\
                 & R2 & 1.6$_{-0.5}^{+0.5}$ & 0.223$_{-0.012}^{+0.012}$ & 4.0$_{-0.5}^{+0.8}$    & 27.7$_{-2.5}^{+2.5}$    & 10.52 & 1.27(285)\\
                 & D1 & 2.0$_{-0.6}^{+0.6}$ & 0.227$_{-0.014}^{+0.021}$ & 2.6$_{-0.3}^{+0.6}$    & 36.7$_{-4.3}^{+4.3}$    & 10.75 & 1.14(250)\\
                 & D2 & 2.0$_{-0.5}^{+0.5}$ & 0.227$_{-0.013}^{+0.013}$ & 2.9$_{-0.5}^{+0.3}$    & 36.9$_{-4.6}^{+3.1}$    & 10.67 & 1.35(290)\\
                 & D3 & 2.1$_{-0.5}^{+0.5}$ & 0.234$_{-0.012}^{+0.012}$ & 2.3$_{-0.4}^{+0.3}$    & 29.5$_{-2.8}^{+4.2}$    & 9.79  & 1.10(282)\\
                 & D4 & 1.1$_{-0.4}^{+0.4}$ & 0.229$_{-0.010}^{+0.015}$ & 2.1$_{-0.2}^{+0.5}$    & 20.9$_{-3.6}^{+2.4}$    & 9.11  & 1.22(290)\\
                 & D5 & 1.6$_{-0.4}^{+0.4}$ & 0.222$_{-0.009}^{+0.009}$ & 2.0$_{-0.2}^{+0.3}$    & 16.7$_{-2.1}^{+2.1}$    & 8.39  & 1.35(315)\\
                 & D6 & 1.6$_{-0.3}^{+0.3}$ & 0.212$_{-0.008}^{+0.008}$ & 2.1$_{-0.3}^{+0.4}$    & 12.5$_{-1.8}^{+1.8}$    & 7.88  & 1.36(338)\\
             (F6)& PF & 1.1$_{-0.1}^{+0.1}$ & 0.173$_{-0.004}^{+0.003}$ & 1.8$_{-0.1}^{+0.2}$    & 11.0$_{-0.9}^{+1.0}$    & 7.29  & 1.20(627)\\
                 & R1 & 1.1$_{-0.2}^{+0.2}$ & 0.169$_{-0.005}^{+0.005}$ & 2.1$_{-0.2}^{+0.4}$    & 12.1$_{-1.7}^{+1.4}$    & 7.44  & 1.11(479)\\
                 & R2 & 1.1$_{-0.2}^{+0.2}$ & 0.170$_{-0.006}^{+0.006}$ & 2.4$_{-0.2}^{+0.2}$    & 19.2$_{-1.6}^{+1.7}$    & 8.28  & 1.10(462)\\
                 & R3 & 1.2$_{-0.2}^{+0.2}$ & 0.182$_{-0.006}^{+0.006}$ & 2.5$_{-0.2}^{+0.2}$    & 22.9$_{-1.7}^{+1.7}$    & 8.80  & 1.11(495)\\
                 & D1 & 1.3$_{-0.2}^{+0.2}$ & 0.184$_{-0.007}^{+0.004}$ & 2.4$_{-0.2}^{+0.2}$    & 25.1$_{-1.8}^{+1.8}$    & 8.96  & 1.11(527)\\
                 & D2 & 1.1$_{-0.2}^{+0.2}$ & 0.191$_{-0.006}^{+0.006}$ & 2.0$_{-0.2}^{+0.2}$    & 20.9$_{-1.7}^{+1.8}$    & 8.51  & 0.93(476)\\
                 & D3 & 0.7$_{-0.2}^{+0.2}$ & 0.188$_{-0.006}^{+0.006}$ & 2.0$_{-0.2}^{+0.3}$    & 14.7$_{-1.7}^{+1.7}$    & 8.03  & 1.13(457)\\
                 & D4 & 0.9$_{-0.2}^{+0.2}$ & 0.184$_{-0.007}^{+0.007}$ & 1.7$_{-0.3}^{+0.3}$    & 11.6$_{-1.9}^{+1.9}$    & 7.55  & 1.09(440)\\
                 & D5 & 1.1$_{-0.2}^{+0.2}$ & 0.174$_{-0.007}^{+0.007}$ & 1.6$_{-0.2}^{+0.3}$    & 11.3$_{-1.9}^{+1.9}$    & 7.28  & 1.26(430)\\
                 & D6 & 1.1$_{-0.2}^{+0.2}$ & 0.170$_{-0.005}^{+0.010}$ & 1.5$_{-0.1}^{+0.2}$    &  9.8$_{-1.7}^{+1.7}$    & 7.14  & 1.16(424)\\
                 & D7 & 1.1$_{-0.2}^{+0.2}$ & 0.165$_{-0.010}^{+0.009}$ & 1.3$_{-0.2}^{+0.2}$    & 10.4$_{-2.3}^{+2.5}$    & 7.02  & 0.96(423)\\
EI Eri       (F7)& R1 & 0.8$_{-0.4}^{+0.4}$ & 0.178$_{-0.011}^{+0.011}$ & 4.9$_{-1.1}^{+1.6}$    & 21.0$_{-2.8}^{+2.8}$    & 10.37 & 0.98(322)\\
                 & R2 & 1.3$_{-0.5}^{+0.5}$ & 0.217$_{-0.014}^{+0.014}$ & 3.7$_{-0.6}^{+0.6}$    & 45.7$_{-3.7}^{+5.6}$    & 13.64 & 1.02(319)\\
                 & D1 & $<0.8$              & 0.216$_{-0.014}^{+0.014}$ & 3.3$_{-0.5}^{+0.7}$    & 31.0$_{-4.0}^{+4.0}$    & 11.89 & 1.02(294)\\
                 & D2 & 0.5$_{-0.4}^{+0.4}$ & 0.215$_{-0.011}^{+0.011}$ & 3.2$_{-0.5}^{+0.8}$    & 16.6$_{-2.8}^{+2.8}$    & 10.09 & 0.90(323)\\
                 & D3 & 0.7$_{-0.5}^{+0.3}$ & 0.182$_{-0.010}^{+0.010}$ & 3.0$_{-0.8}^{+1.1}$    & 12.9$_{-3.6}^{+3.6}$    &  8.85 & 1.14(341)\\
                 & D4 & 0.9$_{-0.4}^{+0.4}$ & 0.174$_{-0.009}^{+0.009}$ & 3.1$_{-1.1}^{+2.5}$    &  8.2$_{-2.8}^{+2.8}$    &  8.25 & 1.12(312)\\
                 & D5 & 1.2$_{-0.4}^{+0.4}$ & 0.162$_{-0.011}^{+0.011}$ & 1.6$_{-0.4}^{+0.9}$    &  9.2$_{-3.8}^{+4.0}$    &  7.97 & 0.93(352)\\
                 & AF & 0.6$_{-0.2}^{+0.2}$ & 0.165$_{-0.005}^{+0.005}$ & 3.2$_{-0.5}^{+0.7}$    & 10.8$_{-1.4}^{+1.5}$    &  8.55 & 1.16(477)\\
\hline
\end{tabular}
~~\\
{\it Note:} LS stands for light curve segments, FS stands for
``Flare Segments'', $N_H$ is in unit of $10^{20}$ cm$^{-2}$, abundances $Z$ is in solar unit, temperatures (kT) are in keV, emission measures (EM) are in $10^{52}$ cm$^{-3}$, and X-ray luminosity (\lx) is in $10^{30}$ \lum.  $\chi_\nu^2$ is the minimum reduced $\chi^2$ and DOF stands for degrees of freedom. 
~~\\
* Spectral parameters of hot component in APEC 3T fit
\end{table}

\begin{figure}
\includegraphics[width=90mm]{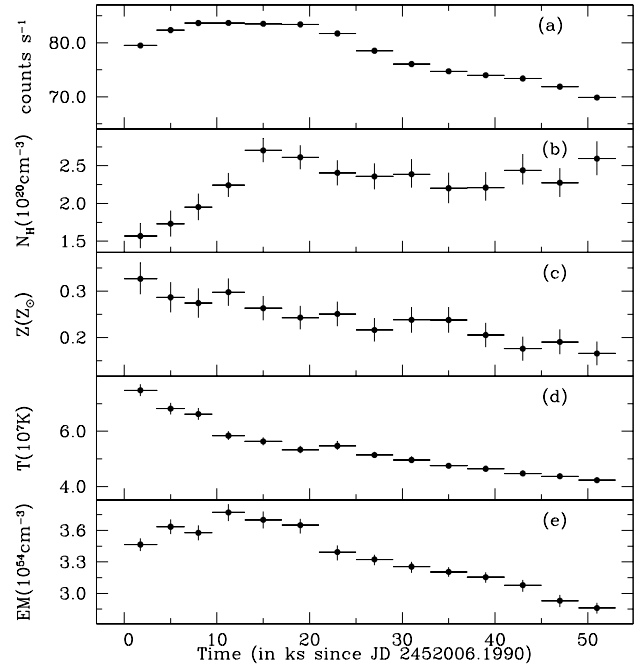}
\caption{ Variation of (a) count rate, (b) hydrogen column density (N$_H$), (c) abundance (Z), (d) temperature (T) and (e) emission measure (EM) during the XMM-Newton observations during the flare F2  from $\sigma$ Gem. The temperature and smission measure are corresponding to the hottest component of the best fit 3T-plasma model.}
\label{fig:sigmagemfevol}
\end{figure}

\begin{figure}
\includegraphics[width=90mm]{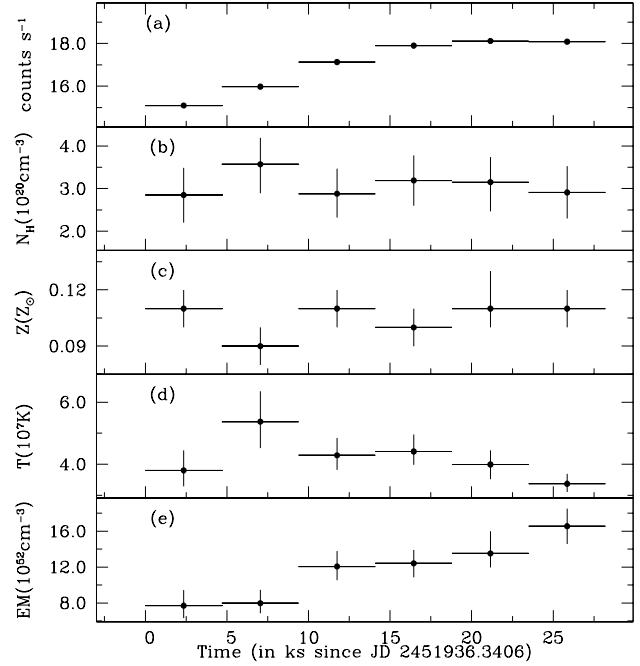}
\caption{Similar to Fig. \ref{fig:sigmagemfevol} but for flare F3 from star $\lambda$ And}
\label{fig:lambdaandfevol}
\end{figure}

\begin{figure}
\includegraphics[width=90mm]{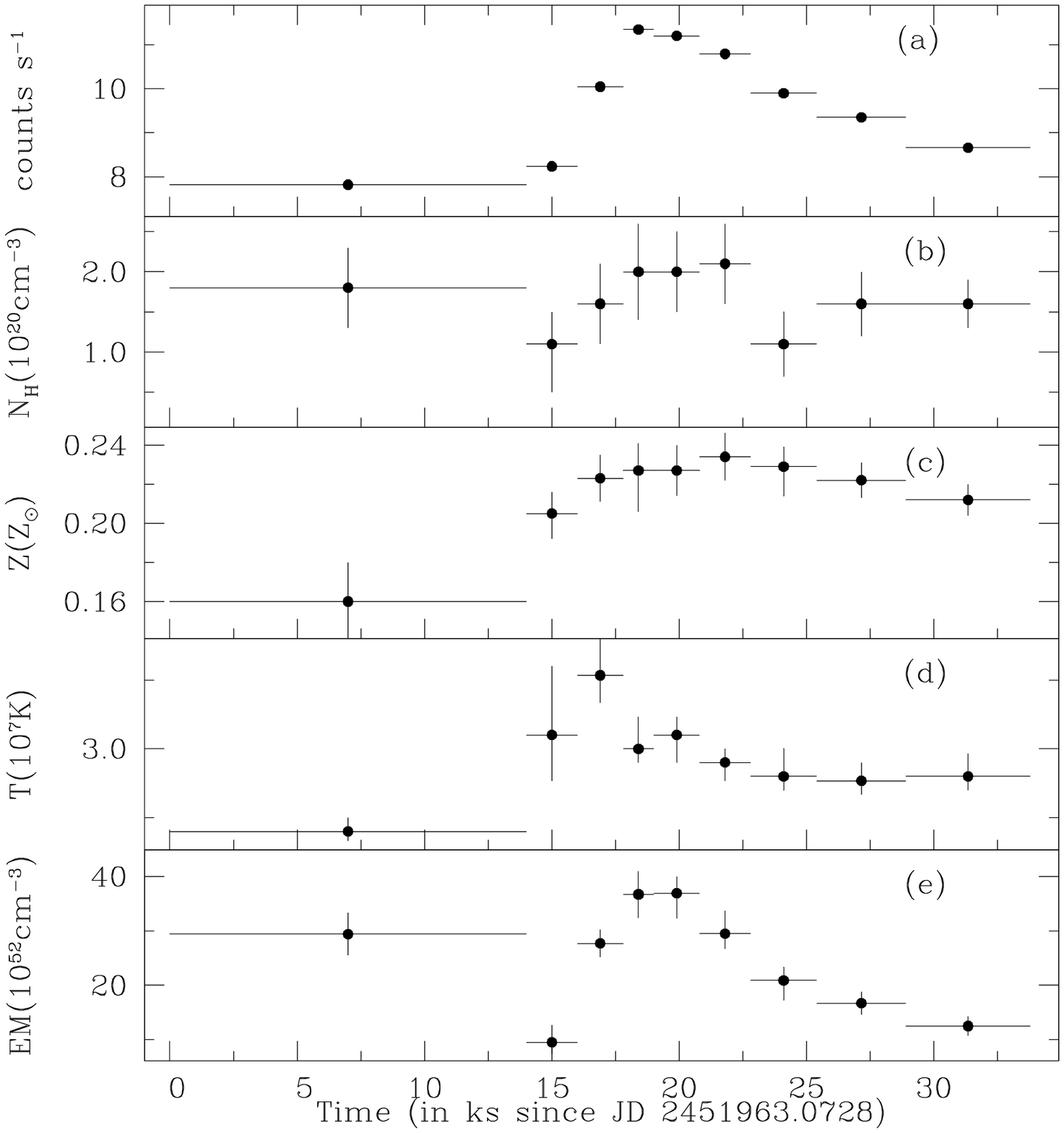}
\caption{Similar to Fig. \ref{fig:sigmagemfevol} but for the flare F5 from star V711 Tau}
\label{fig:v711taufevol1}
\end{figure}

\begin{figure}
\includegraphics[width=90mm]{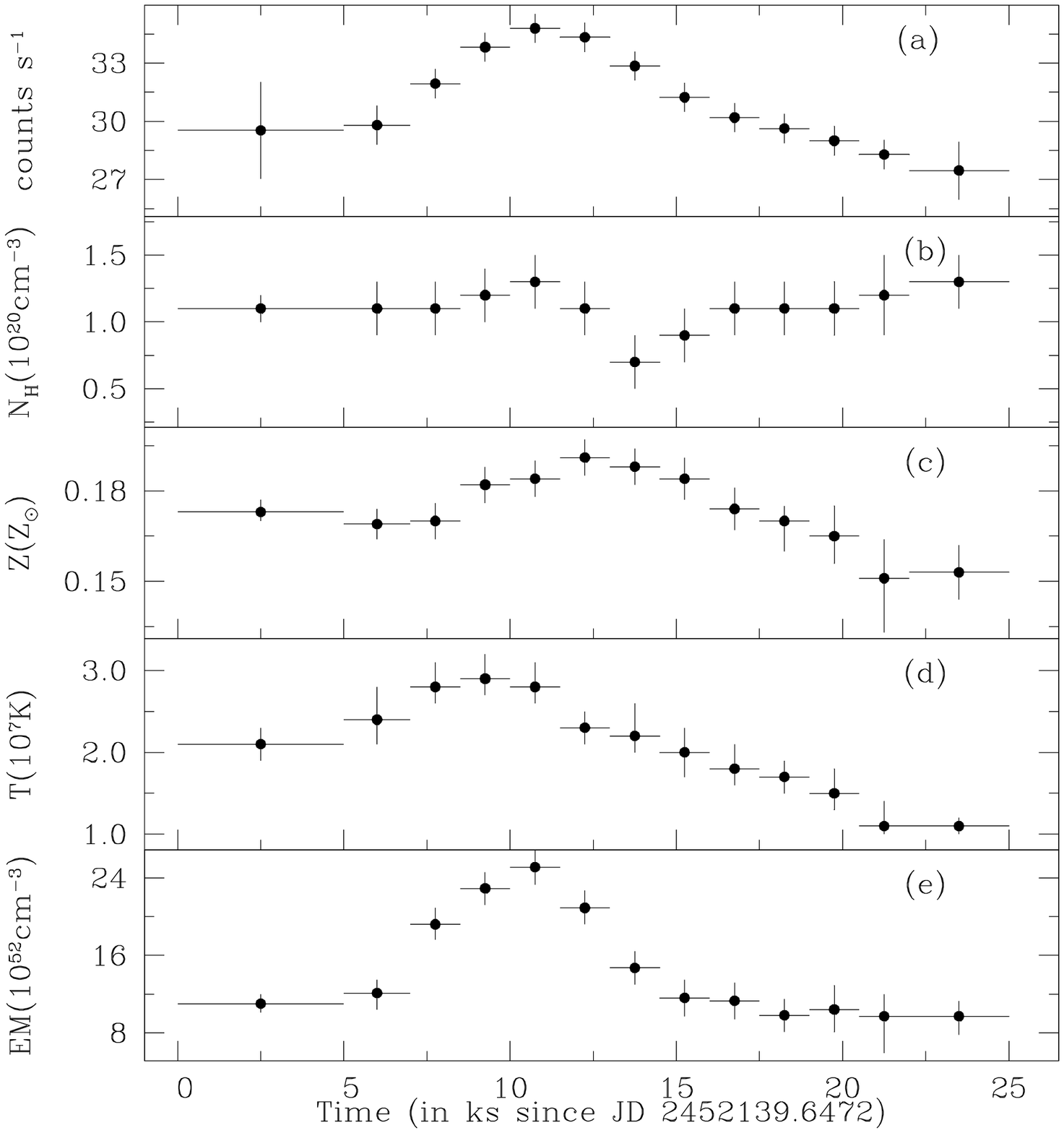}
\caption{Similar to Fig. \ref{fig:sigmagemfevol} but for the flare F6 from star V711 Tau}
\label{fig:v711taufevol}
\end{figure}

\begin{figure}
\includegraphics[width=90mm]{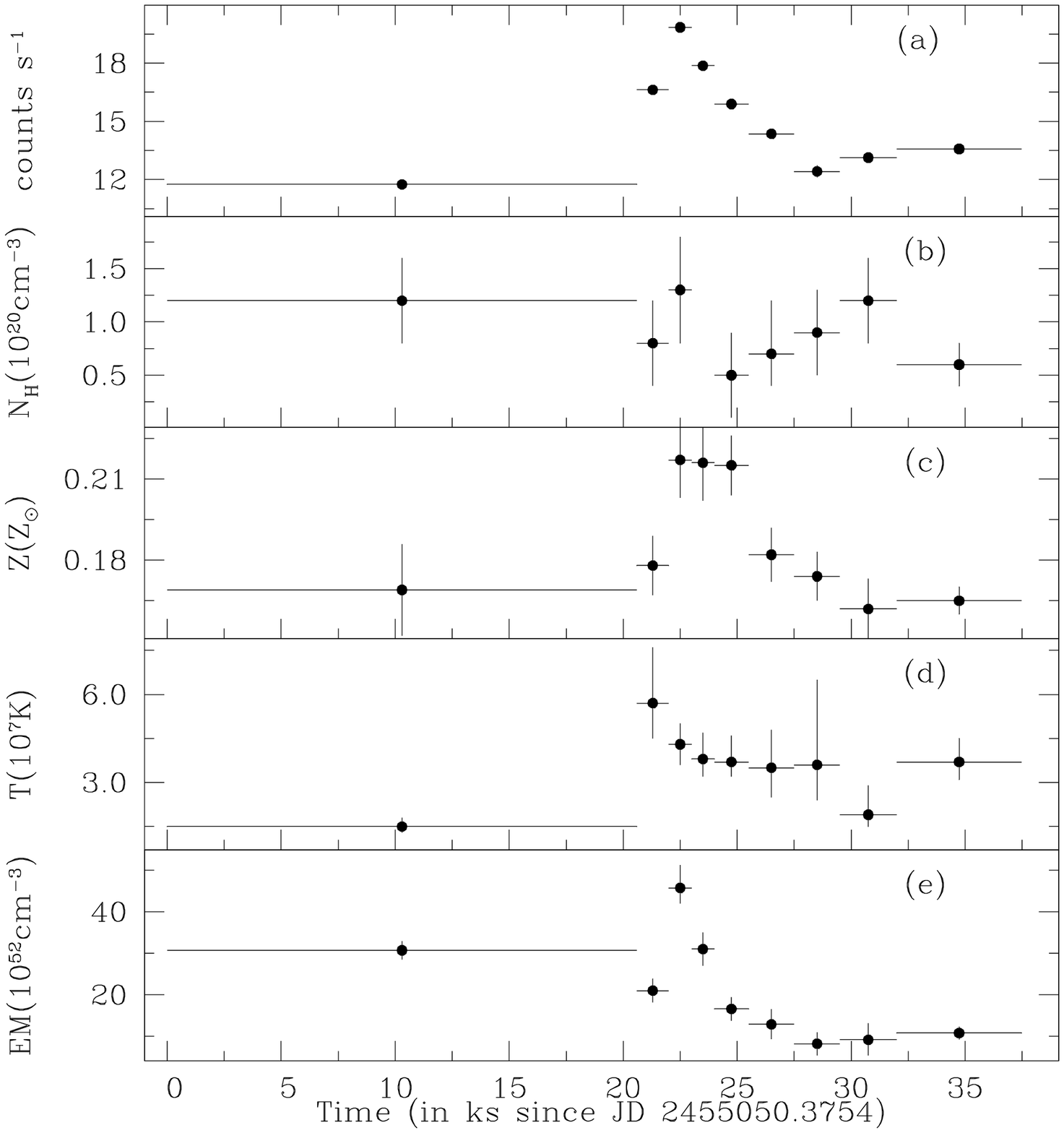}
\caption{Similar to Fig. \ref{fig:sigmagemfevol} but for the flare F7 from star EI Eri}
\label{fig:eierifevol}
\end{figure}

\subsection{Loop modeling}
In order to trace the coronal structure responsible for the flares, we employed time dependent hydrodynamic model of Reale et al. (1997) for a single loop and a simplified approach as given by Haisch (1983).  The details of hydrodynamic flare model (Reale et al. 1997, 2004 and Reale 2007) are given in Pandey \& Singh (2008). This model includes both plasma cooling and the effect of heating during flare decay.  The empirical formula for the estimation of the unresolved flaring loop length has been derived as

\begin{equation}
L_{hmd}  = \frac{\tau_{d}\sqrt{T_{max}}}{3.7\times10^{-4} f(\zeta)}   ~~~~~~~~ f(\zeta) \geq 1 \\
\label{eq:lhmd}
\end{equation}

\noindent
where $L_{hmd}$ is loop half-length in unit of cm,  $T_{max}$ is the maximum
temperature in unit of K, $\tau_{d}$ is the decay time of the light curve, and $f(\zeta)$ is a non-dimensional correction factor larger than one. $\zeta$ is  the slope of the decay path in the density-temperature (n-T) diagram (Sylwester et al. 1993) and is maximum ($\sim 2$) if heating is negligible and minimum ($\sim 0.5$) if heating dominates the decay.  The correction factor $f(\zeta)$  and loop maximum temperature are calibrated for  the XMM-Newton EPIC spectral response and are given as (Reale 2007):
\begin{equation}
f(\zeta) = \frac{0.51}{\zeta - 0.35} + 1.36 ~~~~~~\rm{ (for ~0.35 < \zeta \leq 1.6); }\\
\label{eq:fzeta}
\end{equation}
The average temperatures of the loops, usually lower than the real loop maximum temperatures, are found from the spectral analysis of the data. The loop maximum temperature for EPIC instruments is calibrated as
\begin{equation}
T_{max} = 0.13 T_{obs}^{1.16}
\label{eq:tmax}
\end{equation}

The loop length derived from this method is purely based on the decay phase of the flare. Alternatively, Reale (2007) derive the semiloop length from the rise phase and peak phase of the flare as:

\begin{equation}
L_{hmr} \approx  9.5\times10^2 t_M \frac{T_{max}^{5/2}}{T_M^2}
\label{eq:lhmr}
\end{equation}

\noindent
where $L_{hmr}$ is the loop half-length in unit of cm,  $T_{max}$ is the maximum temperature in unit of K, $T_M$ is the temperature at density maximum and $t_M$ is the time in unit of s at which density maximum occurs.  The time of the maximum emission measure is a good proxy for $T_M$.  Haisch (1983) suggested a simplified approach which is based on  quasi static radiative and conductive cooling during the earliest phases of flare decay. Given an estimate of two measured quantities, the emission measure ($EM_{ave}$) and the decay time scale of the flare ($\tau_d$ ) Haisch's approach (Haisch 1983) leads to the following expression for loop length ($L_{ha}$)

\begin{equation}
L_{ha} = 5\times10^{-6} EM_{ave}^{1/4}\tau_d^{3/4}
\label{eq:lha}
\end{equation}

\noindent

Once the length of the flaring loop has been derived, it is possible to infer additional physical parameters of the flaring plasma. These parameters could be electron density ($n_e$), flaring volume ($V$), maximum pressure ($p$), heating rate per unit volume ($E_H$) and minimum magnetic field (B) to confine the plasma.  For details, we refer to our previous paper (Pandey \& Singh 2008). The results of the model parameters are given in Table \ref{tab:loop} and explained below.

\begin{figure*}
\centering
\subfigure[Flare F1]{\includegraphics[height=8.0cm,angle=-90]{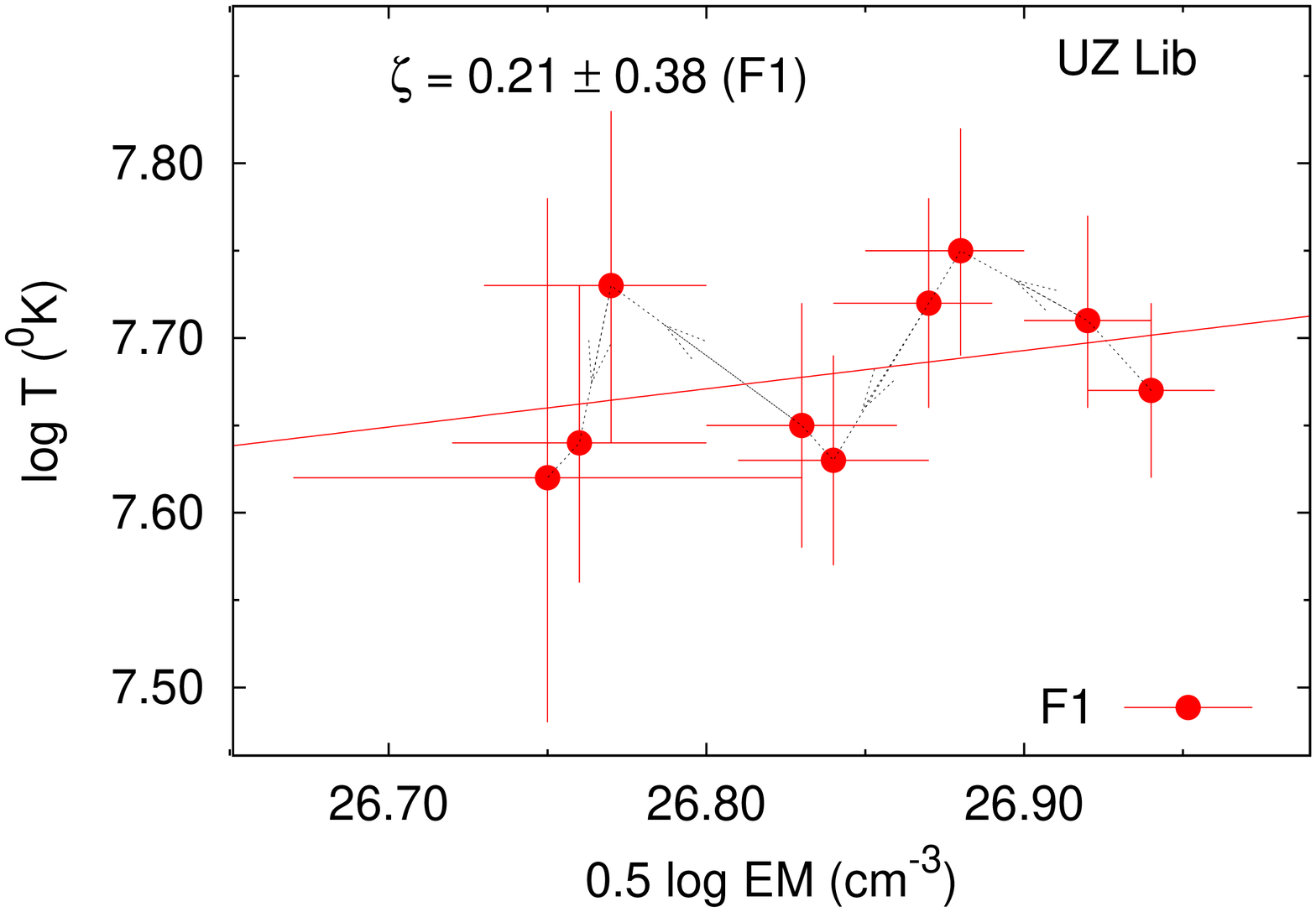}}
\subfigure[Flare F2]{\includegraphics[height=8.0cm,angle=-90]{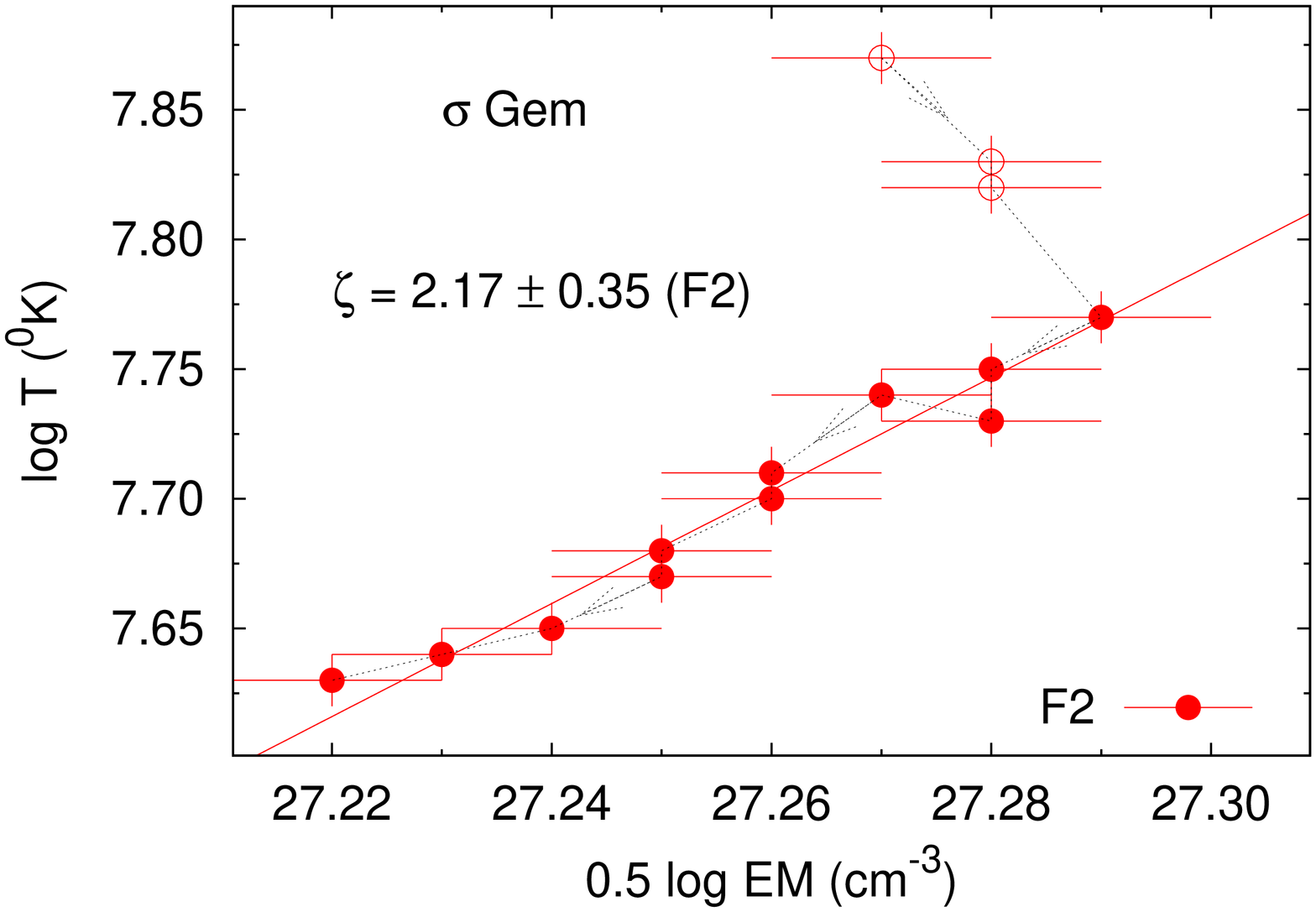}}
\subfigure[Flare F5]{\includegraphics[height=8.0cm,angle=-90]{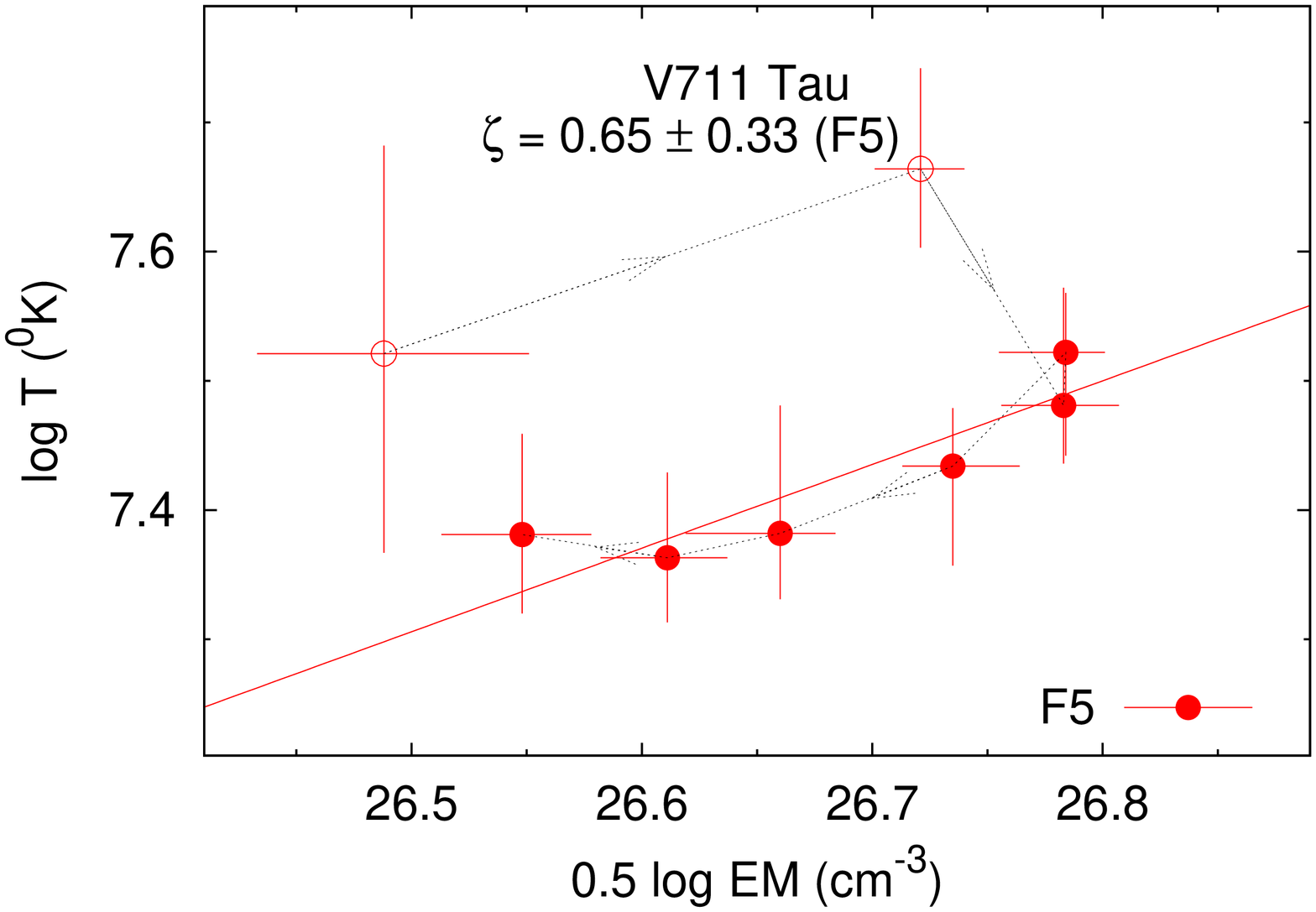}}
\subfigure[Flare F6]{\includegraphics[height=8.0cm,angle=-90]{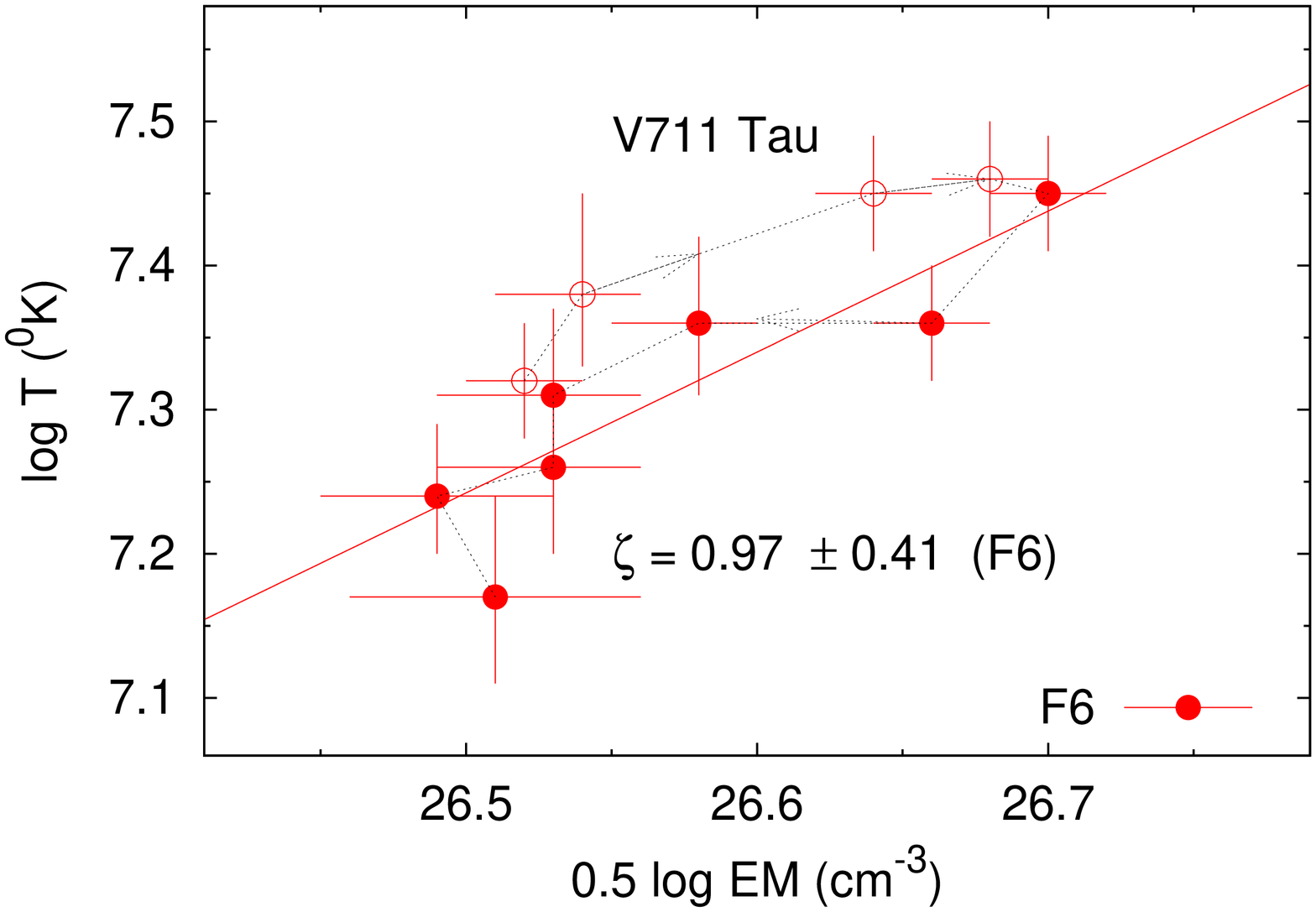}}
\subfigure[Flare F7]{\includegraphics[height=8.0cm,angle=-90]{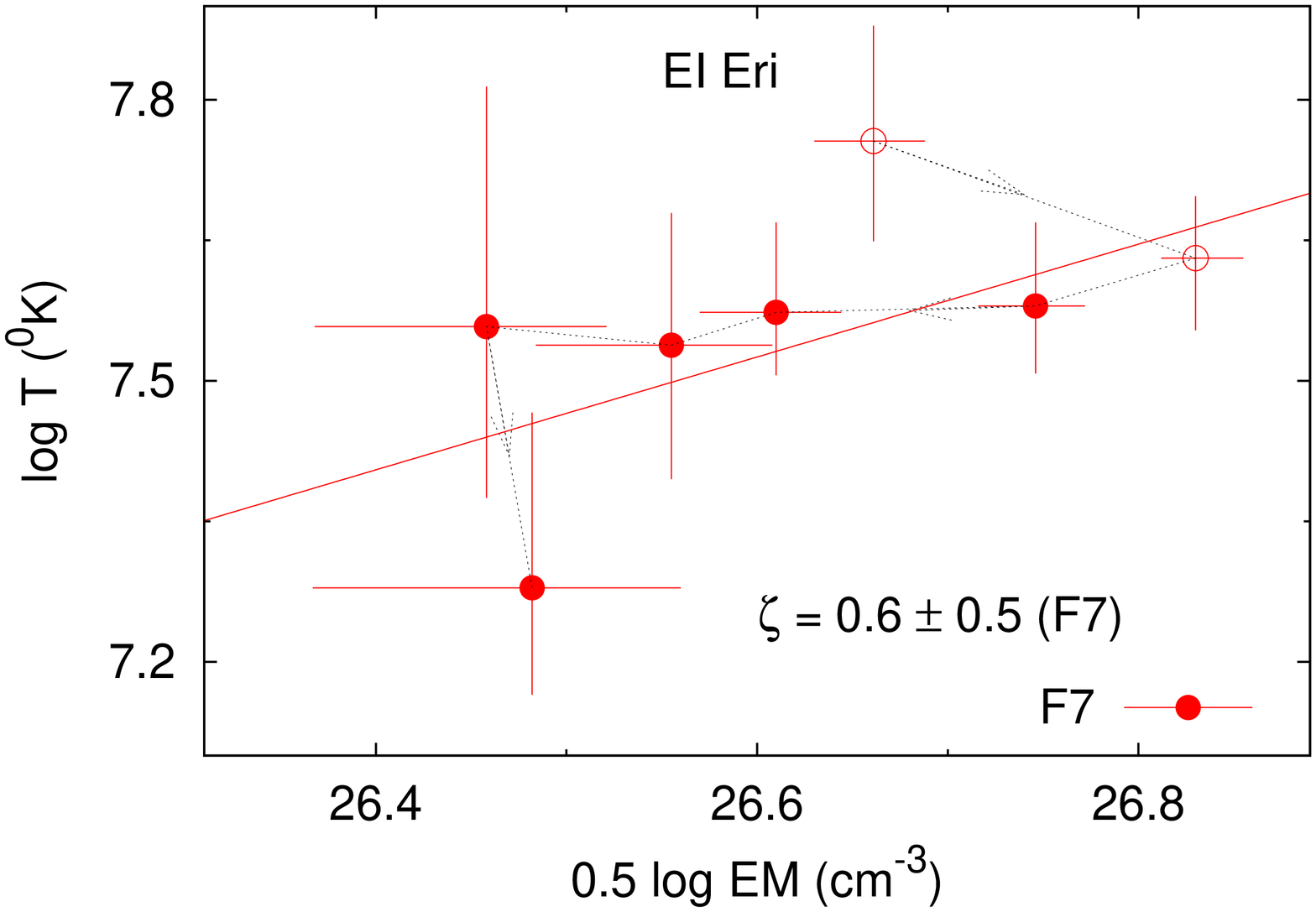}}
\caption{The density-temperature (n-T) diagram during the decay phase of the flares, where EM$^{1/2}$ 
has been used as a proxy of density. Symbols for the each flares are given at the bottom right corner.
Straight lines represent the best linear fit to the data corresponding to the decay phase. $\zeta$ is
slope of n-T diagram. The dotted lines with arrows show the path of n-T diagram.}
\label{fig:nt}
\end{figure*}

\begin{table*}
\centering
\caption{Parameters derived for flares.}
\label{tab:loop}
\begin{tabular}{lrccccccc}
\hline
\hline
Object(FN ~$\rightarrow$) & (1) &UZLib(F1) &$\sigma$ Gem(F2) &$\lambda$ And(F3)&V711 Tau (F4)  &V711 Tau(F5)    &V711Tau(F6)   &EI Eri(F7)    \\
Parameters ($\downarrow$) &                &                 &                 &               &                &              &             \\
\hline                                                                                                          
$\zeta  $   & (2) &$0.21\pm0.38$           &$2.17\pm0.33$    &                 & ...           &$0.65\pm0.33$   &$0.97\pm0.41$ &$0.6\pm0.5$ \\
$  T_{max}$ & (3) &$128\pm21$              &$177\pm6$        &$121\pm24$       & $98\pm32$     &$101\pm17$      &$59  \pm5   $ &$129\pm30$   \\
$  T_{M}$   & (4) & ...                    &$133\pm4    $    &$70\pm7    $     & ...           &$61\pm11$       &$56  \pm5   $ &$93\pm17$    \\
$  EM_{ave}$& (5) &$50.1\pm3.6$            &$336\pm6$        &$11.8\pm1.7$     & $4.2\pm0.8$   &$23.8\pm1.0$    &$14.8\pm0.6$  &$20.7\pm1.3$     \\
$  t_M     $& (6) &...                     &$>11250$         &$25850\pm2350$   & ...           &$5900\pm900$    &$5750\pm750$  &$1900\pm500$\\
$  L_{hmd} $& (7) &$ <18.2    $            &$<64.3      $    &....             & ...           &$<10.6$         &$6.5\pm1.7$   &$<5.5$    \\
$  L_{hmr} $& (8) &...                     &$>25.5\pm4.1$    &$79.7\pm18.3$    & ...           &$11.9\pm3.2$    &$4.6\pm0.8$   &$3.9\pm1.5$     \\
$  L_{ha}  $& (9) &$23.4\pm0.7$            &$51.8\pm0.4$     &...              & $2.8\pm0.3$   &$9.9\pm0.2$     &$7.5\pm0.1$   &$5.2\pm0.1$     \\
$p$         &(10) &$>4.3      $            &$>3.1$           &$0.8\pm0.5$      & $13\pm12$     &$>3.6     $     &$1.2\pm0.4$   &$>14$    \\
$ n_e^*$    &(11) &$>1.2      $            &$>0.6$           &$0.24\pm0.07$    & $4.6\pm1.6$   &$>1.3     $     &$0.7 \pm0.2 $ &$>3.9$      \\
$V^*$       &(12) &$<54$      &            $<919$            &$293$            & $0.3$         &$<22.3$         &$49.8       $ &$<2.9$        \\
$ E_h$      &(13) &$>0.73$                 &$0.18$           &$0.03$           & $12.3$        &0.94            &$0.38       $ &$ 7.8$       \\
$B$         &(14) &$328$                   &$279$            &$141$            & 571           &300             &$173        $ &593          \\
\hline
\end{tabular}
\\
(1) FN is flare name,
(2) $\zeta$ is slope of decay path in density-temperature diagram,
(3) $T_{max}$  is the maximum temperature in the loop at the flare peak in  unit of $10^7$ K based on spectral fit and equation (\ref{eq:tmax}),
(3) $T_{M}$  is the temperature in the loop at the density peak in  unit of $10^7$ K based on spectral fit and equation (\ref{eq:tmax}),
(4) $EM_{ave}$  is the average emission measure during the flare in unit of $10^{52}$ cm$^{-2}$,
(5) $t_M$  is the temperature at which density peaks in unit of seconds
(7,8,9) $L_{hmd}$, $L_{hmr}$ and $L_{ha}$ are lengths of the flaring loop in  unit of $10^{10}$ cm and are determined using equations (\ref{eq:lhmd}), (\ref{eq:lhmr}), and (\ref{eq:lha}),
(10) $p$ is the maximum pressure in the loop at the flare peak in unit of $10^3$ dyne ~cm$^{-2}$ and derived from RTV law, $T_{max} = 1.4 \times 10^3 (pL)^{1/3}$ (Rosner, Tucker \& Vaiana 1978),
(11) $n_e$ is the maximum electron density in the loop at the flare peak in unit of $10^{11}$ cm$^{-3}$, assuming that hydrogen plasma is totally ionised ($p = 2n_ekT_{max}$),
(12) $V$  is the volume of flaring plasma in unit of $10^{30}$ cm$^{3}$ and estimated using the equation $EM = n_e^2 V$, where EM is the peak emission measure.
(13) $E_h$  is the heating rate per unit volume at the flare peak in the unit ergs s$^{-1}$ cm$^{-3}$ estimate from the loop scaling law, $E_H \approx 10^{-6} T_{max}^{3.5} L^{-2}$ .
(14) $B$ is the minimum magnetic field necessary for confinement in Gauss and can be simply estimated as $p = B^2/8\pi$.

(*) The  pressure as derived here is a maximum value, appropriate for the equilibrium condition, which can be reached only for very long-lasting heating. The density and volume values derived thereby are therefore be treated as upper and lower limits, respectively.
\end{table*}

\subsubsection{UZ Lib: Flare F1}
The maximum temperature was determined using the equation \ref{eq:tmax} and found to be $128\pm21$ MK for the flare F1.  Fig. \ref{fig:nt} (a) shows the evolution of the flares  in n-T diagram with  a least-square fit to the decay phase, where the EM$^{1/2}$ has been used as a proxy of density. The path of n-T diagram is shown by dotted line with arrows. The continuous straight line shows the best linear fit to the corresponding data during the decay phase, providing  a slope $\zeta$. The value of the $\zeta$ indicates the presence of sustained heating during the decay phase of the flare F1.  The value of $\zeta$ for the flare F1 is less than the lower asymptotic value for which the equation \ref{eq:lhmd} can be applied. Therefore, an upper limit for the loop length can be derived by using the  value of $\zeta$ at the high extreme of the error bar. The resulting loop length  using the hydrodynamic loop model based on decay phase are estimated to be $<18.2\times10^{10}$ cm ($\sim$ 0.2\rstar) for the flare F1.  The  loop lengths derived from hydrodynamic method based on decay phase is found to be $\sim$ 1.3 times smaller than that derived from Haisch approach (see Table 7). The estimated loop lengths were found to be much smaller than the pressure scale height\footnote{$h_p \sim 5000 T_{max}/(g/g_\odot)$} ($h_p$  $\sim 5\times10^{13}$ cm). The maximum pressure in the loop at the flare peak, estimated from the loop scaling law (Rosner, Tucker \& Vaiana 1978) was found to be $> 4300$ dyne cm$^{-2}$ for the flare F1. Assuming that the hydrogen plasma is totally ionised ($p= 2 n_e k T_{max}$), the maximum plasma density in loop at the flare peak is estimated to be $>1.2 \times 10^{11}$ cm$^{-3}$ for the flare F1.  Using the observed peak emission measure the loop volume was computed to be of the order of $< 5.4\times10^{31}$ cm$^{3}$. A hint for the heat pulse intensity comes from the flare maximum temperature. By applying the loop scaling laws (Rosner, Tucker \& Vaiana 1978; Kuin \& Martens 1982) and loop maximum temperature, the  heating rate per unit volume ($E_h$) for the flare F1 would be $>0.73$ erg cm$^{-3}$ s$^{-1}$, respectively. From the derived pressure of the flare plasma, the minimum magnetic field required to confine the plasma should be 328 G.

\subsubsection{$\sigma$ Gem: Flare F2 }
The maximum temperature of the flare F2 was derived to be $177\pm6$ MK, which is
the highest among all the observed flares. The slope of n-T diagram indicates that the sustained 
heating was negligible during the decay phase of the flare. As seen from Fig. {\ref{fig:mospnlc}}
 (c) and Fig. {\ref{fig:sigmagemfevol}}, the quiescent states before and after the flare were not
 observed. Therefore, the larger value of the $\zeta$ could be due to incomplete flare evolution.
 The value of $\zeta$ for the flare F2 is outside the domain of the validity of method, therefore,
  loop length of $< 64.3 \times 10^{10}$ cm  was derived by  using $\zeta = 1.6$ (the maximum allowable value). Alternatively, the loop length can be derived using the equation ({\ref{eq:lhmr}}). As seen in
 Fig. {\ref{fig:mospnlc}}(c), the entire rise phase of the flare F2 was not observed, therefore,
 the value of $t_M$ was determined to be in  a lower limit. Thus, the loop length derived to be
 $>25.5\pm4.1 \times 10^{10}$ cm, which is $\sim 2.5$ times smaller than that derived from the
 hydrodynamic decay model. The loop length derived from hydrodynamic method based on peak and
 rise phase is also found to be $\sim 2$ times smaller than that from Haisch approach (see Table
 {\ref{tab:loop}}).  The derived loop length from equation ({\ref{eq:lhmd}}) may be over estimated
 as the value of $\zeta$ is at the upper extreme end. However, loop length derived from hydrodynamic
 method based on rise phase may be under estimated as the rise phase of flare was not observed
 completely.  Loop length derived from each method is less than the stellar radius (i.e. L $\sim$ 0.3 - 0.8\rstar). For further parameter estimation, we have used the loop length   derived from hydrodynamic method, which is much
less than the $h_p$ (=$7\times10^{13}$ cm). The estimated loop parameters are
given in Table \ref{tab:loop}.

\subsubsection{$\lambda$ And: Flare F3}
In the case of the flare F3, the loop length could be derived only from the hydrodynamic model based of the rise and peak phases (see equation \ref{eq:lhmr}) as no decay phase was observed.  The derived loop length (=$79.7\pm 18.3 \times 10^{10}$ cm $\approx$ 1.5 \rstar) is less than the pressure scale height (=$5\times10^{13}$ cm). The maximum temperature was estimated  to be $121\pm24$ MK. The maximum pressure, density and volume are estimated to be $800\pm500$ dyne cm$^{-2}$, $2.4\pm0.7\times10^{10}$ cm$^{-3}$ and $2.93\times10^{32}$ cm$^{3}$, respectively. The heating rate per unit volume was estimated to be smaller than that estimated for other observed flares.

\subsubsection{V711 Tau: Flares F4, F5 and F6}
For the flare flares F4, F5 and F6 the loop maximum temperatures were found to be $98\pm32$, $101\pm17$ and $59\pm5$ MK, respectively. The maximum temperature observed during the flare F6  is the lowest among all the other observed flares. The slopes of decay path of n-T diagram for the flares F5 and F6 (see Figs. \ref{fig:nt}c and \ref{fig:nt}d) indicate the presence of a strong sustained heating showing that the observed decay paths were driven by time evolution of heating process. For the flare F4, loop length could be derived only from Haisch approach and the derived loop length was $2.8\pm0.3\times 10^{10}$ cm ($\sim$ 0.2/0.1\rstar). For the flare F5, the loop lengths derived from each of the methods are  consistent within $1\sigma$ level.   Adopting the loop length derived from  hydrodynamic decay method (i.e. $<$0.8/0.4 \rstar) and the  peak emission measure of $37^{+3}_{-5} \times 10^{52}$ cm$^{-3}$ for the flare F5, the maximum pressure, density and volume were estimated to be $>3600$ dyne cm$^{-2}$, $>1.3\times10^{11}$ cm$^{-3}$ and $2.23\times10^{31}$ cm$^{3}$, respectively. A 300 G of minimum magnetic field was estimated to confine the plasma in the flaring loop for the flare F5. For the flare F6, the loop length derived from hydrodynamic decay method is within $ \sim 1\sigma$ level to that derived from hydrodynamic rise method. However,  loop length derived from Haisch approach is more than that from hydrodynamic methods. Loop  parameters for the flare F6 were derived by adopting the loop length of $6.5\pm1.7 \times10^{10}$ cm ($\sim$ 0.24/0.5\rstar) and are given in  Table \ref{tab:loop}.  The loop length derived for each flare is less then the pressure scale height ($\sim 4 - 6 \times 10^{12}$ cm). 

\subsubsection{EI Eri: Flare F7}
Fig. \ref{fig:nt}(e) shows the n-T diagram for the flare F7, where the slope of $0.6\pm0.5$ implies the presence of sustained heating during the decay. The loop length based on   equation \ref{eq:lhmd} was derived by using the upper extreme value of $\zeta=1.1$ as $\zeta$ at lower extreme  of the error is outside the validity of the method. Thus  the loop length was derived to be $<5.5 \times 10^{10}$ cm ($\sim$ 0.23\rstar), which is similar to that derived from equation \ref{eq:lhmr}.   The loop lengths derived from hydrodynamic methods based on rise and peak phases  is smaller than that derived from the Haisch approach (see Table \ref{tab:loop}). The derived loop length is smaller than the pressure scale height. The maximum temperature of the loop was derived to be $129\pm30$ MK. Using the loop length of $5.5\times10^{10}$ cm and peak emission measure of $4.57\times10^{53}$ cm$^{-3}$, the maximum pressure, density and volume of the flaring plasma were estimated to be $>1.4\times10^4$ dyne cm$^{-2}$, $>3.9\times10^{11}$ cm$^{-3}$ and $2.9\times10^{30}$ cm$^3$, respectively. A minimum magnetic field of 593 G was required to confine the plasma at the peak of flare.

\section{Discussion and Conclusions}
\begin{figure}
\includegraphics[width=60mm, angle=-90]{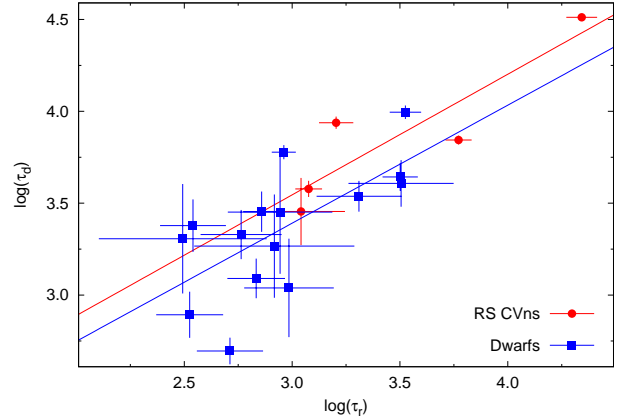}
\caption{Plot between log(\tr) and log(\td) for flares from a sample of RS CVns (solid circles) and G-K dwarfs (solid squares). The lines represents the best fit log-linear correlations.
The values of \td ~and \tr ~for G-K dwarfs are taken from Pandey\& Singh (2008).  }
\label{fig:tdtr}
\end{figure}

We have carried out  a detailed analysis of seven X-ray flares observed with XMM-Newton from five cool  subgiants and giants in RS CVn binaries. Except the flare F4, the decay  time  ($>1$ h) and the total energy released ($1.2-8.0\times 10^{32}$ \lum) indicate that these flares  belong to a class of  long decay flares (Pallavicini et al. 1990). However, excluding the flare F2 of $\sigma$ Gem, a strong sustained heating was found to be present during the decay of all flares.  The strong sustained heating during the flare decay indicates  that the decay light curves is dominated by the temporal heating profile and not by free decay of the heated loop.  Therefore, these flare can not be classified as long decay flares but may be classified as an arcade.  The decay of these observed flares are smaller than that of few pre-main-sequence stars, where the decay time was found upto few days (Favata et al. 2005, Getman et al. 2008). But, the observed flare decay time is comparable or more than that of the flare observed in the late type G-K dwarfs, other RS CVns and the flare dMe stars (e.g. Pandey \& Singh 2008, Schmitt 1994, Osten \& Brown 1999).  Fig. \ref{fig:tdtr} shows that correlation between the $\tau_d$ and $\tau_r$ for the flares from RS CVns and G-K dwarfs. The straight lines show the linear regression fit in the form of $\tau_d \propto \tau_r^{0.67\pm0.18}$ for RS CVns and  $\tau_d \propto \tau_r^{0.64\pm0.20}$ for G-K dwarfs. These relations are similar to the relation predicted for standard magnetic reconnection model i.e. $\tau_d \propto \tau_r^{0.5}$ ( Petschek 1964, Shibata \& Yokoyama 1999,2002, Imanishi et al. 2003), where it is assumed that the decay phase is dominated by radiative cooling.

There were two flares, F2 from $\sigma$ Gem and F5 from V711 Tau that showed a sustained level of approximately constant emission at the peak of the flare. In the past such scenario was  seen only for a small number of flares from V711 Tau, $\sigma^2$ CrB and AT Mic (e.g. Agrawal et al. 1986, Osten \& Brown 1999, Raassen et al., 2003). The appearance of plateaus could be due to the reconnection process, where reconnection balances the loop expansion to sustain a constant luminosity.

 During the flares observed  in a sample of RS CVns, the peak luminosities were in the range $10^{30.8}$ $<$ \lx ~$<$ $10^{31.8}$ \lum ~indicating that these flares are more powerful than the flares observed from dwarf stars ( $10^{28.7}$ $<$ \lx $<$ $10^{30.2}$; Pandey \& Singh 2008) but are less powerful than the flares  from PMS stars ($10^{31}$ $<$ \lx $<$ $10^{33}$ \lum; Getman et al. 2008). This confirms the previous finding that  the large flares on active RS CVn binaries  involve small energies in comparison to the total stellar luminosity. However, few flares have been reported earlier from dMe stars, where peak X-ray flare luminosity reached upto 30 per cent of bolometric luminosity (e.g. GINGA EQ 1839.6+8002 -Pan et al. 1997; EV Lac -Favata et al. 2000c;  AB Dor - Maggio et al. 2000).  The total energy released during the X-ray flares observed in our sample was found to be of the order of $10^{35-37}$ erg. This implies that these flares are $10^{2-4}$ more energetic than the flares observed in dMe and G-K dwarfs (Pandey \& Singh 2008, Pallavicini 1990), but are as energetic as flares observed in some of pre-main-sequence stars (Favata et al 2005; Getman et al. 2008). The energy released during  flares  are also comparable to that from the superflare on the RS CVn binary II Peg ($10^{36.8}$; Osten et al. 2007).

It appears that the quiescent coronae of the stars UZ Lib, $\sigma$ Gem and $\lambda$ And are represented by two temperature plasma. However, the quiescent coronae of V711 Tau and EI Eri are represented by three temperature plasma. The coolest and the hottest  temperature of quiescent coronae were found to be $\sim$ 0.3 - 0.8 and  $\sim$ 1.2-2.5 keV, respectively. These temperatures are more than that derived for RS CVns by Dempesey et al. (1993) using ROSAT PSPC data. They found the bimodal distribution of temperatures centered near 0.17  and 1.38 keV.  The quiescent coronae of RS CVn -type stars appears to be hotter than the  quiescent coronae of cool dwarf stars.  The time resolved spectroscopy of these flares shows the coronal spectrum during the flare can be represented with a 1T model, with quiescent state taken into account as a frozen background contribution.
In the case of  $\sigma$ Gem, the $N_H$ was found to vary along the rise and decay phases of flare light curve. Enhancement of hydrogen column density, associated with flaring events, have been observed in the past in Proxima Cen (Haisch et al. 1983), V773  Tau (Tsuboi et al. 1998), Algol (Ottmann \& Schmitt 1996; Favata \& Schmitt 1999), and LQ Hya (Covino et al. 2001), and usually interpreted as due to Coronal Mass Ejection (CME). As shown in Fig \ref{fig:sigmagemfevol} the hydrogen column density increased as flare rise, peaked and then showed decreasing trend as flare decayed. Such a behavior is compatible with the picture of a cloud of material  being ejected from the flaring region at the beginning of the event, and subsequently thinning out as it expands and cools into the space. Solar observations indicate  a tight relationship between CME and flares (Vr{\v s}nak et al. 2004, Goplaswamy 2010). However, there are large number of solar flares without CME associations (Gopalswamy et al. 2009). Rodon{\`o} et al. (1999) suggesting that extra-absorption might occur in some stars due to the presence of neutral hydrogen in circumstellar environment.

The metal abundances were also found to vary along the flare light curves in the flares observed from $\sigma$ Gem, V711 Tau and EI Eri (see Figs. \ref{fig:sigmagemfevol}, \ref{fig:v711taufevol1}, \ref{fig:v711taufevol} and \ref{fig:eierifevol}). Using the RGS data of flares F2 from $\sigma$ Gem,  Nordon \& Beher (2008) show that abundances enriched during the flare. A change of the metal abundance during a strong flare was detected for the RS CVn like stars II Peg (Mewe et al. 1997), UX Ari (G\"udel et al. 1999) and $\sigma^2$ Crb (Osten et al. 2003), Algol (Ottmann \& Schmitt 1996; Favata \& Schimtt 1999), dwarf stars EV Lac (Favata et al. 1998; 2000a), CN Leo (Liefke et al. 2010) and AB Dor (G\"udel et al. 2001), and  premainsequence star V773 Tau (Tsuboi et al. 1998). In these cases abundances vary with an initial rise from a quiescent state value of $Z \approx 0.3Z_\odot$ to $Z = 1 Z_\odot$, and a subsequent decay to the pre-flare value. Our results are in general consistent with the previously observed behavior. In case of $\sigma$ Gem the metal abundances go up as material is ejected,  as is seen in the simultaneously increasing $N_H$. The most likely interpretation for the  abundances variations observed during the flares is that at the beginning of the event fresh chromospheric material is evaporated in the flaring loops, thus enhancing its abundances; once the material is brought in the flaring coronal structure the fractionation mechanism which is responsible for lower abundances observed in coronal plasma would start operating, bringing the coronal abundances back to quiescent state value (see also Favata \& Micela 2003). 

For most of the flares, the temperature and the emission measure show well-defined trends i.e. the changes in the temperature and the EM are correlated with the variations observed in the light curves during the flares. Similar trends were also seen during the flares detected in the cool pre-main-sequence and main sequence stars (Favata et al. 2005; Giardino et al. 2006; Pandey \& Singh 2008).  The peak flare temperature was found in the range of 59 - 177 MK for all the flares. These values are similar to those observed from active stars like pre-main-sequence stars (T$_{max}$ $\sim$  80-­270 MK; Favata, Micela \& Reale 2001; Favata et al. 2005; Getman et al. 2008), Algol (T$_{max} \sim $ 70-150 MK; van den Oord \& Mewe 1989, Favata et al. 2000c), AB Dor (T$_{max} \sim$  140-170 MK; Maggio et al. 2000), other RS CVns (T$_{max} \sim $ 25-120MK, G\"udel 2004) but are more than  those found in the flares from the dMe star and G-K dwarfs ($\sim$ 10-50 MK; Reale \& Micela 1998; Favata et al. 2000b;  Briggs \& Pye 2003; Pandey \& Singh 2008 ). The flare peak temperatures in our sample were also found to be comparable or more than that derived for a superflare on II Peg (118-152 MK; Osten et al. 2007). In these flares either both the emission measure and the temperature peaked simultaneously or the temperature evolved before the emission measure. A similar delay is often observed both in the solar and stellar flares (Sylwester et al. 1993; van den Oord \& Mewe 1989; Favata et al. 2000a; Maggio et al. 2000; Stelzer et al. 2002). A delay between the temperature and the emission measure peak  indicates a coherent plasma evolution and therefore a flare occurring inside a single loop, or at least the presence of a dominant loop early in the flare. 

 For all the flares, the estimated maximum electron density under assumption of a totally ionized hydrogen plasma was found in the order of 10$^{10-11}$ cm$^{-3}$ . This is comparable to the values expected from a plasma in coronal conditions (Landini et al. 1986).

 The observed flares are modeled using the hydrodynamic model based on the decay time and the rise time, and Haisch's approach. For the flares F5, F6 and F7, the loop lengths derived from the hydrodynamic model based on the decay time are found to be consistent within $1\sigma$ to that from the rise time and  peak phase method. In the case of flare F2 of $\sigma$ Gem, however, the loop lengths from the two methods were found to be inconsistent. This could be due to the involvement of several different coronal loops during the decay of the flare or the heat pulse triggering the flare may not a top hat function as is assumed normally. During the flares F1, F5, F6 and F7, the decay paths are driven by sustained heating, which indicate that  a flare is involving a loop arcade or is a two-ribbon flare. It may then occur that the flare begins in a loop and propagates to other loops only progressively, i.e. at later times. Therefore, we may obtain a significant discrepancy between the estimation made from the rise phase and the one from the decay phase. Generally, the loop lengths derived from Haisch approach were  found be more than that derived from hydrodynamic models. However, in most of the cases difference between the loop lengths obtained from three methods is well within 1$\sigma$ level. In the case of flare F1, the loop length derived from Haisch approach are significantly larger than that derived from the hydrodynamic model. Covino et al. (2001) have applied one version of the Haisch's approach to a number of X-ray flares on the stars. They report that the loop lengths of these flares derived from the simplified Haisch approach is in general agreement with the method based on detailed hydrodynamic calculations. However, Favata et al. (2001) showed that the large loop sizes obtained by quasi-static analysis are in reality likely to be significantly larger than the hydrodynamic model, typically by factors of $\gtrsim$ 3.

The derived loop lengths presented in Table 7 can be compared to the loop lengths derived from  other stars. The present analysis shows that RS CVn-type binaries have larger coronal structure which are comparable to or less than their  radii. The log of loop length (log L) in our sample ranges from 10.4 to 11.9 .  These loop lengths are similar to or more than that observed for the flares from other active  RS CVns like binaries e.g. UX Ari (log L = 11.0-11.7 ;  Franciosini et al. 2001, G\"{u}del et al 1999), $\sigma^2$ Crb (log L = 10-10.17 ; Osten et al. 2003, van den Oord et al. 1988, Agrawal et al. 1986), II Peg (log L=10.46-11 ; Mewe et al. 1997, Doyle et al. 1991, Tagliaferri et al. 1991). However, these flares are comparable to the flares observed from Algol (log L = 10.3-11.9 e.g. Favata et al. 2000c, van den Oord \& Mewe 1989, White et al. 1986). If we compare these loop lengths with the flares from much younger stars, it was noticed that the flaring structures are larger in RS CVn binaries than that  from the main-sequence stars (log L =9.8-10.6; Pandey \& Singh 2008), and are comparable or smaller  than the flaring structures from pre-mainsequence stars (log L$\sim$10-12; Favata et al. 2005, Getman et al. 2008).  This  implies that RS CVns  have large coronal magnetic activity. The high activity level of RS CVns is generally due to their fast rotation, itself a consequence of tidal locking of the stellar rotation period with the orbital period of binary.

The large sizes of flaring loops (larger than the component star) indicate that the coronal structures in active binaries are large, and would typically span both stars. For active RS CVn binaries, Uchida \& Sakurai (1983) suggested  the idea of inter-binary loops, with magnetic structures linking the two stars.  We have compared the loop length with the distance between binary components and found that the flaring loop lengths are much smaller then the  binary separation for the star V711 Tau ($\sim$ 3\rstar ~for circular orbit; \rstar is radius of more active component in binary system; Fekel 1983). For the largest flare from V711 Tau (F6), the loop length is  $\sim$ 13 per cent of  binary separation. In the case of flare F7 from EI Eri, the loop length is one tenth of binary separation ($\sim$ 3 \rstar ~for circular orbit; Washuettl et al. 2009).  For other binary stars, we could not compare the loop lengths with  binary separations as  the separations between two components are not known, only the semi-major axis of one component is known (see Eker et al. 2008).    The origin of flares in the close binary systems having starspots are interpreted in terms of the  reconnection of the magnetic flux tubes of one component with the emerging and submerging pairs of spots on the other component (Uchida \& Sakurai 1985). However, the flares from V711 Tau and EI Eri  appear to originate very close to the stars and extended to only a small fraction of binary separation. In the literature, it has been also noticed that corona of even the most active stars is quite compact and does not extend to large (larger than stellar radius) distance from the star itself (e.g. Brickhouse et al. 2001).

At the flare peak, the maximum X-ray luminosity must be lower than the total energy rate ($H = E_h.V$) to satisfy the energy balance relation for the flaring plasma as a whole. The rest of the input energy is used for thermal conduction, kinetic energy and radiation at lower frequencies. For the flares under study the maximum X-ray luminosity observed is about 40-60 per cent of H.  These values are more than  those reported for the solar flares, where the soft X-ray radiation only accounts for less than 20 per cent of total energy (Wu et al. 1986). In comparison, the fraction of X-ray radiation to the total energy has been found to be 15 to 35 per cent for the G, K and M dwarfs (Favata et al. 2000a,b; Reale et al. 2004, Pandey \& Singh 2008). 

Applying loop scaling law $V = 2\pi \beta^2 L^3$,  and if the detected flares are produced by single loop, their aspect ratio ($\beta$) were estimated in the range 0.1 to 0.2 for all the flares.  Similar aspect ratio was also observed for solar and other stellar coronal loops
for which typical values of $\beta$ are in the range of 0.1 - 0.3 (e.g. Shimizu 1995).

We have derived the flare rates for RS CVn-types and G-K dwarfs. Following Stelzer et al. (2000),  the flare rate (F) is estimated as
\begin{equation}
F = \frac{{\bar{\tau}} N}{T_d}\pm\frac{\sigma_{\bar{\tau}} \sqrt{N}}{T_d}
\end{equation}

\noindent
where ${\bar{\tau}} = \sqrt{{\bar{\tau}}_d^2+{\bar{\tau}}_r^2}$ and $\sigma_{\bar{\tau}} = \sqrt{\sigma_{{\bar{\tau}}_d}^2 + \sigma_{{\bar{\tau}}_r}^2 }$ are the mean value and uncertainty in e-folding decay and rise times derived with ASURV's Kaplan-Meier estimator, N is number of flares, $T_d$ is total duration of observations. For a sample of RS CVn-type stars, the ${\bar{\tau}}_r$ and ${\bar{\tau}}_d$ are determined to be $8709\pm1341$ and $4168\pm917$, respectively. Total number of flares observed from the sample of RS CVn are 7 and total duration of the observations is 462 ks. Thus the flare rates for a sample of RS CVn stars was found to be $19\pm2$ per cent. Similarly, using our previous paper (Pandey \& Singh  2008) the flare rate for the G-K dwarfs was found to be $18.5\pm0.3$ per cent. The flare rates for pre-mainsequence stars in $\rho$ Oph star forming region  and Taurus molecular cloud were determined to be $15\pm1$ and  9 per cent, respectively (Imanishi et al. 2003; Stelzer et al. 2007). Surprisingly, we obtained the larger flare rates for RS CVns and G-K dwarfs. Generally,  it was shown that the younger stars tend to have more frequent flaring activity (see Imanishi et al. 2003).  Larger flare rate for RS CVns with respect to young stars may be result of longer rise and decay times, and limited number of observations.  


\section*{Acknowledgments}
This work uses data obtained by XMM-Newton, an ESA science mission with instruments and contributions directly funded by ESA Member States and the USA (NASA). JCP thanks Rachel Osten for careful reading  of our manuscript and providing  constructive criticisms.  We thank the referee of the paper for his/her useful suggestions.

\end{document}

%% file: def.tex
\def\deg{\hbox{$^\circ$}}              
\def\lum{$\rm{erg}~\rm{s}^{-1}$}       
\def\flu{$\rm{erg}~\rm{s}^{-1}\rm{cm}^{-2}$}       
\def\lx{$\rm{L}_{X}$}                 
\def\rstar{$\rm{R}_{\star}$}              
\def\cts{counts s$^{-1}$}
\def\td{$\tau_{d}$}
\def\tr{$\tau_{r}$}
\def\ften{\hbox{10$^{-11}$}}              

%% file: ms.bbl
\begin{thebibliography}{1000}
\bibitem[\protect\citeauthoryear{Agrawal \& Vaidya}{1988}]{1988MNRAS.235..239A} Agrawal P.~C., Vaidya J., 1988, MNRAS, 235, 239
\bibitem[\protect\citeauthoryear{Agrawal, Rao, \& Riegler}{1986}]{1986MNRAS.219..777A} Agrawal P.~C., Rao A.~R., Riegler G.~R., 1986, MNRAS, 219, 777 
\bibitem[]{}Anders E., Grevesse N., 1989, Geochim. Cosmochim. Acta, 53, 197
\bibitem[]{}Arnaud K. A., 1996, ASPC, 101, 17
\bibitem[\protect\citeauthoryear{Audard et al.}{2001}]{2001A&A...365L.329A} Audard M., Behar E., G{\"u}del M., Raassen A.~J.~J., Porquet D., Mewe R., Foley C.~R., Bromage G.~E., 2001, A\&A, 365, L329
\bibitem[\protect\citeauthoryear{Ayres et al.}{2003}]{2003ApJ...583..963A} Ayres T.~R., Brown A., Harper G.~M., Osten R.~A., Linsky J.~L., Wood B.~E., Redfield S., 2003, ApJ, 583, 963
\bibitem[\protect\citeauthoryear{Bopp \& Stencel}{1981}]{1981ApJ...247L.131B} Bopp B.~W., Stencel R.~E., 1981, ApJ, 247, L131
\bibitem[]{}Briggs K. R.,  Pye J. P., 2003, MNRAS, 345, 714
\bibitem[]{}Brown J. M., Brown A., 2006, ApJ, 638L, 37
\bibitem[]{}Covino S., Panzera M. R., Tagliaferri G., Pallavicini, R., 2001, A\&A, 371, 973
\bibitem[]{}Dempsey R. C., Linsky J. L., Schmitt J. H. M. M., Fleming T. A., 1993, ApJ. 413, 333
\bibitem[]{}Dennis B. R., Schwartz R. A., 1989, SoPh, 121, 75
\bibitem[]{}Dickey J. M., Lockman F. J., 1990, ARAA, 28, 215
\bibitem[]{1999MNRAS.302..457D} Donati J.-F., 1999, MNRAS, 302, 457
\bibitem[]{}Doyle J. G., et al.,  1991, MNRAS, 248, 503 
\bibitem[]{1997A&AS..123..209D} Duemmler R., Ilyin I.~V., Tuominen I., 1997, A\&AS, 123, 209
\bibitem[\protect\citeauthoryear{Favata et al.}{1998}]{1998IAUC.6987....3F} Favata F., Sciortino S., Micela G., Maggio A., 1998, IAUC, 6987, 3
\bibitem[]{}Favata F., Schimtt J. H. M. M., 1999, A\&A, 350, 900
\bibitem[\protect\citeauthoryear{Favata et al.}{2000}]{2000A&A...353..987F} Favata F., Reale F., Micela G., Sciortino S., Maggio A., Matsumoto H., 2000a, A\&A, 353, 987
\bibitem[\protect\citeauthoryear{Favata, Micela, \& Reale}{2000}]{2000A&A...354.1021F} Favata F., Micela G., Reale F., 2000b, A\&A, 354, 1021
\bibitem[\protect\citeauthoryear{Favata et al.}{2000}]{2000A&A...362..628F} Favata F., Micela G., Reale F., Sciortino S., Schmitt J.~H.~M.~M., 2000c, A\&A, 362, 628
\bibitem[]{}Favata F., Micela G., Reale F., 2001, A\&A, 375, 485
\bibitem[]{}Favata F., Micela G., 2003, Space Science Review, 108, 577 
\bibitem[]{}Favata F., Flaccomio E., Reale F., Micela G., Sciortino S., Shang H., Stassun K.~G., Feigelson E.~D., 2005, ApJS, 160, 469
\bibitem[\protect\citeauthoryear{Fekel}{1983}]{1983ApJ...268..274F} Fekel F.~C., Jr., 1983, ApJ, 268, 274
\bibitem[]{}Franciosini E., Pallavicini R., Tagliaferri G., 2001, A\&A, 375, 196
\bibitem[\protect\citeauthoryear{Getman et al.}{2008}]{2008ApJ...688..418G} Getman K.~V., Feigelson E.~D., Broos P.~S., Micela G., Garmire G.~P., 2008, ApJ, 688, 418
\bibitem[]{}Giardino G., Favata F., Silva B., Micela G., Reale F., \& Sciortino S., 2006, A\&A, 453, 241
\bibitem[\protect\citeauthoryear{Gondoin}{2003}]{2003A&A...400..249G} Gondoin P., 2003, A\&A, 400, 249
\bibitem[]{}Gopalswamy, N., 2010, 20th National Solar Physics Meeting, Eds. I.~Dorotovi), pages 108-130,
\bibitem[]{}Gopalswamy, N., Akiyama, S.\& Yashiro, S.,2009, IAU symposium, 257, 283
\bibitem[\protect\citeauthoryear{Grewing, Bianchi, \& Garrido}{1989}]{1989A&A...223..172G} Grewing M., Bianchi L., Garrido R., 1989, A\&A, 223, 172
\bibitem[\protect\citeauthoryear{Griffiths \& Jordan}{1998}]{1998ApJ...497..883G} Griffiths N.~W., Jordan C., 1998, ApJ, 497, 883
\bibitem[\protect\citeauthoryear{G{\"u}del et al.}{1999}]{1999ApJ...511..405G} G{\"u}del M., Linsky J.~L., Brown A., Nagase F., 1999, ApJ, 511, 405
\bibitem[\protect\citeauthoryear{G{\"u}del et al.}{2001}]{2001A&A...365L.336G} G{\"u}del M., et al., 2001, A\&A, 365, L336
\bibitem[\protect\citeauthoryear{G{\"u}del}{2004}]{2004A&ARv..12...71G} G{\"u}del M., 2004, A\&ARv, 12, 71 
\bibitem[\protect\citeauthoryear{G{\"u}del \& Naz{\'e}}{2009}]{2009A&ARv..17..309G} G{\"u}del M., Naz{\'e} Y., 2009, A\&ARv, 17, 309
\bibitem[\protect\citeauthoryear{Haisch et al.}{1983}]{1983ApJ...267..280H} Haisch B.~M., Linsky J.~L., Bornmann P.~L., Stencel R.~E., Antiochos S.~K., Golub L., Vaiana G.~S., 1983, ApJ, 267, 28
\bibitem[\protect\citeauthoryear{Haisch}{1983}]{1983ASSL..102..255H} Haisch B.~M., 1983, in IAU Colloq. 71, Activity in Red-Dwarf Stars, Vol. 102, ed. P. B. Byrne \& M. Rodon{\`o} (Dordrecht: Reidel), 255
\bibitem[\protect\citeauthoryear{Imanishi et al.}{2003}]{2003PASJ...55..653I} Imanishi K., Nakajima H., Tsujimoto M., Koyama K., Tsuboi Y., 2003, PASJ, 55, 653
\bibitem[]{}Jansen F. et al., 2001, A\&A, L365, 1
\bibitem[]{}Kuin N. P. M., Martens P. C. H., 1982, A\&A, 108, 1
\bibitem[]{}Landini M., Monsignori Fossi B. C., Pallavicini R.,  Piro L., 1986, A\&A, 157, 217
\bibitem[\protect\citeauthoryear{Liefke, Fuhrmeister, \& Schmitt}{2010}]{2010A&A...514A..94L} Liefke C., Fuhrmeister B., Schmitt J.~H.~M.~M., 2010, A\&A, 514, A94
\bibitem[]{}Maggio A., Pallavicini R., Reale F., \& Tagliaferri, G., 2000, aap, 356, 627
\bibitem[\protect\citeauthoryear{Majer et al.}{1986}]{1986ApJ...300..360M} Majer P., Schmitt J.~H.~M.~M., Golub L., Harnden F.~R., Jr., Rosner R., 1986, ApJ, 300, 360
\bibitem[]{} Martens P.~C.~H., Kuin N.~P.~M., 1989, SoPh, 122, 263
\bibitem[\protect\citeauthoryear{Mewe et al.}{1997}]{1997A&A...320..147M} Mewe R., Kaastra J.~S., van den Oord G.~H.~J., Vink J., Tawara Y., 1997, A\&A, 320, 147
\bibitem[\protect\citeauthoryear{Nordgren et al.}{1999}]{1999AJ....118.3032N} Nordgren T.~E., et al., 1999, AJ, 118, 3032
\bibitem[\protect\citeauthoryear{Nordon, Behar, {\ G&uuml}del}{2006}]{2006A&A...446..621N} Nordon R., Behar E., G{\"u}del M., 2006, A\&A, 446, 621
\bibitem[\protect\citeauthoryear{Nordon \& Behar}{2008}]{2008A&A...482..639N} Nordon R., Behar E., 2008, A\&A, 482, 639
\bibitem[\protect\citeauthoryear{Ol{\'a}h, Strassmeier, \& Weber}{2002}]{2002A&A...389..202O} Ol{\'a}h K., Strassmeier K.~G., Weber M., 2002, A\&A, 389, 202
\bibitem[\protect\citeauthoryear{Ortolani et al.}{1997}]{1997A&A...325..664O} Ortolani A., Maggio A., Pallavicini R., Sciortino S., Drake J.~J., Drake S.~A., 1997, A\&A, 325, 664
\bibitem[]{} Osten R. A., Brown A., 1999, ApJ, 515, 746
\bibitem[\protect\citeauthoryear{Osten et al.}{2000}]{2000ApJ...544..953O} Osten R.~A., Brown A., Ayres T.~R., Linsky J.~L., Drake S.~A., Gagn{\'e} M., Stern R.~A., 2000, ApJ, 544, 953
\bibitem[\protect\citeauthoryear{Osten et al.}{2003}]{2003ApJ...582.1073O} Osten R.~A., Ayres T.~R., Brown A., Linsky J.~L., Krishnamurthi A., 2003, ApJ, 582, 1073
\bibitem[]{} Osten R. A., Drake S., Tueller J., Cummings J.,  Perri M., Moretti A., Covino S., 2007, ApJ, 654, 1052O
\bibitem[]{} Ottmann R.,  Schmitt J. H. M. M., 1996, A\&A, 307, 813
\bibitem[]{} Padmakar, Pandey S. K., 1999, A\&AS, 138, 203
\bibitem[]{} Pallavicini R., Tagliaferri G.,  Stella L., 1990, A\&A, 228, 403
\bibitem[]{} Pallavicini R., Serio S.,  Vaiana G. S., 1977, ApJ, 216, 108
\bibitem[]{} Pan H. C., Jordan C., Makishima K., Stern R. A., Hayashida K., Inda-Koide M., 1997, MNRAS, 285, 735
\bibitem[\protect\citeauthoryear{Pandey \& Srivastava}{2009}]{2009ApJ...697L.153P} Pandey J.~C., Srivastava A.~K., 2009, ApJ, 697, L153
\bibitem[\protect\citeauthoryear{Pandey \& Singh}{2008}]{2008MNRAS.387.1627P} Pandey J.~C., Singh K.~P., 2008, MNRAS, 387, 1627
\bibitem[\protect\citeauthoryear{Pasquini, Schmitt, \& Pallavicini}{1989}]{1989A&A...226..225P} Pasquini L., Schmitt J.~H.~M.~M., Pallavicini R., 1989, A\&A, 226, 225
\bibitem[\protect\citeauthoryear{Petschek}{1964}]{1964NASSP..50..425P}
Petschek H.~E., 1964, The Physics of Solar Flares, Proceedings of the AAS-NASA Symposium, Eds Wilmot N. Hess. Washington, D. C.,Vol. 50, page 425
\bibitem[\protect\citeauthoryear{Raassen et al.}{2003}]{2003A&A...411..509R} Raassen A.~J.~J., Mewe R., Audard M., G{\"u}del M., 2003, A\&A, 411, 509
\bibitem[]{} Reale F., Betta R., Peres G., Serio S.,  McTiernan J.,  1997, A\&A, 325, 782
\bibitem[]{} Reale F.,  Micela G., 1998, A\&A, 334, 1028
\bibitem[]{} Reale F., G\"{u}del M., Peres G.,  Audard M., 2004, A\&A, 416, 733
\bibitem[]{} Reale F., 2007, A\&A, 471, 271
\bibitem[\protect\citeauthoryear{Rodon{\`o} et al.}{1999}]{1999A&A...346..811R} Rodon{\`o} M., Pagano I., Leto G., Walter F., Catalano S., Cutispoto G., Umana G., 1999, A\&A, 346, 811
\bibitem[]{} Rosner R., Tucker W. H., Vaiana G. S., 1978, ApJ, 220, 643
\bibitem[\protect\citeauthoryear{Schmitt et al.}{1990}]{1990ApJ...365..704S} Schmitt J.~H.~M.~M., Collura A., Sciortino S., Vaiana G.~S., Harnden F.~R., Jr., Rosner R., 1990, ApJ, 365, 704
\bibitem[]{} Schimtt J. H. M. M., 1994, ApJS, 90, 735
\bibitem[\protect\citeauthoryear{Shi et al.}{1998}]{1998A&A...339..840S} Shi J.~R., Zhao G., Zhao Y.-H., You J.-H., 1998, A\&A, 339, 840
\bibitem[\protect\citeauthoryear{Shibata \& Yokoyama}{2002}]{2002ApJ...577..422S} Shibata K., Yokoyama T., 2002, ApJ, 577, 422
\bibitem[\protect\citeauthoryear{Shibata \& Yokoyama}{1999}]{1999ApJ...526L..49S} Shibata K., Yokoyama T., 1999, ApJ, 526, L49
\bibitem[]{} Shimizu,~T., 1995, PASJ, 47, 251
\bibitem[\protect\citeauthoryear{Singh et al.}{1987}]{1987MNRAS.224..481S} Singh K.~P., Slijkhuis S., Westergaard N.~J., Schnopper H.~W., Elgaroy O., Engvold O., Joras P., 1987, MNRAS, 224, 481
\bibitem[]{} Smith R. K., Brickhouse N. S., Liedahl D. A.,  Raymond J. C., 2001, ApJ, 556, L91
\bibitem[]{} Stelzer B., Neuh{\"a}user R., Hambaryan V., 2000, A\&A, 356, 949
\bibitem[]{} Stelzer B., Burwitz V., Audard M., G\"{u}del M., Ness J.-U., Grosso N., Neuh\"{a}user R., Schmitt J. H. M. M., Predehl P., Aschenbach, B., 2002, A\&A, 392, 585
\bibitem[\protect\citeauthoryear{Stelzer et al.}{2007}]{2007A&A...468..463S} Stelzer B., Flaccomio E., Briggs K., Micela G., Scelsi L., Audard M., Pillitteri I., G{\"u}del M., 2007, A\&A, 468, 463
\bibitem[]{} Strassmeier K. G., Hall D. S., Fekel F. C.,  Scheck M., 1993, A\&AS, 100, 173
\bibitem[]{} Str\"{u}der L.  et al., 2001, A\&A, 365, L18
\bibitem[]{} Sylwester B., Sylwester J., Serio S., Reale F., Bentley R. D., Fludra A., 1993, A\&A, 267, 586
\bibitem[]{} Tagliaferri G., White N. E., Doyle J. G., Culhane J. L., Hassall B. J. M., Swank J. H., 1991, A\&A, 251, 161
\bibitem[]{} Tsuboi Y., Koyama K., Murakami H., Hayashi M., Skinner S., Ueno S., 1998, ApJ, 503, 894
\bibitem[]{} Turner M. J. L. et al., 2001, A\&A, 365, L27
\bibitem[]{} van den Oord G. H. J.,  Mewe R., 1989, A\&A, 213, 245
\bibitem[\protect\citeauthoryear{van den Oord, Mewe, \& Brinkman}{1988}]{1988A&A...205..181V} van den Oord G.~H.~J., Mewe R., Brinkman A.~C., 1988, A\&A, 205, 181 
\bibitem[]{}Vr{\v s}nak, B. and Ru{\v z}djak, D., Sudar, D., \& Gopalswamy, N., 2004, A\&A, 423, 717
\bibitem[]{} Washuettl A., Strassmeier K. G., Granzer T., Weber M., Oláh K.,  2009, AN, 330, 27
\bibitem[]{} White N. E., Culhane J. L., Parmar A. N., Kellett B. J., Kahn S., van den Oord G. H. J., Kuijpers, J., 1986, ApJ, 301, 262
\bibitem[]{} Wu S. T. et al., 1986, in Energitic Phenomenon on the Sun, eds. M. Kundu\& B. Woodgate, No. 2439 in NASA Conference Publication, NASA, p. 5
\bibitem[\protect\citeauthoryear{Yi et al.}{1997}]{1997A&A...318..791Y} Yi Z., Elgaroy O., Engvold O., Westergaard N.~J., 1997, A\&A, 318, 791
\end{thebibliography}
